\documentclass[prl,aps,twocolumn,superscriptaddress,fleqn]{revtex4-1}
\usepackage{bm}
\usepackage{graphicx}

\graphicspath{{Figures/}}

\usepackage{amsmath,float}
\usepackage{amssymb,xcolor}
\usepackage[utf8]{inputenc}
\usepackage[T1]{fontenc}
\usepackage{color}
\usepackage{upgreek,physics}
\usepackage{subfigure}
\usepackage[unicode=true,colorlinks=true,citecolor=blue,urlcolor=black]{hyperref}
\usepackage{placeins}
\usepackage{lipsum}
\usepackage{epsfig}
\usepackage{makecell,microtype,fix-cm}
\usepackage[utf8]{inputenc}
\usepackage{wrapfig}
\usepackage[exponent-product=\cdot]{siunitx}
\usepackage{braket}

\usepackage[normalem]{ulem}

\newcommand{\nix}[1]{}

\usepackage{nicefrac} 			

\usepackage{tikz}

\usepackage{mhchem}
\usepackage{siunitx}

\usepackage{chngcntr,printlen}

\begin{document}

	\title{Terahertz spin ratchet effect in magnetic metamaterials}

	\author{M. Hild}
	\affiliation{Terahertz Center, University of Regensburg, 93040 Regensburg, Germany}
	
		\author{L. E. Golub}
	\affiliation{Terahertz Center, University of Regensburg, 93040 Regensburg, Germany}
	
	\author{A. Fuhrmann}
	\affiliation{Terahertz Center, University of Regensburg, 93040 Regensburg, Germany}
	
	\author{M. Otteneder}
	\affiliation{Terahertz Center, University of Regensburg, 93040 Regensburg, Germany}
	
	\author{M. Kronseder}
	\affiliation{Terahertz Center, University of Regensburg, 93040 Regensburg, Germany}
	
	\author{M. Matsubara}
	\affiliation{Department of Physics, Tohoku University, Sendai 980-8578, Japan}
	\affiliation{Center for Science and Innovation in Spintronics, Tohoku University, Sendai 980-8577, Japan}
	
	\author{T.~Kobayashi}
	\affiliation{Department of Physics, Tohoku University, Sendai 980-8578, Japan}
	
	\author{D. Oshima}
\affiliation{Department of Electronics, Nagoya University, Furo-cho, Chikusa-ku, Nagoya 464-8603, Japan}

	\author{A. Honda}
\affiliation{Institute of Materials and Systems for Sustainability, Nagoya University, Furo-cho, Chikusa-ku, Nagoya 464-8603, Japan}

	\author{T. Kato}
		\affiliation{Department of Electronics, Nagoya University, Furo-cho, Chikusa-ku, Nagoya 464-8603, Japan}
		\affiliation{Institute of Materials and Systems for Sustainability, Nagoya University, Furo-cho, Chikusa-ku, Nagoya 464-8603, Japan}

	\author{J. Wunderlich}
	\affiliation{Terahertz Center, University of Regensburg, 93040 Regensburg, Germany}
	\affiliation{Institute of Physics ASCR, v.v.i., Cukrovarnická 10, 162 00 Prague 6, Czech Republic}
			
	\author{C. Back}
	\affiliation{Technical University Munich, 85748 Garching, Germany}

	\author{S. D. Ganichev}
	\affiliation{Terahertz Center, University of Regensburg, 93040 Regensburg, Germany}
	\affiliation{CENTERA, Institute of High Pressure Physics PAS, 01142 Warsaw, Poland}

	\begin{abstract}
We report on  spin ratchet currents driven by terahertz radiation electric fields in a Co/Pt magnetic  metamaterial formed by   triangle-shaped holes forming an antidots lattice and subjected to an external magnetic field applied perpendicularly to the metal film plane. We show that for a radiation wavelength substantially larger than the period of the antidots array the radiation causes a polarization-independent spin-polarized ratchet current.  The  current is generated by the periodic asymmetric radiation intensity distribution caused by the near-field diffraction at the edges of the antidots, which  induces spatially inhomogeneous periodic electron gas heating, and a phase-shifted periodic asymmetric electrostatic force. The developed microscopic theory shows that the magnetization of the Co/Pt film results in a spin ratchet current caused by  both  the anomalous Hall and the anomalous Nernst effects. Additionally, we observed a polarization-dependent trigonal spin photocurrent, which is caused by the scattering of electrons at the antidot boundaries resulting in a spin-polarized current due to the magnetization. Microscopic theory of these effects reveals that the trigonal photocurrent is generated at the boundaries of the triangle antidots, whereas the  spin ratchet is generated due to the spatially periodic temperature gradient  over the whole film. This difference causes substantially different hysteresis widths of these two currents. 
\end{abstract}
	
	\maketitle
	
	\section{Introduction}
	\label{introduction}
	
In a system with spatial asymmetry, carriers can perform a directed motion in response to ac electric fields or thermal/quantum fluctuations, i.e. when it is driven away from thermal equilibrium by an additional deterministic or stochastic perturbation. This directed transport, generally known as ratchet effect, has a long history and is relevant for different fields of physics, chemistry and biology, for reviews see e.g. Refs.~\cite{Juelicher1997,Linke2002,Reimann2002,Haenggi2009, Ivchenko2011,Budkin2016a,Lau2020}. Ratchet effects, whose prerequisites are simultaneous breaking of both thermal equilibrium and spatial inversion symmetry, can be realized in a great variety of forms. Examples range from mechanical systems and molecular motors to electric transport in 1D semiconductor systems or metamaterials. Currently, ratchet systems already have fascinating ramifications in engineering and natural sciences. In metamaterials, conversion of high-frequency radiation into direct electric current due to the ratchet effect  has been demonstrated in various  two-dimensional semiconductor systems with periodic grating gate structures having an asymmetric configuration of the gate electrodes  with  period $d \ll \lambda$, where $\lambda$ is the radiation wavelength~\cite{Olbrich2009,Kannan2011,Olbrich2011,Nalitov2012, Otsuji2013,Drexler2013,Kurita2014,Budkin2014, Faltermeier2015,Olbrich2016,Popov2016,Fateev2019,Hubmann2020,BoubangaTombet2020, DelgadoNotario2020,Moench2022a,Tamura2022,Moench2022b}. Electric currents in response to a high-frequency electric field are detected in a wide range of temperatures (from room to helium temperatures) and frequencies (from tens of GHz to tens of THz) and can be excited both without and in  presence of an external magnetic field. In the latter case, it is called the magneto-ratchet effect.
		
It has been shown  theoretically~\cite{Scheid2006} and demonstrated experimentally~\cite{Costache2010} that ratchet effects can also drive pure spin currents and spin-polarized electric currents. These types of ratchet effects were named spin ratchets. Spin ratchet effects, which can open novel ways for efficient generation and control of spin fluxes~\cite{Flatte2008}, attracted growing attention. Since the first work, several different origins of spin ratchet currents have been suggested and discussed~\cite{Demikhovskii2006,Matsuno2007,Scheid2007b,Braunecker2007,Lin2008,Smirnov2008a,Liang2009,Smirnov2009,Scheid2010,Lu2010,Abdullah2014,Ang2015,Gomonay2016,Chen2019,Zangara2019,Huang2022,Velez2022}, for review see~\cite{Bercioux2015}. For example, it has been shown that a stationary spin current can be generated by applying an ac driving current to a symmetric or asymmetric periodic structure with Rashba spin-orbit coupling. Recently, the  magneto-ratchet effect resulting in a spin-polarized electric current has been excited by applying terahertz radiation to  structures with an asymmetric double grating gate (DGG), lateral superlattices fabricated on diluted magnetic semiconductor (DMS), and GaN two-dimensional electron systems~\cite{Faltermeier2017,Faltermeier2018,Sai2021}. The spin magneto-ratchet is caused by spin-orbit interaction together with the action of a magnetic field resulting in the Zeeman effect and is expected to be substantially enhanced  in DMS materials with a lateral superlattice made of ferromagnetic materials. Note that in the latter case, a novel concept for a spin-transistor has been realized~\cite{Betthausen2012}. These works show that spin ratchet and spin transport in lateral superlattices may yield a potentially important basis for practical devices. Hereby, magnetic metamaterials made of ferromagnetic metals are good candidates for realizing   spin ratchets. 
	
Most recently,  spin-sensitive currents have been observed in Co/Pt-based magnetic films with a lateral superlattice consisting of triangular antidots arising from excitation with visible and near-infrared light with $\lambda \approx d$~\cite{Matsubara2022}. The dc current has been excited in individual triangles due to  the magneto-photogalvanic effect~\cite{Belkov2008}. Arrays of triangular- and semidisc-antidots in nonmagnetic materials fabricated on GaAs-based high-electron-mobility transistor structures  were previously  used to demonstrate the ratchet effect excited by high-frequency radiation with fulfilled condition $d \ll \lambda$~\cite{Lorke1998,Chepelianskii2007,Sassine2008,Kannan2012,Motta2021}.	In this work, we used Co/Pt-based magnetic films with a lateral superlattice and extend the frequency range to terahertz (THz) radiation with a wavelength substantially larger than the period of the triangles and demonstrate that spin ratchets are efficiently excited in such structures  in the absence, as well as in the presence, of an external magnetic field. We demonstrate that  THz laser radiation results in a spin-polarized direct electric current consisting of two contributions: a polarization-independent ratchet current and a contribution  whose direction and magnitude are determined by the orientation of the THz electric field (linear magnetic ratchet effect). We show that the application of an external  magnetic field normal to the metal film results in a hysteretic behavior of both contributions to the  ratchet current. We develop a theory which fully describes all experimental findings. We demonstrate that the polarization-dependent spin ratchet current appears due to the trigonal symmetry of the individual antidots. The polarization-insensitive photocurrent caused by the Seebeck ratchet~\cite{Ivchenko2011} and spin ratchet effects become possible due to the reduced symmetry of the periodic structure as a whole and are absent in experiments with shorter wavelength satisfying the condition  $\lambda \approx d$~\cite{Matsubara2022}. The experimental data and the theoretical model are discussed by taking the asymmetric electrostatic potential profile and near-field effects explicitly into account.  We show that the THz radiation-induced spin ratchet current appears in the triangular antidots superlattice due to the anomalous Nernst effect (ANE) and the anomalous Hall effect (AHE).

\section{Sample and methods}
\label{samples_methods}

\begin{figure}[b]
\centering
\includegraphics[width=\linewidth]{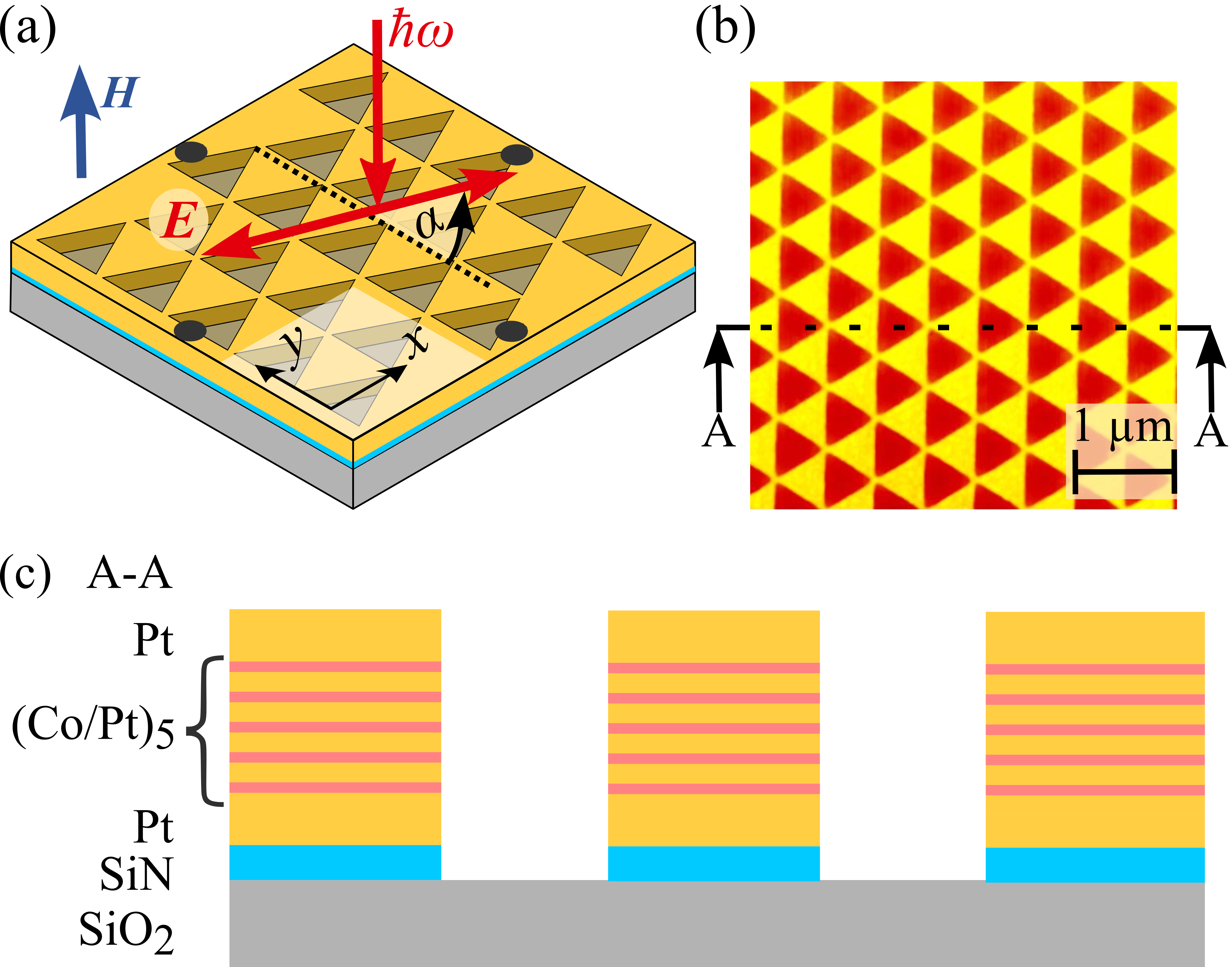}
\caption{Experimental setup (a), AFM image of the equilateral triangles (b) and cross-section (c) of the magnetic metamaterial formed by equilateral triangle-shaped antidots. Panel (a): the sample is irradiated with linearly polarized light along $z$-direction, i.e., normal to the sample surface. The red double arrow illustrates the radiation electric field vector $\bm E$ for linearly polarized radiation rotated anticlockwise by the azimuth angle $\alpha$ from the $y$-axis parallel to the triangle basis. Two pairs of ohmic contacts allow us to probe the photocurrent excited in $x$- and $y$-directions~\cite{Footnote_Fig_1}. Note that the sketched triangles are enlarged for better visibility.  A magnetic field $\bm H$ is applied along the $z$-axis. Panel (b) shows an AFM image of the array of the triangle-shaped antidots  with a period $0.55\,\mu$m. Panel (c) sketches the structure cross-section by A-A plane.  It shows the layer  stacking of the antidot lattice. 
}
\label{fig1}
\end{figure}

Our samples were fabricated   on a $\SI{4}{mm}\times \SI{4}{mm}$   silicon dioxide (\ce{SiO2})  substrate and a silicon nitride (SiN) layer of \SI{5}{nm} thickness. On top, a metallic cobalt/platinum (Co/Pt) multilayer film was deposited   with a fivefold repeated stacking sequence of \SI{0.5}{nm} \ce{Co} and \SI{0.9}{nm} \ce{Pt}. This multilayer is embedded between two \SI{2}{nm} \ce{Pt} layers. We structured all Co and Pt layers with magnetron sputtering. The metallic layers have ferromagnetic properties  with the easy axis perpendicular to the layers, i.e.  in the growth direction $z$. Consequently, application of a magnetic field $\bm H$ that is greater than the coercive field along the $z$-axis results in a constant magnetization $\bm M \parallel z$ everywhere in the patterned film with perpendicular magnetic anisotropy. The key element is the array of equilateral, triangular holes forming an antidot lattice. We patterned this lattice in an area of $250\,\SI{}{\micro m}\times 250\,\SI{}{\micro m}$ by electron beam lithography and argon ion etching. The holes, about \SI{20}{nm} in depth, have a  side length of \SI{480}{nm}, a period of \SI{550}{nm} and are in the dimension of optical wavelengths.   The cross-section of the  layer sequence and the antidots of the  manufactured sample is given in Fig.\,\ref{fig1}(c). The size of one antidot is slightly smaller than the spacing between them; so  almost triangle-shaped metal parts are formed, see atomic force microscope image presented in Fig.\,\ref{fig1}(b). For photoelectric and transport measurements we made two pairs of ohmic contacts orientated along the height ($x$-axis)  and baseline ($y$-axis) of  the triangles, see Figs.\,\ref{fig1}(a) and \ref{figA1} in  Appendix~\ref{appendixA}. The  measured magnetic field and temperature dependence of the  two-point resistance is shown in Fig.\,\ref{figA2} in Appendix\,\ref{appendixA}. It shows that the
resistance at fixed temperature is independent of magnetic field in the field range used in the experiments ($\abs{\mu_0H}\leq \SI{2}{T}$, where $\mu_0$ is  the vacuum permeability) and  slightly increases by increasing the temperature from \SI{120}{K} to \SI{300}{K}.

Figure~\ref{fig1}(a) shows the experimental setup. The sample was excited by normally incident, linearly polarized radiation with a frequency of $f=\SI{2.54}{THz}$, which corresponds to a wavelength of $\lambda = \SI{118.8}{\micro\meter}$ and  photon energy of $\hbar\omega= \SI{10.5}{meV}$. The THz radiation was generated by a cw  methanol molecular gas laser optically pumped by a carbon dioxide laser~\cite{Dantscher2017}. Laser radiation with power $P\approx\SI{50}{mW}$ was modulated by an optical chopper. 
The laser beam spot at the sample position was measured with a pyroelectric camera: it had an almost Gaussian beam profile with a full width at half maximum of  $\SI{1.8}{mm}$.
Subsequently, the radiation intensity on the sample was about $\rm{2~W/cm}^2$.  In two complementary experiments, we used high-power pulsed THz radiation with $\lambda=\SI{90.5}{\micro m}$  ($f=\SI{3.3}{THz}$) and cw infrared radiation $\lambda=\SI{0.8}{\micro m}$. In the latter case we used a cw Ti:Sapphire laser with $\SI{200}{W.cm^{-2}}$. In the former one, we used a pulsed optically pumped \ce{NH3} laser with a pulse length of \SI{100}{ns} and a peak intensity of \SI{100}{kW.cm\tothe{-2}}~\cite{Ganichev1993,Ganichev1995}, both parameters are analyzed by photon drag~\cite{Ganichev1985} and photogalvanic~\cite{Danilov2009} detectors. 
Depending on the wavelength, the beam spot diameter of the pulsed THz laser was from 1.5 to $\SI{3}{\milli\metre}$,  which was controlled by a pyroelectric camera~\cite{Ganichev1999}. 

In the experiments the orientation of the radiation electric field vector $\bm{E}$ was rotated counterclockwise by the azimuth angle $\alpha$ with $\alpha=0$ corresponding to $\bm{E}\parallel y$. For that, we used $x$-cut crystalline quartz  half-wave plates. The  photovoltage $U_\mathrm{ph}$ generated in the  unbiased samples as a result of the excitation with cw radiation was amplified by a factor of 100 and measured applying   standard lock-in technique. The photocurrent was calculated as $J=U_\mathrm{ph}/R_\mathrm{s}$, where $R_\mathrm{s}$ is  the sample's resistance. In experiments with single-pulsed laser radiation the signal was detected with a digital oscilloscope as a voltage drop over a load resistance $R_\mathrm{L}=\SI{50}{\ohm}$.
 
 For studying  photoelectric effects related to  the magnetization  in the magnetic metamaterial, a magnetic field $\bm{H}$ was applied normal to the structure plane.  The measurements were performed in a wide range of temperatures from \SI{100}{K} to \SI{300}{K}.  For room temperature measurements, a water-cooled electromagnet with $\abs{\mu_0 H}\leq \SI{0.4}{T}$ was used.  For low-temperature measurements the samples were placed in an  optical, temperature-variable magneto-cryostat with $z$-cut crystal quartz windows. Here a magnetic field of $\abs{\mu_0 H} = \pm \SI{2}{T}$  was obtained with a superconducting magnet. Additionally, we studied the magnetization curve of the unpatterned and patterned Co/Pt multilayer films using Faraday rotation measurements under illumination of 0.8~$\mu$m laser light~\cite{Matsubara2022}.
 
  \begin{figure}[tb]
 	\centering
 	\includegraphics[width=\linewidth]{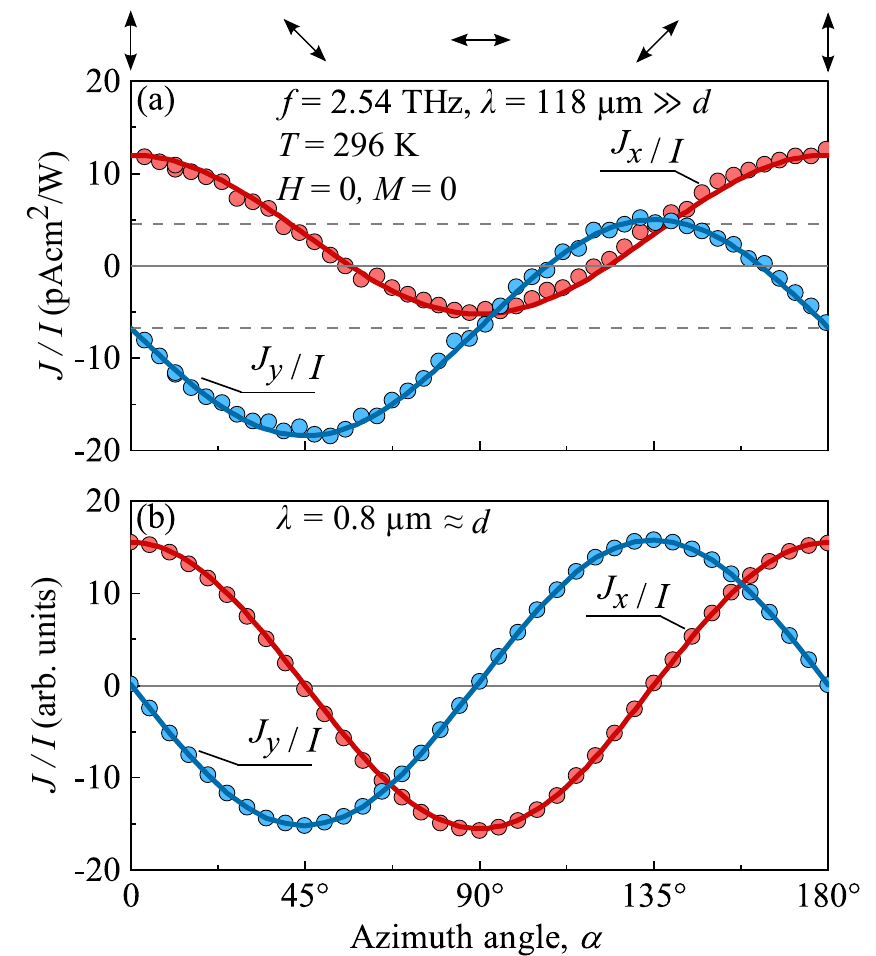}
 	\caption{Polarization dependencies of the photocurrents $J_x$ (red circles) and $J_y$ (blue circles) normalized to the radiation intensity $I$  obtained for zero external magnetic field and magnetization, $\mu_0H =M=0$. Panel (a): photocurrents in response to the radiation with wavelength $\lambda = \SI{118}{\micro m}$ much larger than the period of the antidots array $d= \SI{0.55}{\micro m}$. The red solid line is a fit after Eq.\,\eqref{x-B0} with fitting parameters $A_1 = \SI{8,6}{pAcm^2/W}$, and $C_1 = \SI{3,4}{pAcm^2/W}$. The blue solid line is a fit after Eq.\,\eqref{y-B0} with fitting parameters $\tilde{A}_1 = \SI{11,7}{pAcm^2/W}$, and $\tilde{C}_1 = \SI{-6,7}{pAcm^2/W}$. Dashed horizontal lines show the magnitudes of the polarization-independent offsets. Arrows on top illustrate the orientation of the radiation electric field vector for several values of $\alpha$. Panel (b): photocurrents excited by radiation with $\lambda = \SI{0.8}{\micro m} \approx d.$ Red and blue curves are fits after Eqs.\,\eqref{x-B0} and~\eqref{y-B0} with the fit parameters $A_1= \tilde{A}_1$ and zero  amplitudes of the polarization-independent contributions, $C_1 = \tilde{C}_1 =0$. 
 	}
 	\label{fig2}
 \end{figure}
 
\section{Results}
 \label{results}
 
We begin with the results obtained in the sample in a magnetic multidomain state, which was not subjected to any applied magnetic field before. Since the patterned film does not exhibit a globally constant magnetization, we assign the global magnetization $M = 0$ to this state. Irradiating the sample by THz radiation with  $\lambda=\SI{118}{\micro m}\gg d$ ($f=2.54$\,THz)  we detected a signal which varies upon rotation of the radiation electric field vector $\bm E$. Figure~\ref{fig2}(a) shows a polarization dependence  for the photocurrent measured in $x$- and $y$-directions.  The data can be well fitted by 
\begin{align} 
	J_x &= A_1\,\cos2\alpha +C_1  \,,\label{x-B0} \\
	J_y& =- \tilde{A}_1\,\sin2\alpha +\tilde{C}_1 \,. \label{y-B0}
\end{align}
The signal was also detected for a substantially smaller   wavelength of $\lambda=\SI{0.8}{\micro m}\approx d$ and its polarization dependence is also described by Eqs.\,\eqref{x-B0}, \eqref{y-B0},  see Fig.\,\ref{fig2}(b). However, in this case no offsets were observed, so that $C_1 = \tilde{C}_1 =0$. The photocurrent excited by $\lambda=\SI{0.8}{\micro m}\approx d$ has recently been studied in detail in Ref~\cite{Matsubara2022}. Therefore, in this paper we only briefly discuss this result and focus in the following on the photocurrent excited with  $\lambda=\SI{118}{\micro m}\gg d$. 

\begin{figure} 
	\centering
	\includegraphics[width=\linewidth]{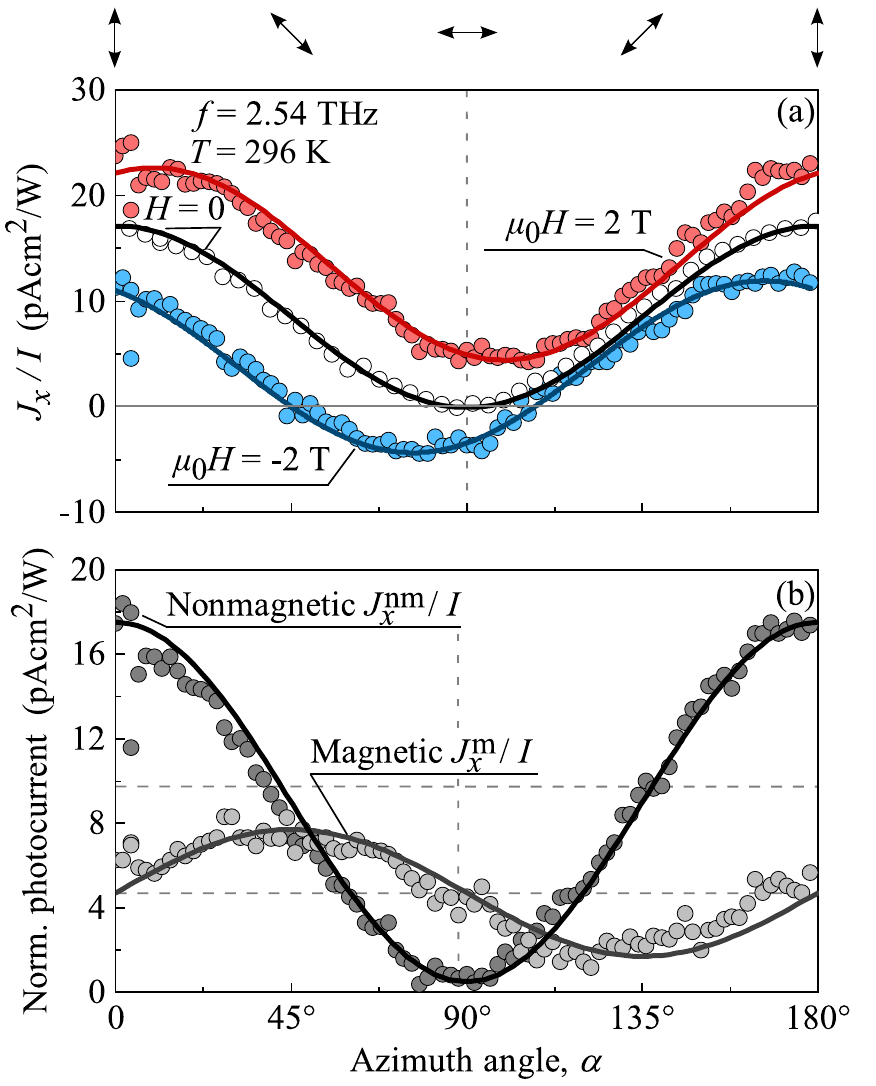}
	\caption{Panel (a): Polarization dependencies of the photocurrent $J_x$ normalized to radiation intensity $I$ obtained for an external magnetic field $\mu_0H = \pm \SI{2}{T}$, red and blue circles, respectively. Black circles show $J_x/I$ obtained for $\mu_0H =M=0$.  Arrows on top illustrate the orientation of the radiation electric field vector for several values of $\alpha$. The curves show fits after Eq.\,\eqref{eq:j_x-B_not_0}. Panel (b): Polarization dependence of the normalized nonmagnetic ($J_x^{\mathrm{nm}}$) and magnetic ($J_x^{\mathrm{m}}$) contributions obtained as the odd and even part with respect to the magnetization $M_z$ induced by the magnetic field $\mu_0H_z$, see Eqs.\,\eqref{eq:NM} and \eqref{eq:M}. The curves are fits after the first and second terms in the right hand side of the Eq.\,\eqref{eq:j_x-B_not_0} (nonmagnetic photocurrent $J_x^{\mathrm{nm}}$) and the two last terms of the right-hand side of the same equation (magnetic photocurrent $J_x^{\mathrm{m}}$). Fit parameters are: $A_1 = \SI{8.5}{pAcm^2/W}$, $C_1 = \SI{9}{pAcm^2/W}$, $A_2 = \SI{3}{pAcm^2/W}$, and $C_2 = \SI{5}{pAcm^2/W}$. Dashed horizontal lines show the magnitudes of the polarization-independent offsets.
	}
	\label{fig3}
\end{figure}

The application of an external magnetic field $\bm H$ larger than the coercive field and normal to the metal layers results in a constant magnetization $\bm M$ everywhere in the patterned magnetic lattice and changes the polarization dependence of the signal~\footnote{Note that application of an in-plane magnetic field does not affect the signals in the whole studied range $|\mu_0H|\leq \SI{2}{T}$.}. It is shown in Fig.\,\ref{fig3}(a)  that the photocurrent is now described by 
 \begin{align} 
J_x = A_1\,\cos2\alpha +C_1 + M_z\left(A_2\,\sin2\alpha +C_2 \right)\,. \label{eq:j_x-B_not_0}
\end{align}
Decomposing the polarization dependence into the odd and even part with respect to the magnetization $M$, we obtain the nonmagnetic  $J_x^{\mathrm{nm}}$ and magnetic $J_x^{\mathrm{m}}$ contributions to the photocurrent:
\begin{align} 
J_x^{\mathrm{nm}}& = \frac12 \left[ J_x(+M_z) + J_x(-M_z) \right] \,,\label{eq:NM}	\\
J_x^{\mathrm{m}}& = \frac12 \left[J_x(+M_z) - J_x(-M_z) \right]\,.\label{eq:M}
\end{align}
The corresponding dependences are shown in Fig.\,\ref{fig3}(b) and confirm that the nonmagnetic and magnetic photocurrent contributions indeed vary as $\cos2\alpha$ and $\sin2\alpha$ functions, respectively. We emphasize that the $J_x^{\mathrm{m}}$ contribution has a magnetization-dependent offset, whereas the offset of  $J_x^{\mathrm{nm}}$ does not depend on $M_z$.

\begin{figure}
	\centering
	\includegraphics[width=\linewidth]{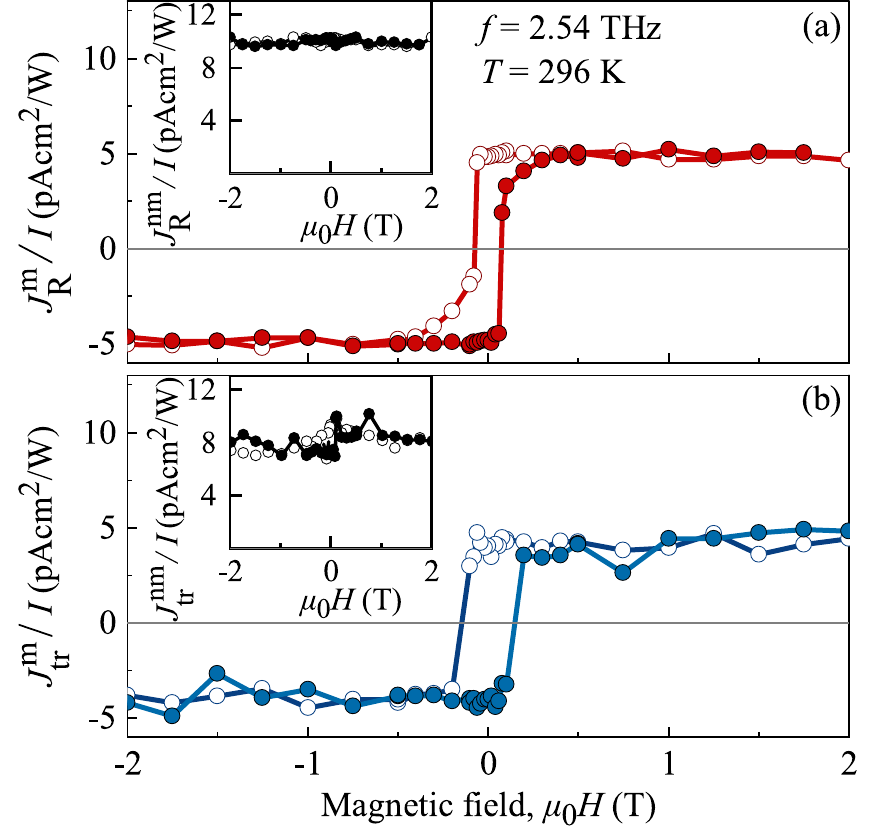}
	\caption{Magnetic field dependence of the magnetization-induced polarization-independent (panel (a)) and polarization-dependent (panel (b)) photocurrent contributions measured in $x$-direction and normalized to the radiation intensity. Following Eq.\,\eqref{eq:j_x-B_not_0} these contributions are defined as  $J_\mathrm{R}^{\mathrm{m}} = M_zA_2\,\sin2\alpha$ and  $J_\mathrm{tr}^{\mathrm{m}}= M_zC_2$. Each point has been extracted from a measured $\alpha$-dependence at a constant magnetic field. Full and empty circles show the forward and backward magnetic field sweeps, respectively. Insets in panels (a) and (b) show magnetic field dependencies of the normalized $J_{\mathrm{R},x}^\mathrm{nm} = C_1$ and $J_{\mathrm{tr}, x}^\mathrm{nm} = A_1\,\cos2\alpha$, respectively. The inset demonstrates that they are independent of the magnetic field for both magnetic field sweep directions. 
}
	\label{fig4-new}
\end{figure}

Figure~\ref{fig4-new} shows the magnetic field dependence of the photocurrent. To describe individual contributions, we introduce the following notation. To distinguish  the nonmagnetic and magnetic photocurrent contributions, we used the superscripts ``nm'' and ``m'', respectively. The polarization-independent offsets  are quoted as $J_\mathrm{R}^{\mathrm{nm}}$ and    $J_\mathrm{R}^{\mathrm{m}}$, whereas the polarization-dependent parts are given  by $J_\mathrm{tr}^{\mathrm{nm}}$ and $J_\mathrm{tr}^{\mathrm{m}}$. The  subscripts ''$\rm{R}$'' and ''$\rm{tr}$'' indicate the ratchet  and the trigonal  photocurrent mechanisms, which as we show below, are responsible  for the polarization-independent offset and the polarization-dependent part, respectively. The polarization-dependent and -independent parts of the photocurrent were obtained from azimuth angle dependences measured for each magnetic field.

The insets in Figs.\,\ref{fig4-new}(a),(b) demonstrate that the magnetization-even contributions are independent of the magnetic field in the studied range from \SI{-2}{T} to  \SI{2}{T}. In contrast, the  magnetization-odd parts exhibit a clear hysteresis for $\abs{\mu_0H}<\SI{0.3}{T}$, whereas at higher $H$-fields they do not depend on magnetic field and have opposite polarities for positive and negative $H$, see Figs.\,\ref{fig4-new}(a),(b)~\footnote{Note that the sample resistance does not depend on magnetic field, see Fig.\,\ref{figA2} in Appendix~\ref{appendixA}.}. The amplitude of the photocurrent at high magnetic fields is almost independent of temperature for $T\gtrsim \SI{100}{K}$ (not shown).

\begin{figure} 
	\centering 
	\includegraphics[width=\linewidth]{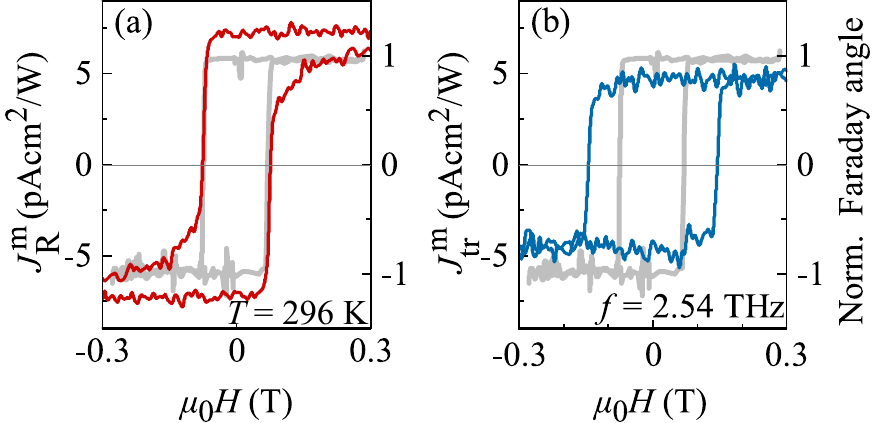}
	\caption{Zoom of the hysteretic parts of the magnetic field dependencies of  $J_\mathrm{R}^{\mathrm{m}}/I$ (panel (a)) and $J_\mathrm{tr}^{\mathrm{m}}/I$ (panel (b)) presented in Fig.\,\ref{fig4-new}. The gray curves show the  magnetic field dependence of the Faraday rotation angle measured in the unpatterned Co/Pt multilayer film. The angles are normalized to the maximum, see right axis. 
}
	\label{fig5-new}
\end{figure}

\begin{figure} 
	\centering 
	\includegraphics[width=\linewidth]{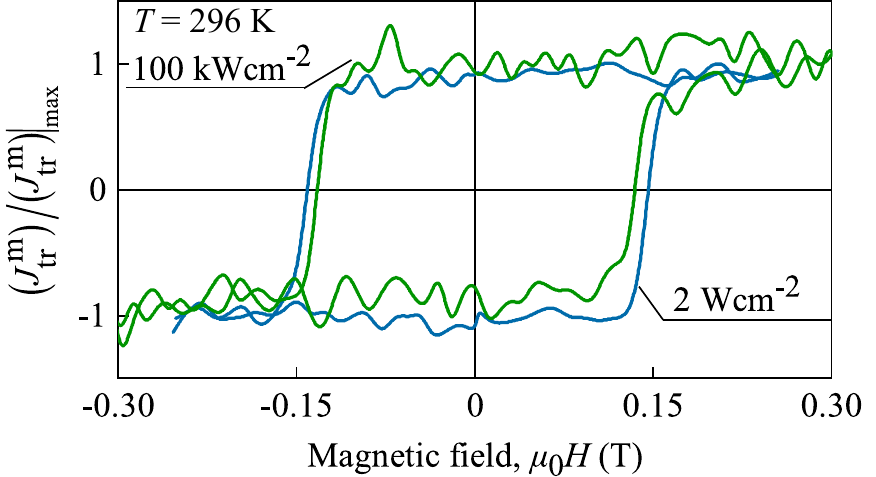}
	\caption{Magnetic field dependence of the magnetization-induced  polarization-dependent  photocurrent contribution measured in $x$-direction at low (blue trace) and high (green trace) radiation intensities. Each curve is normalized to the signal maximum, $(J_\mathrm{tr}^{\mathrm{m}}) / (J_\mathrm{tr}^{\mathrm{m}})|_{\rm max}$.  The blue curve has been obtained with THz radiation of cw laser operating at a frequency $f=$\SI{2.54}{THz} and intensity $I=$\SI{2}{W.cm\tothe{-2}},  and the green one with  radiation from a pulsed laser  operating at a frequency $f=$\SI{3.33}{THz} and  $I=$\SI{100}{kW.cm\tothe{-2}}. The plot reveals that both curves have almost the same hysteresis width despite of almost five orders of magnitude difference in the radiation intensity. 
}
	\label{fig6-new}
\end{figure}

Figures~\ref{fig5-new}(a),(b) show the hysteretic part of the magnetic field dependence for the photocurrent contributions $J_\mathrm{R}^{\mathrm{m}}$ and  $J_\mathrm{tr}^{\mathrm{m}}$, respectively. They also display the magnetic field dependence of the Faraday rotation angle $\theta\propto M$, see gray lines in Fig.\,~\ref{fig5-new}. These  measurements were performed on the unpatterned Co/Pt multilayer film using linearly polarized radiation of $\lambda =  \SI{118}{\micro m}$ passing through the magnetic material. Figure~\ref{fig5-new} demonstrates that the hysteresis width of the polarization-independent contribution $J_\mathrm{R}^{\mathrm{m}}$ is the same as the one of the magnetization, whereas the hysteresis width of the polarization-dependent part  $J_\mathrm{tr}^{\mathrm{m}}$ is  somewhat larger than that of $M$. This is most clearly seen in panel (d) of Fig.\,\ref{fig:300K-hys} in Appendix~\ref{appendixA2} where the photocurrent  contributions   $\propto J_\mathrm{R}^{\mathrm{m}}$ and  $J_\mathrm{tr}^{\mathrm{m}} \sin2\alpha$  at $\alpha = \ang{135}$ have opposite signs. At last but not least, performing measurements at almost five orders of magnitude higher intensities we obtained that while the magnitude of the photocurrent drastically increases  the hysteresis width of  the polarization-dependent photocurrent  $J_\mathrm{tr}^{\mathrm{m}}$ remains unchanged, see Fig.\,\ref{fig6-new}. 

\section{Theory of photocurrents formed in isolated equilateral triangle shaped antidots}
\label{theory}

In the metal layers forming the studied samples the normally incident radiation can excite the photocurrent only in the antidots array because a photocurrent in the unpatterned part of the sample is forbidden by symmetry. For a wavelength $\lambda$ smaller or comparable to the period $d$ the photocurrent generation should be considered for  isolated triangles, whereas for the opposite limit ($\lambda \gg d$) the antidots array should be treated as a metamaterial. As we show below (Sec.~\ref{theo:Seebeck spin ratchet}) in the latter case the photocurrent is caused by the ratchet effect, which additionally gives rise to polarization-independent photocurrents detected in our experiments applying radiation with $\lambda = \SI{118}{\micro m}$. Below we develop the phenomenological and microscopic theory for the polarization-dependent photocurrent excited in the isolated triangles. 

\begin{figure*}[tb]
	\centering
	\includegraphics[]{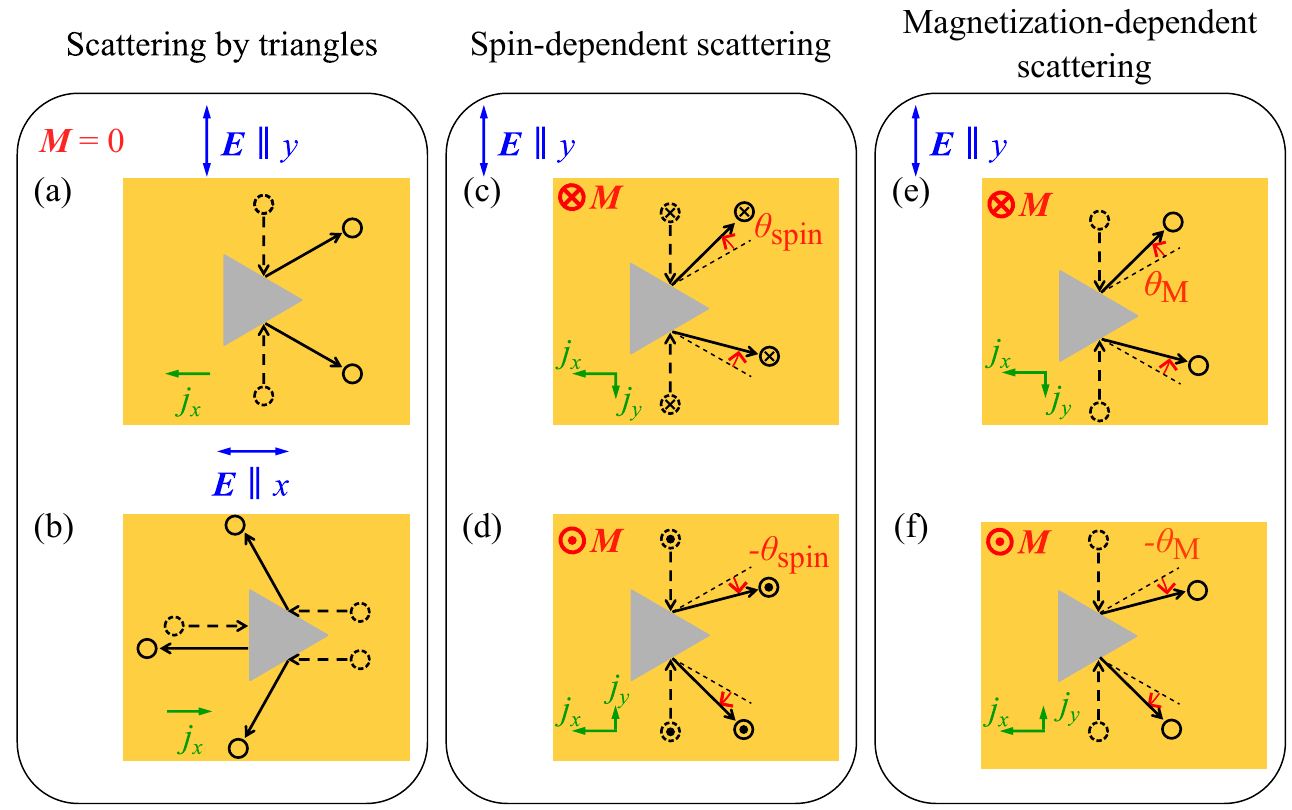}
	\caption{Microscopic models for the trigonal photocurrents excited in equilateral triangle-shaped antidots. The field $\bm E$ results in a directed motion of carriers shown by the dashed arrows.  This is described by the stationary correction to the distribution function 
$f_{\bm p}^{(2)} \propto |\bm E|^2$ being \textit{second order}  in the electric field. Due to asymmetric scattering a directed carrier flow and, therefore, an electric current $j$ is generated. Panels (a) and (b) show the dc current generation for zero magnetization and two directions of the radiation electric field resulting in the momentum alignment along $y$- and $x$-directions, respectively, see Sec.~\ref{micro_trigonal}. Each column represents a different mechanism, as headed on top.  In (a), as a consequence of the external electric field $\bm{E}\parallel y$, electrons predominantly scatter along the $x$-axis (solid arrows), which results in an electric current  in $x$-direction. Rotation of the electric field by \ang{90}, panel (b), reverses the direction of the predominantly scattered electrons; consequently, the electric current changes its sign. The electric field orientation dependence of the dc current $j_x$ is given by $\cos(2\alpha)$, see Eqs.\,\eqref{phenom_C3v} and \eqref{alpha}.	Panels (c) and (d) illustrate the deflection of the electron trajectory by an angle $\theta_{\rm spin}$ caused by the  spin polarization of the carriers due to the out-of-plane magnetization $M_z$, see Sec.~\ref{spin_trigonal}. This process results in the emergence of a spin-polarized photocurrent $j_y\propto M_z$, which reverses its sign upon switching the magnetization direction, as shown in panels (c) and (d) for $\pm M_z$. Panels (e) and (f) show the deflection of the electron trajectory by an angle $\theta_{\rm M}$ due to  magnetization-dependent scattering, see Sec.~\ref{magnetization_trigonal3}. This mechanism does not require spin polarization and also results in a photocurrent $j_y\propto M_z$.
	}
	\label{fig7-new}
\end{figure*}

\subsection{Phenomenology}
\label{isolated}

For the photocurrent formed due to scattering by individual triangular antidots,  the actual point group symmetry is $C_{3v}$. This is the group symmetry of an equilateral triangle with three reflection planes oriented at $120^\circ$ to each other. This symmetry consideration yields the following relations between the photocurrent density~\footnote{We note that while in the theoretical consideration the current density $j$ is used, in the experiments, the electric current $J$ is 	measured, which is proportional to the current density $j$.}, magnetization and radiation polarization parameters:
\begin{align}
	\label{phenom_C3v}
	&	j_x =  \chi P_{\rm lin} E_0^2 + M_z \Phi P_{\rm lin}' E_0^2, 	\\ 
	&	j_y = - \chi P_{\rm lin}' E_0^2 + M_z  \Phi P_{\rm lin} E_0^2. \nonumber
\end{align}
Here $E_0$ is the amplitude of the radiation electric field $\bm E = E_0 \bm e \exp(-i\omega t)+c.c.$ with $\bm e$ being the polarization unit vector,  
$\Phi$ and $\chi$ are constants,  
and the axes $(x,y)$ are related to the main axes of the $C_{3v}$ point group: one of the three reflection planes is chosen as $(zx)$. 
The Stokes parameters of radiation $P_{\rm lin}$ and $P_{\rm lin}'$  are the linear degrees:
%
\begin{equation}
\label{Stokes}
P_{\rm lin} =\abs{e_x}^2-\abs{e_y}^2 , \quad P_{\rm lin}' = e_xe_y^* + e_x^*e_y.
\end{equation}
%
In our experiments applying linearly polarized radiation they vary with the azimuth angle $\alpha$ as
\begin{equation}
	\label{alpha}
P_{\rm lin} =  \cos 2\alpha, \quad P_{\rm lin}' = \sin 2 \alpha.
\end{equation}
We see that the polarization-dependent photocurrent components in $C_{3v}$ symmetry have $\pi$-periodic dependence on the polarization plane orientation. This is caused by the measurement setup where the $j_x$ and $j_y$ components are detected, i.e., the currents in the reflection plane $(zx)$ and perpendicular to it. If, otherwise, the contacts are deposited along the pair of axes rotated by an angle $\Psi$ to $x$, $y$ then the additional phase $3\Psi$ appears in the expressions for the photocurrents which reflects the trigonal symmetry of the system~\cite{Olbrich2014,Danilov2021}.

\subsection{Magnetization-independent trigonal photocurrent}
\label{micro_trigonal}

The magnetization-independent photocurrent  described by the constant $\chi$ in Eqs.\,\eqref{phenom_C3v} is the ``trigonal'' linear photogalvanic effect (LPGE) current $\bm j^{\rm tr}$.  It is present owing to asymmetrical scattering and has been studied in detail in intrinsically trigonal systems~\cite{Belinicher1980,Weber2008,Olbrich2014,Otteneder2020,Danilov2021}. The system under study is extrinsically  trigonal because of the presence of the artificially made macroscopic triangle antidots. Assuming rare electron scattering on triangle antidot boundaries, the trigonal current can be derived from the Boltzmann kinetic equation which reads
\begin{equation}
\label{kin_eq}
\dv{f_{\bm p}}{t} + e\bm E \cdot \dv{f_{\bm p}}{\bm p} = \sum_{\bm p'} \qty(W_{\bm p \bm p'} f_{\bm p'} - W_{\bm p' \bm p} f_{\bm p}).
\end{equation}
Here $f_{\bm p}$ is the  electron distribution function, $\bm p$ is the electron momentum, and $W_{\bm p' \bm p}$ is the probability of  the elastic scattering process $\bm p \to \bm p'$. The scattering probability is conveniently decomposed into the symmetric and asymmetric parts with respect to the interchange of initial and final momenta:
\begin{equation}
	W_{\bm p' \bm p} = W^{(s)}_{\bm p' \bm p} + W^{(a)}_{\bm p' \bm p}, \qquad W^{(s,a)}_{\bm p \bm p'} = \pm W^{(s,a)}_{\bm p' \bm p}.
\end{equation}
The symmetrical part $W^{(s)}_{\bm p' \bm p}$ determines the relaxation times of different Fourier-harmonics of the distribution function, $\tau_{n}$ ($n=1,2,\ldots$):
\begin{equation}
\tau_{n}^{-1}=\sum_{\bm p'} W^{(s)}_{\bm p' \bm p} (1-\cos{n\theta_{\bm p \bm p'}}),
\end{equation}
where $\theta_{\bm p \bm p'}$ is the angle between $\bm p$ and $\bm p'$~\cite{Olbrich2014}.
In particular, $\tau_1$ is the transport relaxation time, and $\tau_2$ is the relaxation time for the momentum-aligned electron distribution.

Taking into account the asymmetric part in  linear order we obtain the trigonal photocurrent described by the constant $\chi$ in Eqs.\,\eqref{phenom_C3v} where~\cite{Otteneder2020}
\begin{align}
\label{eq:chi}
\chi &=   - {2Ne^3 \tau_1 \over m \varepsilon_{\rm F} (1+\omega^2\tau_1^2)} \nonumber \\
&\times \qty[\dv{(\xi_{\rm tr}v_{\rm F}\varepsilon_{\rm F} \tau_2)}{\varepsilon_{\rm F}} - \dv{\tau_1}{\varepsilon_{\rm F}}{\tau_2\over \tau_1}\xi_{\rm tr}v_{\rm F}{1-\omega^2\tau_1\tau_2\over 1+\omega^2\tau_2^2}].
\end{align}
Here $N$ is the  2D  electron concentration, $m$ is the effective mass, $\varepsilon_{\rm F}$ and $v_{\rm F}$ are the Fermi energy and velocity, $\tau_{1,2}$ are taken at the Fermi energy,
and the dimensionless parameter $\xi_{\rm tr}$ accounts for the trigonal scattering asymmetry:
\begin{equation}
\xi_{\rm tr}=\tau_1 \sum_{\bm p'} \left< W^{a}_{\bm p \bm p'} \cos{2\varphi_{\bm p'}}\cos{\varphi_{\bm p}} \right>_{\varphi_{\bm p}}.
\end{equation}
Here angular brackets denote averaging over directions of the momentum $\bm p$ at a fixed energy $\varepsilon_{\rm F}$, and $\varphi_{\bm p}$, $\varphi_{\bm p'}$ are polar angles of the corresponding momenta. For a particular case of scattering by short-range impurities we have
\begin{equation}
\label{eq:chi_sr}
\xi_{\rm tr} = 2\pi g_0 V_0  \left<  F_{\bm p \bm k \bm p'}\cos{2\varphi_{\bm p'}}\cos{\varphi_{\bm p}} \right>_{\varphi_{\bm p},\varphi_{\bm p'},\varphi_{\bm k}}.
\end{equation}
Here $g_0$ is the density of states at the Fermi energy, $V_0$ is the scattering amplitude by impurities, and $F_{\bm p \bm k \bm p'}=\langle u_{\bm p} | u_{\bm p'}\rangle \langle u_{\bm p'} | u_{\bm k} \rangle \langle u_{\bm k} | u_{\bm p} \rangle$ is a product of the Bloch amplitude overlaps.

The considered microscopic mechanism of the trigonal photocurrent $\bm j^{\rm tr}$ formation is illustrated in Figs.\,\ref{fig7-new}(a) and (b). In general, scattering of electrons in a metal at the antidot boundaries violates the detailed equilibrium giving rise to the asymmetrical part of the scattering probability. In equilibrium, i.e. without any external influence, the scattering, obviously,   does not result in an electric current. In the presence of radiation, however, the electron distribution is aligned in momenta along the radiation electric field.  Formally, it means that a correction $\delta f_{\bm p} \propto \tau_2 E_0^2\cos{2\varphi_{\bm p}}$ appears as a solution of the Boltzmann Eq.\,\eqref{kin_eq} in the second order in the radiation electric field where only the symmetrical part of the scattering probability $W^{(s)}_{\bm p \bm p'}$ is taken into account. Then accounting for the asymmetrical part, i.e. scattering by the antidots, we see from Figs.\,\ref{fig7-new}(a),(b) that a directed flow of electrons is formed with a direction governed by the radiation polarization. The resulting current direction depends on the relative orientation of the radiation electric field and the antidot: e.g., the field parallel to the triangle base, see Fig.\,\ref{fig7-new}(a), yields a current flowing in  $x$-direction while rotation of the electric field by $90^\circ$ reverses the current direction, see Fig.\,\ref{fig7-new}(b). The corresponding dependencies of the photocurrents on the electric field vector orientation are given for $j_x$ and $j_y$ by the first terms on the right-hand sides of Eqs.\,\eqref{phenom_C3v}. 

\subsection{Magnetization-induced trigonal photocurrent}
\label{magnetic_trigonal}

A magnetic field applied along the $z$-direction, which causes the magnetization $M_z$, gives rise to new photocurrent contributions, which can be excited by  linearly as well as circularly polarized radiation, see the second and the third terms on the right-hand sides of Eqs.\,\eqref{phenom_C3v}, respectively. We develop a microscopic theory of the former effect and find the corresponding constant $\Phi$. We propose and analyze three  microscopic mechanisms  based on  i)~the anomalous Hall effect (AHE), ii)~spin-dependent scattering, and iii)~magnetization-dependent scattering. Since all detected photocurrents have hysteretic behaviour and saturate at large positive and negative magnetic fields, see Fig.\,\ref{fig4-new}, they are determined solely by the magnetization $M_z$ while effects of the Lorentz force, resulting in the linear in $H_z$ dependence, are negligible.

\subsubsection{Anomalous-Hall-effect--induced trigonal photocurrent}
\label{AHE_trigonal}

First, we consider the photocurrent contribution caused by the Anomalous Hall effect.  As well known in the presence of magnetization any dc electric current acquires a perpendicular component. In the experiments considered this results in a component $\bm j \propto \bm j^{\rm tr}\times \bm M $  perpendicular to the trigonal LPGE current $\bm j^{\rm tr}$. Microscopically, there are both intrinsic and extrinsic contributions to the AHE, and the most efficient one is usually caused by skew-scattering on impurities. In this mechanism, the magnetization-dependent photocurrent is given by
\begin{equation}
	\label{j_AHE}
	\bm j = \xi_{\rm AHE}  P_s \bm j^{\rm tr} \times \hat{\bm z}.
\end{equation}
Here $P_s$ is the electron spin polarization which appears due to the Zeeman effect
\begin{equation}
\label{eq:Ps}
P_s = - {\Delta_Z \over 2\varepsilon_{\rm F}} \propto M_z
\end{equation}
with $\Delta_Z$ being the Zeeman splitting, and the dimensionless skew scattering efficiency $\xi_{\rm AHE}$  given by~\cite{Glazov2020}
\begin{equation}
	\label{xi_AHE}
	\xi_{\rm AHE} = \tau_1 \left< \sum_{\bm p'} \sin{(\varphi_{\bm p}-\varphi_{\bm p'})}W^{(a,{\rm SO})}_{\bm p' \bm p}\right>_{\varphi_{\bm p}}.
\end{equation}
%
Equation~\eqref{j_AHE} agrees with the phenomenological Eqs.\,\eqref{phenom_C3v} yielding $\Phi M_z = \xi_{\rm AHE} P_s \chi$. Note that this mechanism requires double account for the scattering asymmetry: at the first stage where the constant $\chi$ is obtained, Eq.\,\eqref{eq:chi}, and at the second stage, at a deflection of the trigonal current. Despite each asymmetrical scattering probability can be obtained beyond the Born approximation only, in the Co-based structure with strong AHE this photocurrent formation mechanism can be of equal importance or even dominate over the other mechanisms considered below.

\subsubsection{Spin-dependent scattering-induced trigonal photocurrent}
\label{spin_trigonal}

The skew scattering considered above is also responsible for a further contribution which comes from  spin-dependent electron scattering on triangle boundaries. Due to   spin-orbit interaction, spin-up and spin-down electrons scatter off the antidot boundaries with an angle $\pm \theta_{\rm spin}$ with respect to the spin-less case, see in Figs.\,\ref{fig7-new}(c) and (d).
Due to the spin polarization caused by the magnetization, Eq.\,\eqref{eq:Ps}, the majority of electrons having spin-up (spin-down) will be preferentially deflected up (down)   by the angle $\theta_{\rm spin}$ ($-\theta_{\rm spin}$) relative to the non-magnetized situation, see Fig.\,\ref{fig7-new}(c) (Fig.\,\ref{fig7-new}(d)).
%
%
This process results in the generation of a magnetic-field--dependent photocurrent which is rotated relative to the previously considered trigonal LPGE current $\bm j^{\rm tr}$  formed at $M_z=0$. For example, at vertical polarization the component $j_y \propto M_z$ emerges while in the nonmagnetic case the trigonal current $\bm j^{\rm tr}$ flows parallel to the $x$-direction, cf. Figs.\,\ref{fig7-new}(c) and~(d). Reversing the magnetization $M$ changes the sign of the $j_y$ photocurrent component. This consideration fully agrees with the phenomenological Eqs.\,\eqref{phenom_C3v} yielding the photocurrent given by the constant $\Phi$.

Microscopically, the photocurrent density $j_{x,y} \propto M_z \Phi$ can be derived from the kinetic Eq.\,\eqref{kin_eq} similarly to the trigonal photocurrent. Assuming rare electron scattering on the triangle boundaries we  obtain that $M_z\Phi$ is given by 
Eq.\,\eqref{eq:chi} with two substitutions:
\begin{equation}
N \to P_s N, \qquad \xi_{\rm tr} \to \xi_{\rm SO}.
\end{equation}
Here 
the spin-orbit scattering asymmetry factor $\xi_{\rm SO}$ is given by
%
\begin{equation}
	\xi_{\rm SO}= \tau_1 \left< \sum_{\bm p'} \sin{\varphi_{\bm p}}\cos{2\varphi_{\bm p'}}W^{(a,{\rm SO})}_{\bm p' \bm p}\right>_{\varphi_{\bm p}},
\end{equation}
where $W^{(a,{\rm SO})}_{\bm p' \bm p}$ is the asymmetric scattering probability calculated with an account for the spin-orbit interaction in the electron Bloch amplitudes, cf. Eq.\,\eqref{eq:chi_sr}.

\subsubsection{Magnetization-dependent scattering-induced trigonal photocurrent}
\label{magnetization_trigonal3}

Magnetization does not only result in the spin polarization of electrons  considered above  but does also affect the electron orbital motion. Thus,  electrons are scattered off the antidot boundaries at an additional angle $\theta_M$ which is the same for spin-up and spin-down electrons, but reverses its sign upon switching the magnetization direction, see Figs.\,\ref{fig7-new}(e) and~(f). As a result, an additional contribution to the photocurrent $j\propto M_z$ is generated. For the vertical polarization it corresponds to $\pm j_y$, see Figs.\,\ref{fig7-new}(e) and~(f).

Microscopically this contribution is obtained if accounting for the magnetization in the Bloch amplitudes. This yields the asymmetrical scattering probability $W^{(a,M)}_{\bm p' \bm p}$ dependent on $M_z$. The value of $M_z\Phi$ in this mechanism is given by Eq.\,\eqref{eq:chi} with the magnetization-induced factor
\begin{equation}
	\xi_M= \tau_1 \left< \sum_{\bm p'} \sin{\varphi_{\bm p}}\cos{2\varphi_{\bm p'}}W^{(a,M)}_{\bm p' \bm p}\right>_{\varphi_{\bm p}}.
\end{equation}
The value $\xi_M$ is odd in $M_z$ and it is linear in the magnetization to the lowest order.

\section{Theory of ratchet current in antidots magnetic metamaterials}
\label{theo:Seebeck spin ratchet}

Previously we considered the photocurrent generation caused by  the individual triangle antidots. Each triangle has the $C_{3v}$ symmetry with three reflection planes and the $C_3$ axis. 
As we demonstrated above, in this case the photocurrents $j_x$ and $j_y$ are solely defined by the degree of linear polarization given by the corresponding Stokes parameters $P_{\rm lin}$ and $P_{\rm lin}'$.
For a wavelength larger than the period of the array of the triangle antidots forming a lateral superlattice, however, 
large polarization-independent photocurrent contributions are detected. This observation indicates the symmetry reduction
due to the absence of two or even all three reflection planes meaning the point symmetry group $C_s$ or $C_1$, respectively. In the former case, the reflection plane $(zx)$ is the only nontrivial symmetry operation whereas in the latter one we have only trivial symmetry operation, i.e. identity.

\subsection{Phenomenology and the microscopic model of the ratchet current}
\label{Seebeck_phenom}

At normal light incidence, the $C_s$ symmetry allows for the following relations between the photocurrent density $\bm j$, magnetization $\bm M$ and radiation Stokes parameters:
\begin{align}
\label{phenom_Cs}
&	j_x = 	 \Xi \qty(\chi_1 P_{\rm lin} + \chi_0) + M_z \Xi \qty(\Phi_1 P_{\rm lin}'  + \gamma P_{\rm circ}),
\\
&	 j_y= \Xi \qty(- \tilde{\chi}_1 P_{\rm lin}' + \tilde{\gamma} P_{\rm circ} ) + M_z \Xi \qty(\tilde{\Phi}_1 P_{\rm lin} + \Phi_0)\, ,\nonumber
\end{align}
where $\Xi$ is proportional to the square of the radiation electric field $E_0$,
and $P_{\rm circ}= i(e_xe_y^* - e_x^*e_y)$ is the circular polarization degree. 
Comparing with the phenomenological Eqs.\,\eqref{phenom_C3v} obtained for the individual triangles with higher symmetry we see two differences. First, the linear-polarization sensitive contributions, instead of single coefficients $\chi$ and $\Phi$, are now described by independent constants ${\tilde{\chi}_1 \neq \chi_1}$ and  $\tilde{\Phi}_1 \neq \Phi_1$. Second, and crucial, the basically new photocurrent contributions appear which are: the polarization-independent ones described by the constants $\Phi_0$ and $\chi_0$, and the helicity-driven photocurrents  given by the constants $\gamma$ and $\tilde{\gamma}$.
In the case that the symmetry is reduced further to $C_1$, phenomenological equations become lengthy because now $j_x$ as well as $j_y$ are given by a sum of all contributions present in both Eqs.~\eqref{phenom_Cs} with independent weights.
Below, to be specific, we consider for simplicity the microscopic model for the $C_s$ symmetry.

\begin{figure}[tbh]
	\centering
	\includegraphics[width=\linewidth]{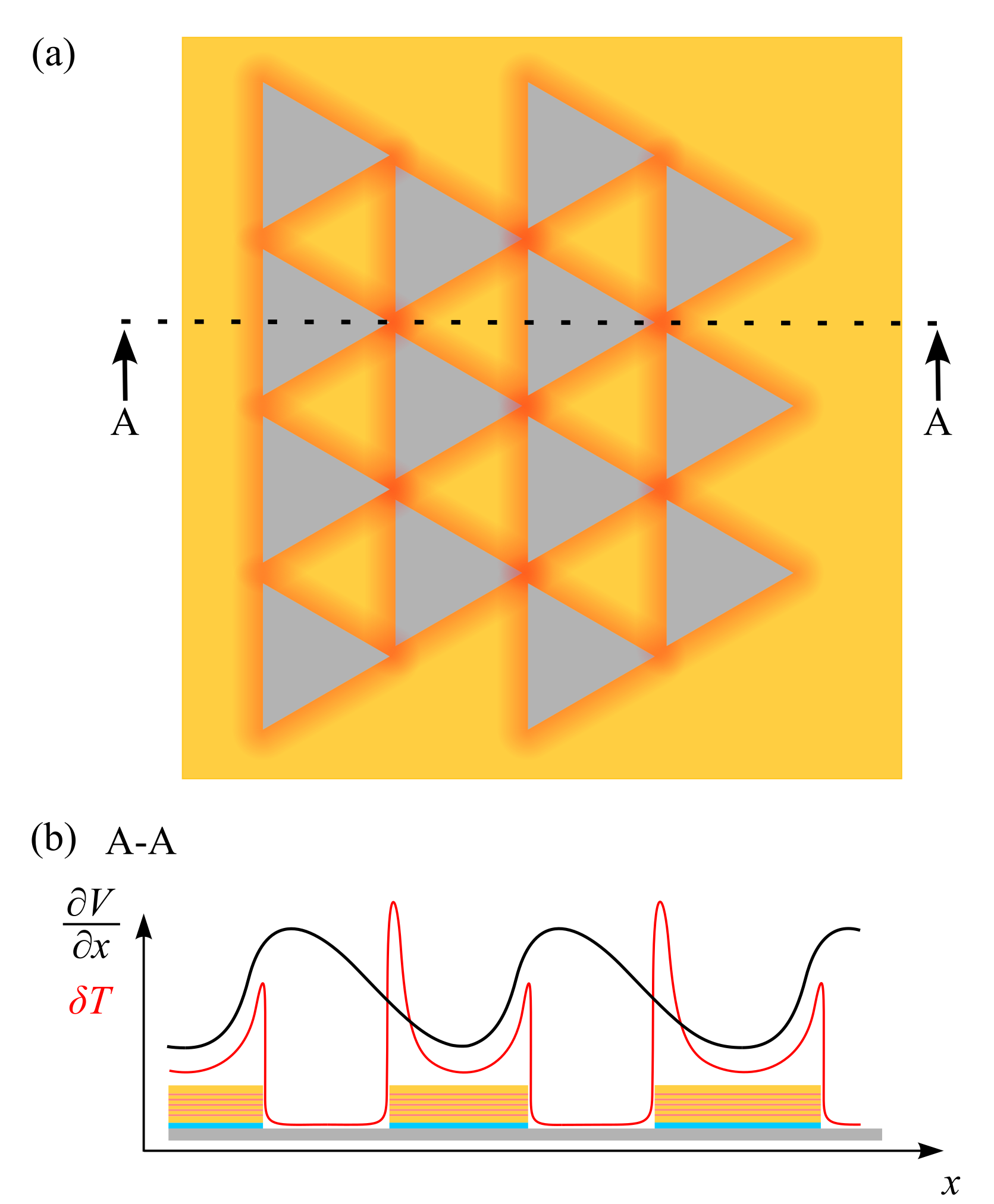}
	\caption{Model for the Seebeck ratchet current. Panel (a) depicts the ensemble of equilateral-triangle shaped antidots 
arranged in a slightly distorted hexagonal lattice having $C_s$-symmetry	
	and the electron temperature profile due to heating of the near-field, schematically highlighted by the red contour. Panel (b) shows the  temperature profile $\delta T(\bm r)\propto E^2_0(\bm r)$ caused by  radiation near-field diffraction at the edges of antidots and the electrostatic force proportional to  $\partial V/\partial x$. The Figure shows that both terms are asymmetric  due to  the triangular shape of the antidots. 
	}
	\label{fig8-new}
\end{figure}

Microscopically, electrons feel the low symmetry of the whole system via the inhomogeneous near-field formed by radiation diffraction from the triangle's boundaries. This near-field $\bm E(\bm r)$ has a profile periodic in two-dimensional space with the period of the lateral superlattice, see Fig.\,\ref{fig8-new}. Another field  acting on electrons is the periodic potential $V(\bm r)$ with  barriers on the borders of the antidots, see Fig.\,\ref{fig8-new}(b). Despite both the near-field and the periodic potential are zero on average, the following  lateral asymmetry parameter is finite
\begin{equation}
	\label{Xi}
	\Xi =\overline{E_0^2(\bm r) \nabla_x {V}(\bm r)},
\end{equation}
where the line denotes averaging over the structure period. The parameter $\Xi$, already introduced in the phenomenological Eqs.\,\eqref{phenom_Cs}, reflects the $C_s$ symmetry of the system giving rise to the corresponding photocurrent contributions. It is clear that  $\Xi=0$  if both the profile of $V(\bm r)$ and $E_0^2(\bm r)$ have identical coordinate dependence. However, the near-field profile is different from that of the triangles, see Fig.\,\ref{fig8-new}(b), and subsequently the parameter $\Xi$ is nonzero.
Note that in the case of $C_1$ point group, an additional lateral asymmetry parameter $\overline{E_0^2(\bm r) \nabla_y {V}(\bm r)}$ emerges.

While  in metamaterials with $\lambda \gg d$ the microscopic mechanism of the  polarization-dependent photocurrent contributions proportional to $P_{\rm lin}$ and $P_{\rm lin}^\prime$  are similar to that obtained above for the isolated triangle antidots the polarization-independent one is new. Below we develop the theory of ratchet effects yielding these contributions. We show that they are caused by a radiation-induced electron gas heating, which results in coordinate-dependent temperature increase, see  $\delta T(\bm r)$ in Fig.\,\ref{fig8-new}(b), which follows the  spatially dependent electric field squared $\abs{\bm E(\bm r)}^2$ formed due to the near-field of diffraction of THz radiation. 

\subsection{Seebeck ratchet current at $\bf M=0$}
\label{Seebeck_M0}

We begin with the Seebeck ratchet photocurrent formed at zero magnetization. As addressed above, the model of its generation is based on  electron gas heating by the near-field $\bm E(\bm r)$. The heating results in the emergence of an inhomogeneous profile of electron temperature $\delta T(\bm r) \propto \abs{\bm E(\bm r)}^2$. It is found from the energy balance:
\begin{equation}
	\label{dT}
	N {k_{\rm B}\delta T(\bm r) \over \tau_T}={2\sigma_0  \over 1+(\omega\tau_1)^2}\abs{\bm E(\bm r)}^2,
\end{equation}
where 
$k_{\rm B}$ is the Boltzmann constant, $\tau_T$ is the temperature relaxation time, $\sigma_0$ is the dc conductivity, and $\tau_1$ taken at the Fermi energy is the transport relaxation time. This inhomogeneous heating leads to a spatial modulation of the dc conductivity $\delta\sigma(\bm r)=\delta T(\bm r)\partial\sigma_0/\partial T$ provided by the temperature dependence $\sigma_0(T)$. As a result, the acceleration of electrons by the periodic electric field $(-1/e)\bm \nabla V(\bm r)$ is decompensated,
and
the electric current is generated with the density~\cite{Nalitov2012}
\begin{equation}
	\label{j_Seebeck}
	j_x =-{1\over e}\pdv{\sigma_0}{T} \overline{\delta T(\bm r) \nabla_x {V}(\bm r)}.
\end{equation}
As the near-field profiles, formed at the borders of neighboring antidots, overlap, we obtain different electric field amplitudes. Consequently, temperatures differ at the triangle bases and apexes along $x$-axis, see  Fig.\,\ref{fig8-new}(b). This figure also demonstrates that the electrostatic force proportional to $\partial V/\partial x$ along this axis, being caused by the triangular shape of the antidots, is asymmetric as well. Therefore, the above average is nonzero giving rise to the Seebeck ratchet current, which is described by the constant $\chi_0$ in Eqs.\,\eqref{phenom_Cs} 
\begin{equation}
	j_x = \Xi\chi_0 =  -\Xi\frac{2\sigma_0\tau_T\partial \sigma_0/\partial T}{eN k_{\rm B} (1+\omega^2\tau_1^2)}.
	\label{eq:chi_0}
\end{equation}

Note that there is an additional contribution to the photocurrent not related to  electron heating.  It is caused by the Dynamic Carrier-Density Redistribution (DCDR)~\cite{Ivchenko2011,Nalitov2012,Faltermeier2018}. 
This mechanism yields a ratchet current density in the form~\cite{Faltermeier2018}
\begin{equation}
	\label{eq:j_non-heat}
	j_\alpha^{\rm DCDR}={i\over 2e\omega}\Xi \sum_{\beta,\eta=x,y}\pdv{\sigma_{\alpha\beta}}{N}\pdv{\sigma_{x\eta}(\omega)}{\varepsilon_{\rm F}} e_\eta e_\beta^* + c.c.,
\end{equation}
where $\hat{\sigma}$ and $\hat{\sigma}(\omega)$ are the tensors of the dc and ac conductivities.

\subsection{Spin ratchet current caused by the magnetization}
\label{Seebeck_spin_th}

Now we turn to the magnetization-induced ratchet effect. Similarly to the  Seebeck contribution considered above, the photocurrent $j_y= M_z\Phi_0\Xi$ microscopically also appears due to inhomogeneous heating of the electron gas resulting in the spatially oscillating temperature correction profile $\delta T(\bm r)$. We denote this contribution the anomalous Nernst ratchet current.

In the presence of magnetization, the electron dc conductivity is a tensor with the off-diagonal conductivity component $\sigma_{yx}$. It is odd in $M_z$ and $\sigma_{yx} \propto M_z$ at low magnetization. The presence of the Hall conductivity is crucial for the magnetization-induced polarization-independent contribution. In the presence of magnetization, the static periodic force of the potential $V(\bm r)$ results in the perpendicular electric current component $j_y(\bm r)=(-1/e)\sigma_{yx}\nabla_x V(\bm r)$. Due to the radiation near-field, a spatially oscillating part of the Hall conductivity $\delta \sigma_{yx}(\bm r)=\delta T(\bm r)\partial \sigma_{yx}/\partial T$ emerges which is odd in $M_z$. As a result, we obtain the Anomalous Nernst ratchet current in the form
\begin{equation}
	\label{Seebeckspin1}	
	j_y (M_z) =-{1\over e}\pdv{\sigma_{yx}}{T} \overline{\delta T(\bm r) \nabla_x {V}(\bm r)}.
\end{equation}
The corresponding term in Eq.\,\eqref{phenom_Cs} is given by
\begin{equation}
	\label{Seebeckspin2}
	j_y =M_z\Xi \Phi_0=  -\Xi\frac{2\sigma_0\tau_T\partial \sigma_{yx}/\partial T}{eN k_{\rm B} (1+\omega^2\tau_1^2)}.
\end{equation}

Alike  the zero-magnetization polarization-independent ratchet current, there is also a magnetization-dependent contribution to the photocurrent caused by the DCDR mechanism and not related to electron heating.
%
Due to the AHE, we have the off-diagonal conductivity component $\sigma_{yx} \propto M_z$. 
Taking for unpolarized radiation in Eq.\,\eqref{eq:j_non-heat}  $e_\eta e_\beta^*=\delta_{\eta\beta}/2$, 
and the ac conductivity in the form $\sigma_{xx}(\omega)=\sigma_0/(1-i\omega\tau_1)$ with $\sigma_0$ being linear in the Fermi energy,
we obtain the following polarization-independent spin ratchet current density
\begin{equation}
	\label{DCDRspin}
	j_y^{\rm DCDR} = -\Xi{\sigma_0\tau_1\partial\sigma_{yx}/\partial N\over 2e\varepsilon_{\rm F}(1+\omega^2\tau_1^2)} .
\end{equation}
This expression shows that the DCDR mechanism is based on the electron concentration dependence of the AHE conductivity rather than on its dependence on temperature.

\section{Discussion}
\label{discussion}

\subsection{Seebeck spin ratchet}
\label{discussion_Seebeck}

The  THz radiation-driven Seebeck ratchet and spin ratchet effects considered above manifest themselves in the presence of the polarization-independent photocurrents $J_\mathrm{R}^{\mathrm{nm}}$ and $J_\mathrm{R}^{\mathrm{m}}$, respectively. They are clearly detected in experiments applying radiation with $\lambda = \SI{118}{\micro m}$, see Figs.\,\ref{fig2}(a) and \,\ref{fig3}, but absent in results obtained with two orders of magnitude smaller wavelength  $\lambda = \SI{0,8}{\micro m}$. This observation is an important consequence of the symmetry reduction from trigonal $C_{3v}$-symmetry to the lower one (Sec.\,\ref{theo:Seebeck spin ratchet}) and is fully in line with the phenomenological Eqs.\,\eqref{phenom_C3v} and \eqref{phenom_Cs} obtained for the individual triangles with symmetry $C_{3v}$ (probed at $\lambda \leq d$) and metamaterial with symmetry $C_s$ or $C_1$ (probed at $\lambda \gg d$), respectively. 
The reduction of symmetry may be caused by different factors such as deviation of the antidot positions in the superlattice, non-ideal and different shapes of the triangles as well as 
by the fact that
the lateral structure is smaller than the beam spot and, consequently, 
this structure can not be considered as infinite. 
Equations~\eqref{phenom_C3v} and~\eqref{phenom_Cs} show that while the polarization dependence of the photocurrents in both cases is defined by the Stokes parameters $P_\mathrm{lin}$ and $P^\prime_\mathrm{lin}$ the polarization-independent contributions are present only in the case $\lambda \gg d$ and forbidden for the photoexcitation of the individual triangles realized for $\lambda \leq d$.

For the $C_s$-symmetry  describing an infinite superlattice,  the zero-magnetization contribution $J_\mathrm{R}^{\mathrm{nm}}$ becomes possible in  $x$-direction, i.e. along the height of the triangles. This photocurrent was indeed observed in the experiment, see red trace in Fig.\,\ref{fig2}(a). The microscopic derivation of the polarization-independent photocurrent density $j_x$ presented in Sec.\,\ref{theo:Seebeck spin ratchet}, which is obtained for the $C_s$ symmetry, gives an intuitive picture. Indeed,  the spatial temperature profile due to the near-field--induced heating is insensitive to the electric field orientation, see Eq.\,\eqref{dT}. The corresponding asymmetric profile of the temperature $\delta T(x)$ together with the asymmetric electrostatic force $\partial V /\partial x$, see Fig.\,\ref{fig8-new}(b), yields a polarization-independent Seebeck ratchet current in  $x$-direction. Note that without magnetization the spin of electrons is not involved in the photocurrent formation.  In the experiment we also detected a polarization-independent photocurrent in  $y$-direction,  see blue trace in Fig.\,\ref{fig2}(a). This observation indicates a further symmetry reduction from $C_s$ with a reflection plane ($xz$) to  $C_1$ which has no nontrivial symmetry operations. 

The magnetization induced in the layers because of the magnetic field applied in $z$-direction results -- in agreement with the theoretical consideration -- in  photocurrents proportional to  $M_z$: they exhibit a hysteresis at small magnetic fields, become constant at high magnetic fields and change sign at reversal of the magnetization $M_z$. Note that, also in line with theory, the magnetic-field even parts do not change upon  variation of  magnetic field, see the insets in Fig.\,\ref{fig4-new}. Comparing the hysteresis traces of the spin ratchet current $J_\mathrm{R}^{\mathrm{m}}$ with the Faraday rotation angle, measured in the unpatterned Co/Pt multilayer film, shows that both have equal widths, see Fig.\,\ref{fig5-new}. This demonstrates  that the $J_\mathrm{R}^{\mathrm{m}}$ photocurrent is formed in the bulk of the material, which is in agreement with the mechanism of the spin ratchet current originating from the spatially periodic temperature gradient caused by the near-field of diffraction, see Sec.~\ref{Seebeck_phenom} and Fig.\,\ref{fig8-new}. The spin ratchet is caused by the AHE, see Sec.~\ref{Seebeck_spin_th}. It is generated in the direction normal to the Seebeck ratchet effect. For $C_s$ symmetry the latter one is given by $j_x = \Xi \chi_0$ (see Eqs.\,\eqref{phenom_Cs}, \eqref{j_Seebeck} and \eqref{eq:chi_0}) and the spin ratchet is described by $M_z\Xi\Phi_0$, see  Eqs.\,\eqref{phenom_Cs}, \eqref{Seebeckspin1} and \eqref{Seebeckspin2}. 
Note that in both zero-magnetization and spin ratchet effect a photocurrent due to the Dynamic Carrier-Density Redistribution can yield additional contributions, see Eqs.\,\eqref{eq:j_non-heat} and 	\eqref{DCDRspin}. Both mechanisms behave equally upon variation of polarization, lateral asymmetry parameter $\Xi$ and frequency. Therefore, experimental discrimination between them is a challenging task and out of scope of the present paper.  

\subsection{Trigonal spin photocurrent}
\label{discussion_trigonal}

Now we consider the polarization-dependent photoresponse. Figures~\ref{fig2} and~\ref{fig3} reveal that the photocurrent exhibits a characteristic dependence on the orientation of the radiation electric field vector relative to the bases of the triangle antidots. Figure~\ref{fig2}  shows that the photocurrent contributions excited in $x$- and $y$-directions for both $\lambda\approx d$ and $\lambda \gg d$, vary respectively as $P_\mathrm{lin}$ and $P_\mathrm{lin}^\prime$, which are given by  Eq.\,\eqref{alpha}.  This observation is fully in line with the phenomenological theory obtained for the isolated triangles, see Eqs.\,\eqref{phenom_C3v},  as well as for the metamaterial, see   Eqs.\,\eqref{phenom_Cs}. The microscopic mechanism of the zero-magnetization polarization-dependent photocurrent contributions is described in Sec.\,\ref{micro_trigonal}. It is based on asymmetric scattering on the triangle antidot boundaries and yields a spin-independent photocurrent. As expected, this mechanism also yields  the polarization dependence observed in the experiment. For $\lambda \approx d$, the amplitudes of the photocurrents $j_x$ and $j_y$ are defined by the same parameter $\chi$ given by Eq.\,\eqref{eq:chi}. Equal amplitudes for both photocurrents are indeed observed in experiments with $\lambda= \SI{0.8}{\micro m}$, see Fig.\,\ref{fig2}(b). For $\lambda \gg d$, the corresponding parameters $\chi_1$ and $\tilde{\chi}_1$ become independent, i.e.,  may have different values. In the corresponding experiment with $\lambda= \SI{118}{\micro m}$ they differ, however, only slightly (by about factor 1.4), see Fig.\,\ref{fig2}(a)~\footnote{Note that the traces were obtained from two separated arrays with identical parameters, which, however, may have slightly different characteristics, see Fig.\,\ref{figA1} in Appendix\,\ref{appendixA}.}. This observation shows that also in this case the microscopic origin of the photocurrent can be well described by the mechanism introduced in Sec.\,\ref{micro_trigonal}, which considers  $C_{3v}$ symmetry and yields equal factors for  $j_x$ and $j_y$.  Microscopically, it stems from the asymmetric scattering at the antidots boundaries, see Figs.\,\ref{fig7-new}(a) and\,(b).

The magnetization $M_z$ results in a phase shift of the polarization-dependent contributions, see Fig.\,\ref{fig3}(a). It is caused by the emergence of the magnetization-induced photocurrent which is 45$^\circ$ phase shifted with respect to the zero-magnetization one, see Eqs.\,\eqref{phenom_Cs}. The experimental traces for  $J_{x}^{\mathrm{nm}}$ and $J_{x}^{\mathrm{m}}$ in Fig.\,\ref{fig3}(b) are in full agreement with the phenomenological Eqs.\,\eqref{phenom_Cs} and \eqref{alpha} as well as with the microscopic theory of the trigonal photocurrents presented in Sec.\,\ref{theory}. Alike the spin ratchet, the magnetization-driven trigonal photocurrent $J_\mathrm{\rm tr}^{\mathrm{m}}$  exhibits a hysteresis at small magnetic fields, becomes constant at high  fields and changes its sign  switching the sign of  $M_z$, see Figs.\,\ref{fig4-new}(b) and \ref{fig5-new}(b). Microscopically, the trigonal photocurrent proportional to $M_z$ is caused by  skew scattering (see Sec.\,\ref{AHE_trigonal} and \ref{spin_trigonal}) and  magnetization-dependent scattering, see Sec.~\ref{magnetization_trigonal3}.  

While the above mechanisms are considered theoretically we note that an additional microscopic mechanism may give rise to  the polarization-dependent photocurrent. One could explain it by the polarization-dependent lateral anisotropy of the laser heating which leads to the anomalous Nernst effect. The anisotropy may be obtained considering the patterned structure as a near-field antenna which results in the laser heating around small central regions, i.e., increased temperature in the vicinity of the triangle's vertices. This yields a maximum of the radiation absorption, when the light polarization is parallel to the heights of triangles, and creates the polarization-dependent temperature gradients $\bm \nabla T(\alpha)$ with triangular space symmetry. Due to the magnetization, this  spatial temperature profile results in the trigonal ANE current $\bm j^{\rm ANE} \propto \bm M \times \bm \nabla T$. This contribution also has a $j^{\rm ANE}_x \propto \sin{2\alpha}$ dependence on the polarization orientation as the current $J^{\rm m}_{\rm tr}$ detected in the experiment, Fig.\,\ref{fig3}(b). A theoretical treatment of this mechanism is a task for future research. Note that in the above mechanism, as well as in the mechanisms considered theoretically in Secs.~\ref{AHE_trigonal} and \ref{spin_trigonal}, the photocurrent $J_\mathrm{\rm tr}^{\mathrm{m}}$  is spin polarized and can be classified as a trigonal spin current.

Finally, we discuss the hysteresis width of the trigonal spin current $J_\mathrm{\rm tr}^{\mathrm{m}}$. Surprisingly, it is about two times larger than that of the spin ratchet current, as well as of the hysteresis of the Faraday rotation angle in the unpatterned Co/Pt films, see photocurrent traces (red and blue lines) and Faraday angle traces (gray lines) in  Figs.\,\ref{fig5-new}(a) and (b). We emphasize that both photocurrent traces were obtained from one experiment and were extracted using the difference in the polarization dependence. Different hysteresis widths for the spin ratchet and trigonal contributions are clearly seen in the original data sets obtained for different azimuth angles $\alpha$, see Fig.\,\ref{fig:300K-hys}; in particular for $\alpha = 135^\circ$ at which $J_\mathrm{\rm R}^{\mathrm{m}}$ and $J_\mathrm{\rm tr}^{\mathrm{m}}$ have opposite signs. This difference clearly indicates that the trigonal photocurrent and the spin ratchet current are formed in different regions of the superlattice. This is in line with the microscopic theory developed above. We have shown that the trigonal photocurrent are formed at the boundaries of the triangle antidots, see Sec.~\ref{theory},  whereas  the spin ratchet current is  generated in the film bulk, see Sec.~\ref{Seebeck_spin_th}. While our paper is primarily aimed at the  spin ratchet, the larger hysteresis width of the trigonal photocurrent needs an additional discussion.  As we addressed above, the hysteresis width of the spin ratchet current coincides  with that of the magnetization measured by  Faraday rotation in the unpatterned films, which agrees with the theory developed in Sec.~\ref{Seebeck_spin_th}. The deviation of the  hysteresis width in the trigonal current with respect to the magnetization in the homogeneous film should be caused by the difference between the magnetization of the unpatterned film and the film in the vicinity of  the antidot boundaries. This can be caused by the enhanced electric fields due to the near-field of diffraction and, related to  local heating, inhomogeneity of $M_z$ at the antidot edges, magnetic domain formation in the superlattice, domain wall pinning, etc. Additional experiments  using  a pulsed laser operating at a similar frequency but almost five orders of magnitude higher intensity demonstrated that such a drastic increase in the radiation electric field  scales only the signal magnitude but does not change the hysteresis widths of $J_\mathrm{\rm tr}^{\mathrm{m}}$, see Fig.\,\ref{fig6-new}. This result excludes changes in the magnetization behavior induced by the radiation electric field, e.g. due to 
the heating of the antidot edges. Note that this is not surprising, because in both cases one would expect a narrowing of the magnetization hysteresis. 

\begin{figure}[tbh] 
	\centering
	\includegraphics[width=\linewidth]{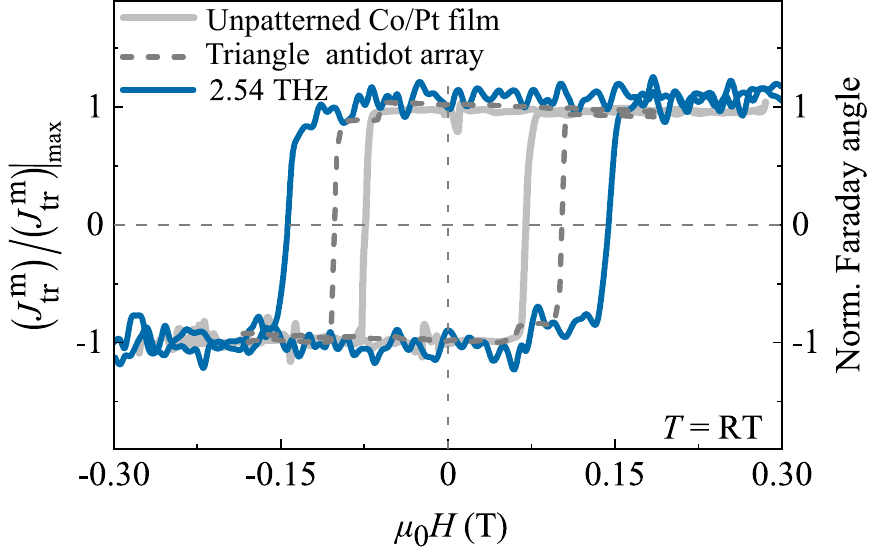}
	\caption{ Magnetic field dependence of the magnetization-induced  polarization-dependent  photocurrent contribution (blue curve) and Faraday angles measured in unpatterned Co/Pt film (solid curve) and the array of the equilateral triangle-shaped antidots  (dashed curve). The photocurrent is measured in $x$-direction and normalized on its maximum, $(J_\mathrm{tr}^{\mathrm{m}}) / (J_\mathrm{tr}^{\mathrm{m}})|_{\rm max}$. Faraday angle traces are also normalized on their maximum values. Coercive fields obtained from the Faraday angle traces are larger for the antidots array ($\mu_0H_\mathrm{c} = 102\,$mT) than that for the unpatterned Co/Pt film with the same layer design (76\,mT). Note that Faraday rotation yields the integrated response from the whole structure. The  coercive field for the trigonal photocurrent is $\mu_0H_\mathrm{c} = 144\,$mT. The photocurrent is formed because of scattering at the triangle boundaries. Consequently, it probes the local magnetization in the vicinity of the antidot edges. 
}
	\label{fig9-new}
\end{figure}

Performing additional experiments on  Faraday rotation on unpatterned films and patterned films with the lateral antidot superlattice we observed that the hysteresis width in the latter case is larger than in the former one, see Fig.\,\ref{fig9-new}. Note that in the magneto-optic experiments we obtain the integrated response from the bulk part and the antidot edges, whereas the trigonal photocurrent gives a signal proportional to the magnetization at the edges only. This result gives a hint that the larger hysteresis width detected in $J_\mathrm{\rm tr}^{\mathrm{m}}$ is most probably related to spatially inhomogeneous switching of the magnetization $M_z$. We can give three simple reasons that may explain the increased width of the magnetic hysteresis in the patterned region. First, we expect that magnetization reversal in perpendicularly magnetized Co/Pt films proceeds via domain wall nucleation and propagation. Owing to the stray field of the perpendicularly magnetized films, the nucleation field is reduced. However, since at the edges of the triangles the magnetic film has been removed,   the stray field is diminished in the edge regions leading to an increased nucleation/switching field. Second, domain walls are pinned at the apexes of triangles so as to reduce the wall area. Third, in the region of the edges defects and pinning sites introduced by lithographic structuring may hinder the propagation of domain walls which are nucleated in the center of the elements towards the edges, leading again to a locally increased switching field.

\section{Summary}
\label{summary}

Our results demonstrate that terahertz radiation with a wavelength substantially larger than the period of the array of the triangle-shaped antidots fabricated from Co/Pt films results in a polarization-independent current shown to be caused by the anomalous Nernst spin ratchet effect. It is generated in the film bulk and exhibits a hysteresis in  magnetic field range comparable to that of the Faraday rotation data measured in unpatterned samples. The polarization-dependent trigonal spin photocurrent, also detected in our experiments, is generated by both  terahertz and infrared radiation.  The trigonal spin photocurrent has a larger hysteresis width because of spatial inhomogeneity of the magnetization probed by the polarization-dependent photocurrent generated  in the region of the antidots edges. This observation provides a novel access to studying the magnetization at the metal films edges.

\section{Acknowledgments}
\label{acknow}

We thank C. Gorini for the fruitful discussions. The support of  the Deutsche Forschungsgemeinschaft (DFG, German Research Foundation) project Ga501/18, RFBR project 21-52-12015, IRAP  Programme  of the Foundation   for   Polish Science   (grant   MAB/2018/9, project CENTERA),  the Volkswagen Stiftung Program (97738),  JSPS KAKENHI (Grant Nos. 21H04649, 22K18962), the Research Foundation for Opto-Science and Technology, the Asahi Glass Foundation, the Support Center for Advanced Telecommunications Technology Research Foundation, and the Tanaka Memorial Foundation is gratefully acknowledged. Sample fabrication was supported by Nanotechnology Platform Program of the Ministry of Education, Culture, Sports, Science and Technology (MEXT), Japan, Grant Number JPMXP09-F21-NU-0046, Nagoya University.

\appendix
\counterwithin{figure}{section}
\setcounter{figure}{0}

\section{Samples characteristics}
\label{appendixA}

\begin{figure}
	\centering
	\includegraphics[width=\linewidth]{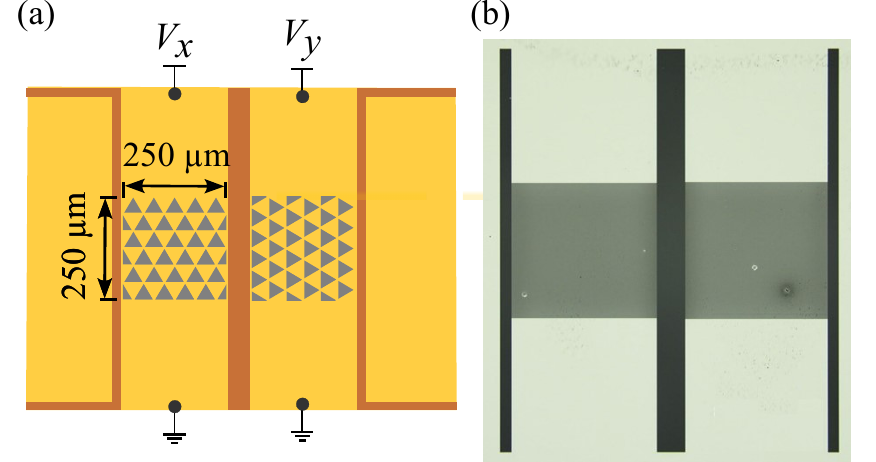}
	\caption{Sketch of the studied sample consisting of  two electrically isolated arrays of the triangle-shaped antidots rotated by \ang{90} with respect to each other. The arrays have a size of $250 \times 250\,\SI{}{\micro m^2}$ and  period of \SI{0.55}{\micro m}. Note that the triangles in the sketch are oversized for better visibility. Two pairs of contacts allow one to probe the photocurrent along the height (left part) and the basis (right side) of the triangles, i.e. in $x$- and $y$-direction. 
	}
	\label{figA1}
\end{figure}

The investigated sample hosts two lateral superlattices formed by the triangle-shaped antidots  which are rotated by \ang{90} with respect to each other, see Fig.~\ref{figA1}. The arrays are electrically isolated using  electron beam lithography and laser cutting (see Ref.\,\cite{Matsubara2022} for more details). This allowed us to probe the photocurrent in $x$- and $y$-direction simultaneously in a single experiment.

\begin{figure}
	\centering
	\includegraphics[width=\linewidth]{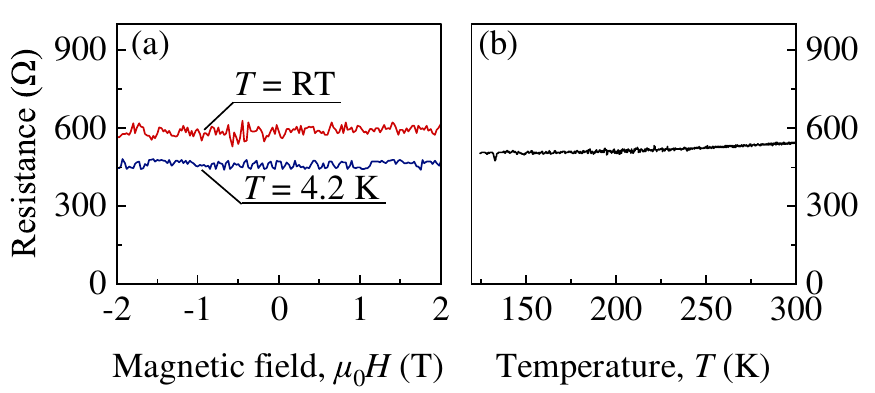}
	\caption{ Panel (a): Dark resistance as a function of magnetic field measured along the triangles' height with a two-point measurement setup. Panel (b): Temperature dependence of the dark resistance at $\mu_0H = 0$. 
	}
	\label{figA2}
\end{figure}

Figure~\ref{figA2}(a) shows the magnetic field of the  dark resistance of the metamaterial measured in a two-point configuration with a bias current of \SI{100}{nA}. It shows that in the studied magnetic field range from \SI{-2}{T} to \SI{2}{T} the resistance remains unchanged. It also exhibits only  a weak dependence on temperature, see Fig.~\ref{figA2}(b). 


\section{Hysteresis of the trigonal photocurrent measured for different azimuth angles}
\label{appendixA2}

\begin{figure}
	\centering
	\includegraphics[width=\linewidth]{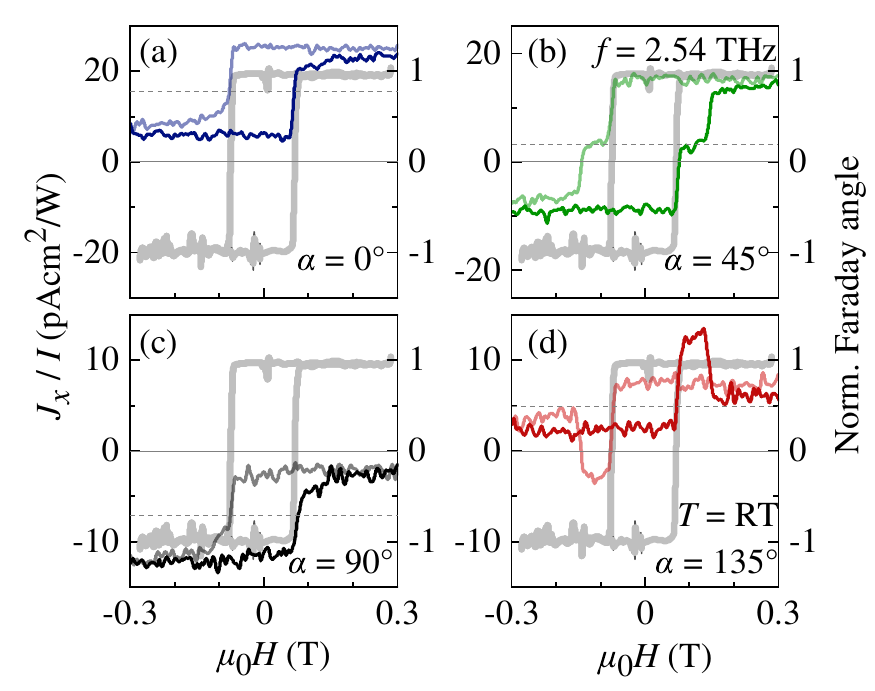}
	\caption{ 
The angle has been chosen such that in the formula  $A_1\cos2\alpha+C_1+M_zA_2\sin2\alpha+MC_2$ either the sine or cosine function vanishes and the remaining part changes its sign by adding \ang{90} (compare panel (a) with (c) and (b) with (d)). The dashed lines indicate the magnetization-independent contribution $A_1\cos2\alpha+C_1$, wich reduces to $C_1$ for $\ang{45}$ and $\ang{90}$. Curves with full and reduced opacity show the forward and	backward magnetic field sweeps, respectively. The gray lines show the magnetic field dependence of the Faraday rotation angle obtained in the unpatterned Co/Pt film normalized on its maximum value.
	}
	\label{fig:300K-hys}
\end{figure}

Figure~\ref{fig:300K-hys} shows the hysteresis of the photocurrent measured for four different azimuth angles $\alpha$. The angles are selected in a way that the polarization-dependent contribution  $J_x = M_zA_2 \sin 2\alpha$ either vanishes and only the polarization-independent part forms the photocurrent ($\alpha = \ang{0}, \ang{90}$) or has a maximum and opposite signs at $\alpha = \ang{45}, \ang{135}$. The figure reveals that in the former case the hysteresis width coincides with the magneto-optical Faraday data of the unpatterned Co/Pt multilayer film (gray traces). For a radiation electric field applied at angles $\alpha = \ang{45}, \ang{135}$ two different hysteresis widths are clearly seen, leading to a step-like change of the widths. It is particularly pronounced for   $\alpha = \ang{135}$ at which the polarization-dependent and polarization-independent parts have opposite signs.

\section{SQUID and magneto-optic Kerr microscopy data}
\label{appendixB_SQUID}

To study the magnetization switching properties of the sample in the patterned and unpatterned regions, we performed additional experiments.

\begin{figure}[tbh] 
	\centering
		\includegraphics[width=\linewidth]{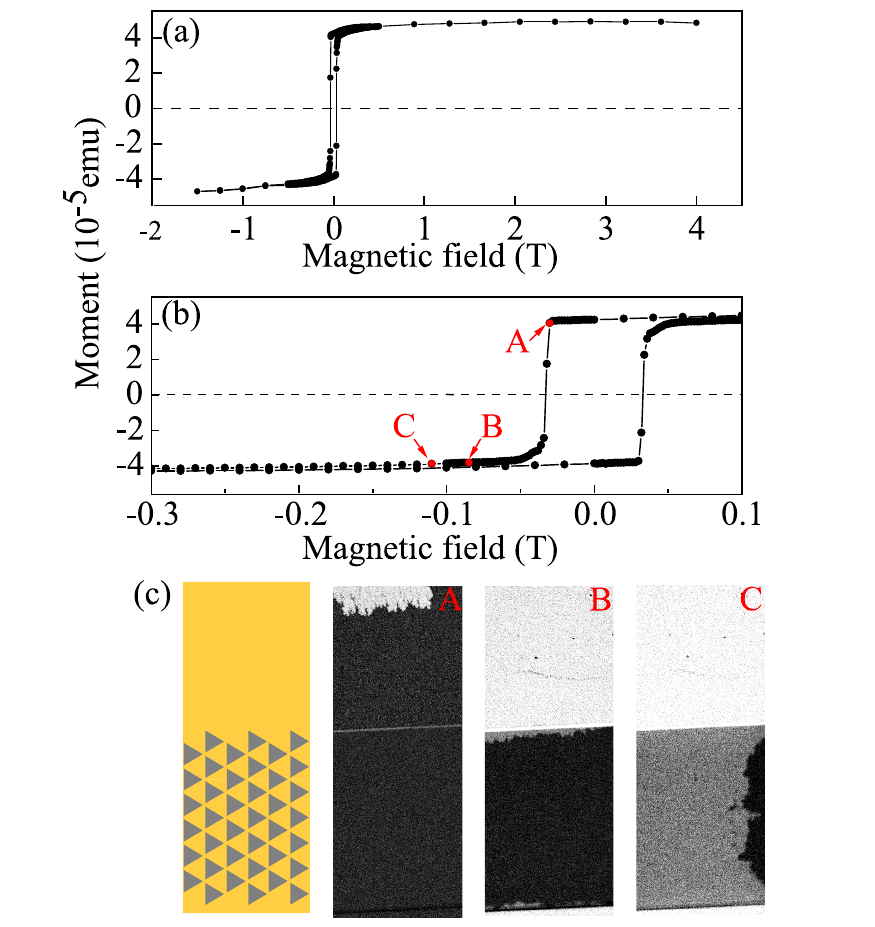}
	\caption{Panels (a) and (b). SQUID loop measured on the full film sample with unpatterned (plane Co/Pt film) and patterned (array of triangular antidots) regions, see the sketch shown in the left side of the panel (c). The loops is measured for magnetic fields in a range from \SI{-1.5}{T} to \SI{4}{T} applied perpendicular to the sample plane, see panel (a). The panel (b) is a zoom-in showing the data for magnetic fields varying from \SI{-0.3}{T} to \SI{0.1}{T}. Panel (c) shows the sample sketch (left) and the Kerr microscopy images taken at the points marked in panel (a) for magnetic fields: \SI{-30}{mT} [point A in the panel (b)];  \SI{-84}{mT} (point B); and \SI{-110}{mT} (point C). In (c) black and white contrast corresponds to upward and downward out-of-plane magnetization, respectively.
	}
	\label{Kronseder}
\end{figure}

Superconducting quantum interference device (SQUID) measurements are used to determine the magnetic properties of the full sample. In Fig.~\ref{Kronseder} we show the SQUID data obtained in a sample with the same  design. In Fig.\,\ref{Kronseder} (a) a SQUID loop measured with magnetic field applied perpendicular to the sample plane and ranging from +4~T to $-1.5$~T is shown. The measurements reveal magnetization saturation at around 1.5~T and a coercive field of around 33~mT, as can be seen in the zoom-in in Fig. \ref{Kronseder}(b). 
Figures~\ref{Kronseder}(c) show polar Kerr-microscopy images in which both patterned and unpatterned parts of the sample were captured simultaneously. The images show the magnetization state during the reversal process (black and white corresponds to upward and downward out-of-plane magnetization, respectively) after application of magnetic fields with different amplitudes. The unpatterned region of the sample shows a coercivity of around 33~mT. Interestingly, to switch the magnetization  in the patterned part of the sample a much larger magnetic field needs to be applied as can be seen in the lower parts of the Kerr-microscopy images. In the measured SQUID loop, a large jump in the signal is observed when the unpatterned part of the magnetic film switches since most of the sample is unpatterned and SQUID is an integrating measurement technique. Note that the spatial resolution of the Kerr microscope does not allow the detection of magnetization reversal of the individual triangles, however, their influence as pinning centers for magnetization reversal can nevertheless be detected.

Magnetization reversal processes in thin films are strongly influenced by crystal defects in the ferromagnetic layer leading to nucleation and pinning sites for domain walls. This is true for both the patterned and the unpatterned part of the sample (see e.g. the fringed domain wall in the upper part of the first image taken at $-30$~mT). Since a single triangle can be subdivided into an inner part untouched by the structuring process and the rim of the triangle, in which crystal defects are very likely to have been introduced by the mechanical treatment during the structuring process, the magnetization reversal of a single perpendicularly magnetized triangle could be stepwise. At lower magnetic fields the inner part of the triangle switches by domain nucleation and propagation since the typical domain wall width for Co/Pt multilayer is much smaller than the size of a triangle. However, the switching process at the rim of a triangle differs for three reasons. First, the missing material reduces the amount of stray field, which aids magnetization switching in perpendicularly magnetized materials, and second, domain walls are pinned at the apexes of triangles so as to reduce the wall area. Third, domain wall propagation towards the rim of the triangles may be hindered by defects and pinned moments. This would also be in line with the SQUID data, which shows saturation of a fraction of the magnetization only at very large magnetic fields.

\bibliography{all_lib.bib}

\begin{thebibliography}{74}%
\makeatletter
\providecommand \@ifxundefined [1]{%
 \@ifx{#1\undefined}
}%
\providecommand \@ifnum [1]{%
 \ifnum #1\expandafter \@firstoftwo
 \else \expandafter \@secondoftwo
 \fi
}%
\providecommand \@ifx [1]{%
 \ifx #1\expandafter \@firstoftwo
 \else \expandafter \@secondoftwo
 \fi
}%
\providecommand \natexlab [1]{#1}%
\providecommand \enquote  [1]{``#1''}%
\providecommand \bibnamefont  [1]{#1}%
\providecommand \bibfnamefont [1]{#1}%
\providecommand \citenamefont [1]{#1}%
\providecommand \href@noop [0]{\@secondoftwo}%
\providecommand \href [0]{\begingroup \@sanitize@url \@href}%
\providecommand \@href[1]{\@@startlink{#1}\@@href}%
\providecommand \@@href[1]{\endgroup#1\@@endlink}%
\providecommand \@sanitize@url [0]{\catcode `\\12\catcode `\$12\catcode
  `\&12\catcode `\#12\catcode `\^12\catcode `\_12\catcode `\%12\relax}%
\providecommand \@@startlink[1]{}%
\providecommand \@@endlink[0]{}%
\providecommand \url  [0]{\begingroup\@sanitize@url \@url }%
\providecommand \@url [1]{\endgroup\@href {#1}{\urlprefix }}%
\providecommand \urlprefix  [0]{URL }%
\providecommand \Eprint [0]{\href }%
\providecommand \doibase [0]{http://dx.doi.org/}%
\providecommand \selectlanguage [0]{\@gobble}%
\providecommand \bibinfo  [0]{\@secondoftwo}%
\providecommand \bibfield  [0]{\@secondoftwo}%
\providecommand \translation [1]{[#1]}%
\providecommand \BibitemOpen [0]{}%
\providecommand \bibitemStop [0]{}%
\providecommand \bibitemNoStop [0]{.\EOS\space}%
\providecommand \EOS [0]{\spacefactor3000\relax}%
\providecommand \BibitemShut  [1]{\csname bibitem#1\endcsname}%
\let\auto@bib@innerbib\@empty
\bibitem [{\citenamefont {Jülicher}\ \emph {et~al.}(1997)\citenamefont
  {Jülicher}, \citenamefont {Ajdari},\ and\ \citenamefont
  {Prost}}]{Juelicher1997}%
  \BibitemOpen
  \bibfield  {author} {\bibinfo {author} {\bibfnamefont {F.}~\bibnamefont
  {Jülicher}}, \bibinfo {author} {\bibfnamefont {A.}~\bibnamefont {Ajdari}}, \
  and\ \bibinfo {author} {\bibfnamefont {J.}~\bibnamefont {Prost}},\ }\href
  {\doibase 10.1103/revmodphys.69.1269} {\bibfield  {journal} {\bibinfo
  {journal} {Rev. Mod. Phys.}\ }\textbf {\bibinfo {volume} {69}},\ \bibinfo
  {pages} {1269} (\bibinfo {year} {1997})}\BibitemShut {NoStop}%
\bibitem [{\citenamefont {Linke}(2002)}]{Linke2002}%
  \BibitemOpen
  \bibfield  {author} {\bibinfo {author} {\bibfnamefont {H.}~\bibnamefont
  {Linke}},\ }\href {\doibase 10.1007/s003390201401} {\bibfield  {journal}
  {\bibinfo  {journal} {Appl. Phys. A}\ }\textbf {\bibinfo {volume} {75}},\
  \bibinfo {pages} {167} (\bibinfo {year} {2002})}\BibitemShut {NoStop}%
\bibitem [{\citenamefont {Reimann}(2002)}]{Reimann2002}%
  \BibitemOpen
  \bibfield  {author} {\bibinfo {author} {\bibfnamefont {P.}~\bibnamefont
  {Reimann}},\ }\href {\doibase 10.1016/s0370-1573(01)00081-3} {\bibfield
  {journal} {\bibinfo  {journal} {Phys. Rep.}\ }\textbf {\bibinfo {volume}
  {361}},\ \bibinfo {pages} {57} (\bibinfo {year} {2002})}\BibitemShut
  {NoStop}%
\bibitem [{\citenamefont {Hänggi}\ and\ \citenamefont
  {Marchesoni}(2009)}]{Haenggi2009}%
  \BibitemOpen
  \bibfield  {author} {\bibinfo {author} {\bibfnamefont {P.}~\bibnamefont
  {Hänggi}}\ and\ \bibinfo {author} {\bibfnamefont {F.}~\bibnamefont
  {Marchesoni}},\ }\href {\doibase 10.1103/revmodphys.81.387} {\bibfield
  {journal} {\bibinfo  {journal} {Rev. Mod. Phys.}\ }\textbf {\bibinfo {volume}
  {81}},\ \bibinfo {pages} {387} (\bibinfo {year} {2009})}\BibitemShut
  {NoStop}%
\bibitem [{\citenamefont {Ivchenko}\ and\ \citenamefont
  {Ganichev}(2011)}]{Ivchenko2011}%
  \BibitemOpen
  \bibfield  {author} {\bibinfo {author} {\bibfnamefont {E.~L.}\ \bibnamefont
  {Ivchenko}}\ and\ \bibinfo {author} {\bibfnamefont {S.~D.}\ \bibnamefont
  {Ganichev}},\ }\href {\doibase 10.1134/s002136401111004x} {\bibfield
  {journal} {\bibinfo  {journal} {JETP Lett.}\ }\textbf {\bibinfo {volume}
  {93}},\ \bibinfo {pages} {673} (\bibinfo {year} {2011})},\ \bibinfo {note}
  {[Pisma v ZhETF \textbf{93}, 752 (2011)]}\BibitemShut {NoStop}%
\bibitem [{\citenamefont {Budkin}\ \emph {et~al.}(2016)\citenamefont {Budkin},
  \citenamefont {Golub}, \citenamefont {Ivchenko},\ and\ \citenamefont
  {Ganichev}}]{Budkin2016a}%
  \BibitemOpen
  \bibfield  {author} {\bibinfo {author} {\bibfnamefont {G.~V.}\ \bibnamefont
  {Budkin}}, \bibinfo {author} {\bibfnamefont {L.~E.}\ \bibnamefont {Golub}},
  \bibinfo {author} {\bibfnamefont {E.~L.}\ \bibnamefont {Ivchenko}}, \ and\
  \bibinfo {author} {\bibfnamefont {S.~D.}\ \bibnamefont {Ganichev}},\ }\href
  {\doibase 10.1134/s0021364016210074} {\bibfield  {journal} {\bibinfo
  {journal} {JETP Lett.}\ }\textbf {\bibinfo {volume} {104}},\ \bibinfo {pages}
  {649} (\bibinfo {year} {2016})}\BibitemShut {NoStop}%
\bibitem [{\citenamefont {Lau}\ and\ \citenamefont {Kedem}(2020)}]{Lau2020}%
  \BibitemOpen
  \bibfield  {author} {\bibinfo {author} {\bibfnamefont {B.}~\bibnamefont
  {Lau}}\ and\ \bibinfo {author} {\bibfnamefont {O.}~\bibnamefont {Kedem}},\
  }\href {\doibase 10.1063/5.0009561} {\bibfield  {journal} {\bibinfo
  {journal} {The Journal of Chemical Physics}\ }\textbf {\bibinfo {volume}
  {152}},\ \bibinfo {pages} {200901} (\bibinfo {year} {2020})},\ \Eprint
  {http://arxiv.org/abs/https://doi.org/10.1063/5.0009561}
  {https://doi.org/10.1063/5.0009561} \BibitemShut {NoStop}%
\bibitem [{\citenamefont {Olbrich}\ \emph {et~al.}(2009)\citenamefont
  {Olbrich}, \citenamefont {Allerdings}, \citenamefont {Bel'kov}, \citenamefont
  {Tarasenko}, \citenamefont {Schuh}, \citenamefont {Wegscheider},
  \citenamefont {Korn}, \citenamefont {Sch\"uller}, \citenamefont {Weiss},\
  and\ \citenamefont {Ganichev}}]{Olbrich2009}%
  \BibitemOpen
  \bibfield  {author} {\bibinfo {author} {\bibfnamefont {P.}~\bibnamefont
  {Olbrich}}, \bibinfo {author} {\bibfnamefont {J.}~\bibnamefont {Allerdings}},
  \bibinfo {author} {\bibfnamefont {V.~V.}\ \bibnamefont {Bel'kov}}, \bibinfo
  {author} {\bibfnamefont {S.~A.}\ \bibnamefont {Tarasenko}}, \bibinfo {author}
  {\bibfnamefont {D.}~\bibnamefont {Schuh}}, \bibinfo {author} {\bibfnamefont
  {W.}~\bibnamefont {Wegscheider}}, \bibinfo {author} {\bibfnamefont
  {T.}~\bibnamefont {Korn}}, \bibinfo {author} {\bibfnamefont {C.}~\bibnamefont
  {Sch\"uller}}, \bibinfo {author} {\bibfnamefont {D.}~\bibnamefont {Weiss}}, \
  and\ \bibinfo {author} {\bibfnamefont {S.~D.}\ \bibnamefont {Ganichev}},\
  }\href@noop {} {\bibfield  {journal} {\bibinfo  {journal} {Phys. Rev. B}\
  }\textbf {\bibinfo {volume} {79}},\ \bibinfo {pages} {245329} (\bibinfo
  {year} {2009})}\BibitemShut {NoStop}%
\bibitem [{\citenamefont {Kannan}\ \emph {et~al.}(2011)\citenamefont {Kannan},
  \citenamefont {Bisotto}, \citenamefont {Portal}, \citenamefont {Murali},\
  and\ \citenamefont {Beck}}]{Kannan2011}%
  \BibitemOpen
  \bibfield  {author} {\bibinfo {author} {\bibfnamefont {E.~S.}\ \bibnamefont
  {Kannan}}, \bibinfo {author} {\bibfnamefont {I.}~\bibnamefont {Bisotto}},
  \bibinfo {author} {\bibfnamefont {J.-C.}\ \bibnamefont {Portal}}, \bibinfo
  {author} {\bibfnamefont {R.}~\bibnamefont {Murali}}, \ and\ \bibinfo {author}
  {\bibfnamefont {T.~J.}\ \bibnamefont {Beck}},\ }\href {\doibase
  10.1063/1.3590255} {\bibfield  {journal} {\bibinfo  {journal} {Applied
  Physics Letters}\ }\textbf {\bibinfo {volume} {98}},\ \bibinfo {pages}
  {193505} (\bibinfo {year} {2011})},\ \Eprint
  {http://arxiv.org/abs/https://doi.org/10.1063/1.3590255}
  {https://doi.org/10.1063/1.3590255} \BibitemShut {NoStop}%
\bibitem [{\citenamefont {Olbrich}\ \emph {et~al.}(2011)\citenamefont
  {Olbrich}, \citenamefont {Karch}, \citenamefont {Ivchenko}, \citenamefont
  {Kamann}, \citenamefont {März}, \citenamefont {Fehrenbacher}, \citenamefont
  {Weiss},\ and\ \citenamefont {Ganichev}}]{Olbrich2011}%
  \BibitemOpen
  \bibfield  {author} {\bibinfo {author} {\bibfnamefont {P.}~\bibnamefont
  {Olbrich}}, \bibinfo {author} {\bibfnamefont {J.}~\bibnamefont {Karch}},
  \bibinfo {author} {\bibfnamefont {E.~L.}\ \bibnamefont {Ivchenko}}, \bibinfo
  {author} {\bibfnamefont {J.}~\bibnamefont {Kamann}}, \bibinfo {author}
  {\bibfnamefont {B.}~\bibnamefont {März}}, \bibinfo {author} {\bibfnamefont
  {M.}~\bibnamefont {Fehrenbacher}}, \bibinfo {author} {\bibfnamefont
  {D.}~\bibnamefont {Weiss}}, \ and\ \bibinfo {author} {\bibfnamefont {S.~D.}\
  \bibnamefont {Ganichev}},\ }\href {\doibase 10.1103/physrevb.83.165320}
  {\bibfield  {journal} {\bibinfo  {journal} {Phys. Rev. B}\ }\textbf {\bibinfo
  {volume} {83}},\ \bibinfo {pages} {165320} (\bibinfo {year}
  {2011})}\BibitemShut {NoStop}%
\bibitem [{\citenamefont {Nalitov}\ \emph {et~al.}(2012)\citenamefont
  {Nalitov}, \citenamefont {Golub},\ and\ \citenamefont
  {Ivchenko}}]{Nalitov2012}%
  \BibitemOpen
  \bibfield  {author} {\bibinfo {author} {\bibfnamefont {A.~V.}\ \bibnamefont
  {Nalitov}}, \bibinfo {author} {\bibfnamefont {L.~E.}\ \bibnamefont {Golub}},
  \ and\ \bibinfo {author} {\bibfnamefont {E.~L.}\ \bibnamefont {Ivchenko}},\
  }\href {\doibase 10.1103/physrevb.86.115301} {\bibfield  {journal} {\bibinfo
  {journal} {Phys. Rev. B}\ }\textbf {\bibinfo {volume} {86}},\ \bibinfo
  {pages} {115301} (\bibinfo {year} {2012})}\BibitemShut {NoStop}%
\bibitem [{\citenamefont {Otsuji}\ \emph {et~al.}(2013)\citenamefont {Otsuji},
  \citenamefont {Watanabe}, \citenamefont {Tombet}, \citenamefont {Satou},
  \citenamefont {Knap}, \citenamefont {Popov}, \citenamefont {Ryzhii},\ and\
  \citenamefont {Ryzhii}}]{Otsuji2013}%
  \BibitemOpen
  \bibfield  {author} {\bibinfo {author} {\bibfnamefont {T.}~\bibnamefont
  {Otsuji}}, \bibinfo {author} {\bibfnamefont {T.}~\bibnamefont {Watanabe}},
  \bibinfo {author} {\bibfnamefont {S.~A.~B.}\ \bibnamefont {Tombet}}, \bibinfo
  {author} {\bibfnamefont {A.}~\bibnamefont {Satou}}, \bibinfo {author}
  {\bibfnamefont {W.~M.}\ \bibnamefont {Knap}}, \bibinfo {author}
  {\bibfnamefont {V.~V.}\ \bibnamefont {Popov}}, \bibinfo {author}
  {\bibfnamefont {M.}~\bibnamefont {Ryzhii}}, \ and\ \bibinfo {author}
  {\bibfnamefont {V.}~\bibnamefont {Ryzhii}},\ }\href {\doibase
  10.1109/tthz.2012.2235911} {\bibfield  {journal} {\bibinfo  {journal} {IEEE
  Trans. Terahertz Sci. Technol.}\ }\textbf {\bibinfo {volume} {3}},\ \bibinfo
  {pages} {63} (\bibinfo {year} {2013})}\BibitemShut {NoStop}%
\bibitem [{\citenamefont {Drexler}\ \emph {et~al.}(2013)\citenamefont
  {Drexler}, \citenamefont {Tarasenko}, \citenamefont {Olbrich}, \citenamefont
  {Karch}, \citenamefont {Hirmer}, \citenamefont {M\"uller}, \citenamefont
  {Gmitra}, \citenamefont {Fabian}, \citenamefont {Yakimova}, \citenamefont
  {Lara-Avila}, \citenamefont {Kubatkin}, \citenamefont {Wang}, \citenamefont
  {Vajtai}, \citenamefont {Ajayan}, \citenamefont {Kono},\ and\ \citenamefont
  {Ganichev}}]{Drexler2013}%
  \BibitemOpen
  \bibfield  {author} {\bibinfo {author} {\bibfnamefont {C.}~\bibnamefont
  {Drexler}}, \bibinfo {author} {\bibfnamefont {S.~A.}\ \bibnamefont
  {Tarasenko}}, \bibinfo {author} {\bibfnamefont {P.}~\bibnamefont {Olbrich}},
  \bibinfo {author} {\bibfnamefont {J.}~\bibnamefont {Karch}}, \bibinfo
  {author} {\bibfnamefont {M.}~\bibnamefont {Hirmer}}, \bibinfo {author}
  {\bibfnamefont {F.}~\bibnamefont {M\"uller}}, \bibinfo {author}
  {\bibfnamefont {M.}~\bibnamefont {Gmitra}}, \bibinfo {author} {\bibfnamefont
  {J.}~\bibnamefont {Fabian}}, \bibinfo {author} {\bibfnamefont
  {R.}~\bibnamefont {Yakimova}}, \bibinfo {author} {\bibfnamefont
  {S.}~\bibnamefont {Lara-Avila}}, \bibinfo {author} {\bibfnamefont
  {S.}~\bibnamefont {Kubatkin}}, \bibinfo {author} {\bibfnamefont
  {M.}~\bibnamefont {Wang}}, \bibinfo {author} {\bibfnamefont {R.}~\bibnamefont
  {Vajtai}}, \bibinfo {author} {\bibfnamefont {P.~M.}\ \bibnamefont {Ajayan}},
  \bibinfo {author} {\bibfnamefont {J.}~\bibnamefont {Kono}}, \ and\ \bibinfo
  {author} {\bibfnamefont {S.~D.}\ \bibnamefont {Ganichev}},\ }\href
  {http://dx.doi.org/10.1038/nnano.2012.231} {\bibfield  {journal} {\bibinfo
  {journal} {Nat. Nanotechnol.}\ }\textbf {\bibinfo {volume} {8}},\ \bibinfo
  {pages} {104} (\bibinfo {year} {2013})}\BibitemShut {NoStop}%
\bibitem [{\citenamefont {Kurita}\ \emph {et~al.}(2014)\citenamefont {Kurita},
  \citenamefont {Ducournau}, \citenamefont {Coquillat}, \citenamefont {Satou},
  \citenamefont {Kobayashi}, \citenamefont {Tombet}, \citenamefont {Meziani},
  \citenamefont {Popov}, \citenamefont {Knap}, \citenamefont {Suemitsu},\ and\
  \citenamefont {Otsuji}}]{Kurita2014}%
  \BibitemOpen
  \bibfield  {author} {\bibinfo {author} {\bibfnamefont {Y.}~\bibnamefont
  {Kurita}}, \bibinfo {author} {\bibfnamefont {G.}~\bibnamefont {Ducournau}},
  \bibinfo {author} {\bibfnamefont {D.}~\bibnamefont {Coquillat}}, \bibinfo
  {author} {\bibfnamefont {A.}~\bibnamefont {Satou}}, \bibinfo {author}
  {\bibfnamefont {K.}~\bibnamefont {Kobayashi}}, \bibinfo {author}
  {\bibfnamefont {S.~B.}\ \bibnamefont {Tombet}}, \bibinfo {author}
  {\bibfnamefont {Y.~M.}\ \bibnamefont {Meziani}}, \bibinfo {author}
  {\bibfnamefont {V.~V.}\ \bibnamefont {Popov}}, \bibinfo {author}
  {\bibfnamefont {W.}~\bibnamefont {Knap}}, \bibinfo {author} {\bibfnamefont
  {T.}~\bibnamefont {Suemitsu}}, \ and\ \bibinfo {author} {\bibfnamefont
  {T.}~\bibnamefont {Otsuji}},\ }\href {\doibase 10.1063/1.4885499} {\bibfield
  {journal} {\bibinfo  {journal} {Appl. Phys. Lett.}\ }\textbf {\bibinfo
  {volume} {104}},\ \bibinfo {pages} {251114} (\bibinfo {year}
  {2014})}\BibitemShut {NoStop}%
\bibitem [{\citenamefont {Budkin}\ and\ \citenamefont
  {Golub}(2014)}]{Budkin2014}%
  \BibitemOpen
  \bibfield  {author} {\bibinfo {author} {\bibfnamefont {G.~V.}\ \bibnamefont
  {Budkin}}\ and\ \bibinfo {author} {\bibfnamefont {L.~E.}\ \bibnamefont
  {Golub}},\ }\href {\doibase 10.1103/physrevb.90.125316} {\bibfield  {journal}
  {\bibinfo  {journal} {Phys. Rev. B}\ }\textbf {\bibinfo {volume} {90}},\
  \bibinfo {pages} {125316} (\bibinfo {year} {2014})}\BibitemShut {NoStop}%
\bibitem [{\citenamefont {Faltermeier}\ \emph {et~al.}(2015)\citenamefont
  {Faltermeier}, \citenamefont {Olbrich}, \citenamefont {Probst}, \citenamefont
  {Schell}, \citenamefont {Watanabe}, \citenamefont {Boubanga-Tombet},
  \citenamefont {Otsuji},\ and\ \citenamefont {Ganichev}}]{Faltermeier2015}%
  \BibitemOpen
  \bibfield  {author} {\bibinfo {author} {\bibfnamefont {P.}~\bibnamefont
  {Faltermeier}}, \bibinfo {author} {\bibfnamefont {P.}~\bibnamefont
  {Olbrich}}, \bibinfo {author} {\bibfnamefont {W.}~\bibnamefont {Probst}},
  \bibinfo {author} {\bibfnamefont {L.}~\bibnamefont {Schell}}, \bibinfo
  {author} {\bibfnamefont {T.}~\bibnamefont {Watanabe}}, \bibinfo {author}
  {\bibfnamefont {S.~A.}\ \bibnamefont {Boubanga-Tombet}}, \bibinfo {author}
  {\bibfnamefont {T.}~\bibnamefont {Otsuji}}, \ and\ \bibinfo {author}
  {\bibfnamefont {S.~D.}\ \bibnamefont {Ganichev}},\ }\href {\doibase
  10.1063/1.4928969} {\bibfield  {journal} {\bibinfo  {journal} {J. Appl.
  Phys.}\ }\textbf {\bibinfo {volume} {118}},\ \bibinfo {pages} {084301}
  (\bibinfo {year} {2015})}\BibitemShut {NoStop}%
\bibitem [{\citenamefont {Olbrich}\ \emph {et~al.}(2016)\citenamefont
  {Olbrich}, \citenamefont {Kamann}, \citenamefont {K{\"o}nig}, \citenamefont
  {Munzert}, \citenamefont {Tutsch}, \citenamefont {Eroms}, \citenamefont
  {Weiss}, \citenamefont {Liu}, \citenamefont {Golub}, \citenamefont
  {Ivchenko}, \citenamefont {Popov}, \citenamefont {Fateev}, \citenamefont
  {Mashinsky}, \citenamefont {Fromm}, \citenamefont {Seyller},\ and\
  \citenamefont {Ganichev}}]{Olbrich2016}%
  \BibitemOpen
  \bibfield  {author} {\bibinfo {author} {\bibfnamefont {P.}~\bibnamefont
  {Olbrich}}, \bibinfo {author} {\bibfnamefont {J.}~\bibnamefont {Kamann}},
  \bibinfo {author} {\bibfnamefont {M.}~\bibnamefont {K{\"o}nig}}, \bibinfo
  {author} {\bibfnamefont {J.}~\bibnamefont {Munzert}}, \bibinfo {author}
  {\bibfnamefont {L.}~\bibnamefont {Tutsch}}, \bibinfo {author} {\bibfnamefont
  {J.}~\bibnamefont {Eroms}}, \bibinfo {author} {\bibfnamefont
  {D.}~\bibnamefont {Weiss}}, \bibinfo {author} {\bibfnamefont {M.-H.}\
  \bibnamefont {Liu}}, \bibinfo {author} {\bibfnamefont {L.~E.}\ \bibnamefont
  {Golub}}, \bibinfo {author} {\bibfnamefont {E.~L.}\ \bibnamefont {Ivchenko}},
  \bibinfo {author} {\bibfnamefont {V.~V.}\ \bibnamefont {Popov}}, \bibinfo
  {author} {\bibfnamefont {D.~V.}\ \bibnamefont {Fateev}}, \bibinfo {author}
  {\bibfnamefont {K.~V.}\ \bibnamefont {Mashinsky}}, \bibinfo {author}
  {\bibfnamefont {F.}~\bibnamefont {Fromm}}, \bibinfo {author} {\bibfnamefont
  {T.}~\bibnamefont {Seyller}}, \ and\ \bibinfo {author} {\bibfnamefont
  {S.~D.}\ \bibnamefont {Ganichev}},\ }\href {\doibase
  10.1103/physrevb.93.075422} {\bibfield  {journal} {\bibinfo  {journal} {Phys.
  Rev. B}\ }\textbf {\bibinfo {volume} {93}},\ \bibinfo {pages} {075422}
  (\bibinfo {year} {2016})}\BibitemShut {NoStop}%
\bibitem [{\citenamefont {Popov}(2016)}]{Popov2016}%
  \BibitemOpen
  \bibfield  {author} {\bibinfo {author} {\bibfnamefont {V.~V.}\ \bibnamefont
  {Popov}},\ }\href {\doibase 10.1063/1.4954948} {\bibfield  {journal}
  {\bibinfo  {journal} {Appl. Phys. Lett.}\ }\textbf {\bibinfo {volume}
  {108}},\ \bibinfo {pages} {261104} (\bibinfo {year} {2016})}\BibitemShut
  {NoStop}%
\bibitem [{\citenamefont {Fateev}\ \emph {et~al.}(2019)\citenamefont {Fateev},
  \citenamefont {Mashinsky}, \citenamefont {Sun},\ and\ \citenamefont
  {Popov}}]{Fateev2019}%
  \BibitemOpen
  \bibfield  {author} {\bibinfo {author} {\bibfnamefont {D.~V.}\ \bibnamefont
  {Fateev}}, \bibinfo {author} {\bibfnamefont {K.~V.}\ \bibnamefont
  {Mashinsky}}, \bibinfo {author} {\bibfnamefont {J.~D.}\ \bibnamefont {Sun}},
  \ and\ \bibinfo {author} {\bibfnamefont {V.~V.}\ \bibnamefont {Popov}},\
  }\href {\doibase 10.1016/j.sse.2019.04.004} {\bibfield  {journal} {\bibinfo
  {journal} {Solid-State Electron.}\ }\textbf {\bibinfo {volume} {157}},\
  \bibinfo {pages} {20} (\bibinfo {year} {2019})}\BibitemShut {NoStop}%
\bibitem [{\citenamefont {Hubmann}\ \emph {et~al.}(2020)\citenamefont
  {Hubmann}, \citenamefont {Bel'kov}, \citenamefont {Golub}, \citenamefont
  {Kachorovskii}, \citenamefont {Drienovsky}, \citenamefont {Eroms},
  \citenamefont {Weiss},\ and\ \citenamefont {Ganichev}}]{Hubmann2020}%
  \BibitemOpen
  \bibfield  {author} {\bibinfo {author} {\bibfnamefont {S.}~\bibnamefont
  {Hubmann}}, \bibinfo {author} {\bibfnamefont {V.~V.}\ \bibnamefont
  {Bel'kov}}, \bibinfo {author} {\bibfnamefont {L.~E.}\ \bibnamefont {Golub}},
  \bibinfo {author} {\bibfnamefont {V.~Y.}\ \bibnamefont {Kachorovskii}},
  \bibinfo {author} {\bibfnamefont {M.}~\bibnamefont {Drienovsky}}, \bibinfo
  {author} {\bibfnamefont {J.}~\bibnamefont {Eroms}}, \bibinfo {author}
  {\bibfnamefont {D.}~\bibnamefont {Weiss}}, \ and\ \bibinfo {author}
  {\bibfnamefont {S.~D.}\ \bibnamefont {Ganichev}},\ }\href {\doibase
  10.1103/physrevresearch.2.033186} {\bibfield  {journal} {\bibinfo  {journal}
  {Phys. Rev. Research}\ }\textbf {\bibinfo {volume} {2}},\ \bibinfo {pages}
  {033186} (\bibinfo {year} {2020})}\BibitemShut {NoStop}%
\bibitem [{\citenamefont {Boubanga-Tombet}\ \emph {et~al.}(2020)\citenamefont
  {Boubanga-Tombet}, \citenamefont {Knap}, \citenamefont {Yadav}, \citenamefont
  {Satou}, \citenamefont {But}, \citenamefont {Popov}, \citenamefont
  {Gorbenko}, \citenamefont {Kachorovskii},\ and\ \citenamefont
  {Otsuji}}]{BoubangaTombet2020}%
  \BibitemOpen
  \bibfield  {author} {\bibinfo {author} {\bibfnamefont {S.}~\bibnamefont
  {Boubanga-Tombet}}, \bibinfo {author} {\bibfnamefont {W.}~\bibnamefont
  {Knap}}, \bibinfo {author} {\bibfnamefont {D.}~\bibnamefont {Yadav}},
  \bibinfo {author} {\bibfnamefont {A.}~\bibnamefont {Satou}}, \bibinfo
  {author} {\bibfnamefont {D.~B.}\ \bibnamefont {But}}, \bibinfo {author}
  {\bibfnamefont {V.~V.}\ \bibnamefont {Popov}}, \bibinfo {author}
  {\bibfnamefont {I.~V.}\ \bibnamefont {Gorbenko}}, \bibinfo {author}
  {\bibfnamefont {V.}~\bibnamefont {Kachorovskii}}, \ and\ \bibinfo {author}
  {\bibfnamefont {T.}~\bibnamefont {Otsuji}},\ }\href {\doibase
  10.1103/physrevx.10.031004} {\bibfield  {journal} {\bibinfo  {journal} {Phys.
  Rev. X}\ }\textbf {\bibinfo {volume} {10}},\ \bibinfo {pages} {031004}
  (\bibinfo {year} {2020})}\BibitemShut {NoStop}%
\bibitem [{\citenamefont {Delgado-Notario}\ \emph {et~al.}(2020)\citenamefont
  {Delgado-Notario}, \citenamefont {Cleric{\`{o}}}, \citenamefont {Diez},
  \citenamefont {Vel{\'{a}}zquez-P{\'{e}}rez}, \citenamefont {Taniguchi},
  \citenamefont {Watanabe}, \citenamefont {Otsuji},\ and\ \citenamefont
  {Meziani}}]{DelgadoNotario2020}%
  \BibitemOpen
  \bibfield  {author} {\bibinfo {author} {\bibfnamefont {J.~A.}\ \bibnamefont
  {Delgado-Notario}}, \bibinfo {author} {\bibfnamefont {V.}~\bibnamefont
  {Cleric{\`{o}}}}, \bibinfo {author} {\bibfnamefont {E.}~\bibnamefont {Diez}},
  \bibinfo {author} {\bibfnamefont {J.~E.}\ \bibnamefont
  {Vel{\'{a}}zquez-P{\'{e}}rez}}, \bibinfo {author} {\bibfnamefont
  {T.}~\bibnamefont {Taniguchi}}, \bibinfo {author} {\bibfnamefont
  {K.}~\bibnamefont {Watanabe}}, \bibinfo {author} {\bibfnamefont
  {T.}~\bibnamefont {Otsuji}}, \ and\ \bibinfo {author} {\bibfnamefont {Y.~M.}\
  \bibnamefont {Meziani}},\ }\href {\doibase 10.1063/5.0007249} {\bibfield
  {journal} {\bibinfo  {journal} {APL Photonics}\ }\textbf {\bibinfo {volume}
  {5}},\ \bibinfo {pages} {066102} (\bibinfo {year} {2020})}\BibitemShut
  {NoStop}%
\bibitem [{\citenamefont {M\"onch}\ \emph {et~al.}(2022)\citenamefont
  {M\"onch}, \citenamefont {Potashin}, \citenamefont {Lindner}, \citenamefont
  {Yahniuk}, \citenamefont {Golub}, \citenamefont {Kachorovskii}, \citenamefont
  {Bel'kov}, \citenamefont {Huber}, \citenamefont {Watanabe}, \citenamefont
  {Taniguchi}, \citenamefont {Eroms}, \citenamefont {Weiss},\ and\
  \citenamefont {Ganichev}}]{Moench2022a}%
  \BibitemOpen
  \bibfield  {author} {\bibinfo {author} {\bibfnamefont {E.}~\bibnamefont
  {M\"onch}}, \bibinfo {author} {\bibfnamefont {S.~O.}\ \bibnamefont
  {Potashin}}, \bibinfo {author} {\bibfnamefont {K.}~\bibnamefont {Lindner}},
  \bibinfo {author} {\bibfnamefont {I.}~\bibnamefont {Yahniuk}}, \bibinfo
  {author} {\bibfnamefont {L.~E.}\ \bibnamefont {Golub}}, \bibinfo {author}
  {\bibfnamefont {V.~Y.}\ \bibnamefont {Kachorovskii}}, \bibinfo {author}
  {\bibfnamefont {V.~V.}\ \bibnamefont {Bel'kov}}, \bibinfo {author}
  {\bibfnamefont {R.}~\bibnamefont {Huber}}, \bibinfo {author} {\bibfnamefont
  {K.}~\bibnamefont {Watanabe}}, \bibinfo {author} {\bibfnamefont
  {T.}~\bibnamefont {Taniguchi}}, \bibinfo {author} {\bibfnamefont
  {J.}~\bibnamefont {Eroms}}, \bibinfo {author} {\bibfnamefont
  {D.}~\bibnamefont {Weiss}}, \ and\ \bibinfo {author} {\bibfnamefont {S.~D.}\
  \bibnamefont {Ganichev}},\ }\href {\doibase 10.1103/PhysRevB.105.045404}
  {\bibfield  {journal} {\bibinfo  {journal} {Phys. Rev. B}\ }\textbf {\bibinfo
  {volume} {105}},\ \bibinfo {pages} {045404} (\bibinfo {year}
  {2022})}\BibitemShut {NoStop}%
\bibitem [{\citenamefont {Tamura}\ \emph {et~al.}(2022)\citenamefont {Tamura},
  \citenamefont {Tang}, \citenamefont {Ogiura}, \citenamefont {Suwa},
  \citenamefont {Fukidome}, \citenamefont {Takida}, \citenamefont {Minamide},
  \citenamefont {Suemitsu}, \citenamefont {Otsuji},\ and\ \citenamefont
  {Satou}}]{Tamura2022}%
  \BibitemOpen
  \bibfield  {author} {\bibinfo {author} {\bibfnamefont {K.}~\bibnamefont
  {Tamura}}, \bibinfo {author} {\bibfnamefont {C.}~\bibnamefont {Tang}},
  \bibinfo {author} {\bibfnamefont {D.}~\bibnamefont {Ogiura}}, \bibinfo
  {author} {\bibfnamefont {K.}~\bibnamefont {Suwa}}, \bibinfo {author}
  {\bibfnamefont {H.}~\bibnamefont {Fukidome}}, \bibinfo {author}
  {\bibfnamefont {Y.}~\bibnamefont {Takida}}, \bibinfo {author} {\bibfnamefont
  {H.}~\bibnamefont {Minamide}}, \bibinfo {author} {\bibfnamefont
  {T.}~\bibnamefont {Suemitsu}}, \bibinfo {author} {\bibfnamefont
  {T.}~\bibnamefont {Otsuji}}, \ and\ \bibinfo {author} {\bibfnamefont
  {A.}~\bibnamefont {Satou}},\ }\href {\doibase 10.48550/ARXIV.2207.00135} {\
  (\bibinfo {year} {2022}),\ 10.48550/ARXIV.2207.00135}\BibitemShut {NoStop}%
\bibitem [{\citenamefont {Mönch}\ \emph {et~al.}(2022)\citenamefont {Mönch},
  \citenamefont {Potashin}, \citenamefont {Lindner}, \citenamefont {Yahniuk},
  \citenamefont {Golub}, \citenamefont {Kachorovskii}, \citenamefont {Bel'kov},
  \citenamefont {Huber}, \citenamefont {Watanabe}, \citenamefont {Taniguchi},
  \citenamefont {Eroms}, \citenamefont {Weiss},\ and\ \citenamefont
  {Ganichev}}]{Moench2022b}%
  \BibitemOpen
  \bibfield  {author} {\bibinfo {author} {\bibfnamefont {E.}~\bibnamefont
  {Mönch}}, \bibinfo {author} {\bibfnamefont {S.~O.}\ \bibnamefont
  {Potashin}}, \bibinfo {author} {\bibfnamefont {K.}~\bibnamefont {Lindner}},
  \bibinfo {author} {\bibfnamefont {I.}~\bibnamefont {Yahniuk}}, \bibinfo
  {author} {\bibfnamefont {L.~E.}\ \bibnamefont {Golub}}, \bibinfo {author}
  {\bibfnamefont {V.~Y.}\ \bibnamefont {Kachorovskii}}, \bibinfo {author}
  {\bibfnamefont {V.~V.}\ \bibnamefont {Bel'kov}}, \bibinfo {author}
  {\bibfnamefont {R.}~\bibnamefont {Huber}}, \bibinfo {author} {\bibfnamefont
  {K.}~\bibnamefont {Watanabe}}, \bibinfo {author} {\bibfnamefont
  {T.}~\bibnamefont {Taniguchi}}, \bibinfo {author} {\bibfnamefont
  {J.}~\bibnamefont {Eroms}}, \bibinfo {author} {\bibfnamefont
  {D.}~\bibnamefont {Weiss}}, \ and\ \bibinfo {author} {\bibfnamefont {S.~D.}\
  \bibnamefont {Ganichev}},\ }\href {\doibase 10.48550/ARXIV.2208.08299} {\
  (\bibinfo {year} {2022}),\ 10.48550/ARXIV.2208.08299}\BibitemShut {NoStop}%
\bibitem [{\citenamefont {Scheid}\ \emph {et~al.}(2006)\citenamefont {Scheid},
  \citenamefont {Wimmer}, \citenamefont {Bercioux},\ and\ \citenamefont
  {Richter}}]{Scheid2006}%
  \BibitemOpen
  \bibfield  {author} {\bibinfo {author} {\bibfnamefont {M.}~\bibnamefont
  {Scheid}}, \bibinfo {author} {\bibfnamefont {M.}~\bibnamefont {Wimmer}},
  \bibinfo {author} {\bibfnamefont {D.}~\bibnamefont {Bercioux}}, \ and\
  \bibinfo {author} {\bibfnamefont {K.}~\bibnamefont {Richter}},\ }\href
  {\doibase https://doi.org/10.1002/pssc.200672835} {\bibfield  {journal}
  {\bibinfo  {journal} {physica status solidi c}\ }\textbf {\bibinfo {volume}
  {3}},\ \bibinfo {pages} {4235} (\bibinfo {year} {2006})},\ \Eprint
  {http://arxiv.org/abs/https://onlinelibrary.wiley.com/doi/pdf/10.1002/pssc.200672835}
  {https://onlinelibrary.wiley.com/doi/pdf/10.1002/pssc.200672835} \BibitemShut
  {NoStop}%
\bibitem [{\citenamefont {Costache}\ and\ \citenamefont
  {Valenzuela}(2010)}]{Costache2010}%
  \BibitemOpen
  \bibfield  {author} {\bibinfo {author} {\bibfnamefont {M.~V.}\ \bibnamefont
  {Costache}}\ and\ \bibinfo {author} {\bibfnamefont {S.~O.}\ \bibnamefont
  {Valenzuela}},\ }\href {\doibase 10.1126/science.1196228} {\bibfield
  {journal} {\bibinfo  {journal} {Science}\ }\textbf {\bibinfo {volume}
  {330}},\ \bibinfo {pages} {1645} (\bibinfo {year} {2010})}\BibitemShut
  {NoStop}%
\bibitem [{\citenamefont {Flatté}(2010)}]{Flatte2008}%
  \BibitemOpen
  \bibfield  {author} {\bibinfo {author} {\bibfnamefont {M.~E.}\ \bibnamefont
  {Flatté}},\ }\href {\doibase https://doi.org/10.1016/j.chemphys.2010.05.001}
  {\bibfield  {journal} {\bibinfo  {journal} {Nature Physics}\ }\textbf
  {\bibinfo {volume} {4}},\ \bibinfo {pages} {587} (\bibinfo {year} {2010})},\
  \bibinfo {note} {stochastic processes in Physics and Chemistry (in honor of
  Peter Hänggi)}\BibitemShut {NoStop}%
\bibitem [{\citenamefont {Demikhovskiǐ}\ and\ \citenamefont
  {Khomitsky}(2006)}]{Demikhovskii2006}%
  \BibitemOpen
  \bibfield  {author} {\bibinfo {author} {\bibfnamefont {V.~Y.}\ \bibnamefont
  {Demikhovskiǐ}}\ and\ \bibinfo {author} {\bibfnamefont {D.~V.}\ \bibnamefont
  {Khomitsky}},\ }\href@noop {} {\bibfield  {journal} {\bibinfo  {journal}
  {Journal of Experimental and Theoretical Physics Letters}\ }\textbf {\bibinfo
  {volume} {83}},\ \bibinfo {pages} {340} (\bibinfo {year} {2006})}\BibitemShut
  {NoStop}%
\bibitem [{\citenamefont {Matsuno}\ \emph {et~al.}(2007)\citenamefont
  {Matsuno}, \citenamefont {Lottermoser}, \citenamefont {Arima}, \citenamefont
  {Kawasaki},\ and\ \citenamefont {Tokura}}]{Matsuno2007}%
  \BibitemOpen
  \bibfield  {author} {\bibinfo {author} {\bibfnamefont {J.}~\bibnamefont
  {Matsuno}}, \bibinfo {author} {\bibfnamefont {T.}~\bibnamefont
  {Lottermoser}}, \bibinfo {author} {\bibfnamefont {T.}~\bibnamefont {Arima}},
  \bibinfo {author} {\bibfnamefont {M.}~\bibnamefont {Kawasaki}}, \ and\
  \bibinfo {author} {\bibfnamefont {Y.}~\bibnamefont {Tokura}},\ }\href
  {\doibase 10.1103/PhysRevB.75.180403} {\bibfield  {journal} {\bibinfo
  {journal} {Phys. Rev. B}\ }\textbf {\bibinfo {volume} {75}},\ \bibinfo
  {pages} {180403} (\bibinfo {year} {2007})}\BibitemShut {NoStop}%
\bibitem [{\citenamefont {Scheid}\ \emph {et~al.}(2007)\citenamefont {Scheid},
  \citenamefont {Pfund}, \citenamefont {Bercioux},\ and\ \citenamefont
  {Richter}}]{Scheid2007b}%
  \BibitemOpen
  \bibfield  {author} {\bibinfo {author} {\bibfnamefont {M.}~\bibnamefont
  {Scheid}}, \bibinfo {author} {\bibfnamefont {A.}~\bibnamefont {Pfund}},
  \bibinfo {author} {\bibfnamefont {D.}~\bibnamefont {Bercioux}}, \ and\
  \bibinfo {author} {\bibfnamefont {K.}~\bibnamefont {Richter}},\ }\href@noop
  {} {\bibfield  {journal} {\bibinfo  {journal} {Physical Review B}\ }\textbf
  {\bibinfo {volume} {76}},\ \bibinfo {pages} {195303} (\bibinfo {year}
  {2007})}\BibitemShut {NoStop}%
\bibitem [{\citenamefont {Braunecker}\ \emph {et~al.}(2007)\citenamefont
  {Braunecker}, \citenamefont {Feldman},\ and\ \citenamefont
  {Li}}]{Braunecker2007}%
  \BibitemOpen
  \bibfield  {author} {\bibinfo {author} {\bibfnamefont {B.}~\bibnamefont
  {Braunecker}}, \bibinfo {author} {\bibfnamefont {D.~E.}\ \bibnamefont
  {Feldman}}, \ and\ \bibinfo {author} {\bibfnamefont {F.}~\bibnamefont {Li}},\
  }\href {\doibase 10.1103/physrevb.76.085119} {\bibfield  {journal} {\bibinfo
  {journal} {Physical Review B}\ }\textbf {\bibinfo {volume} {76}} (\bibinfo
  {year} {2007}),\ 10.1103/physrevb.76.085119}\BibitemShut {NoStop}%
\bibitem [{\citenamefont {Lin}\ \emph {et~al.}(2008)\citenamefont {Lin},
  \citenamefont {Tang},\ and\ \citenamefont {Chang}}]{Lin2008}%
  \BibitemOpen
  \bibfield  {author} {\bibinfo {author} {\bibfnamefont {C.-H.}\ \bibnamefont
  {Lin}}, \bibinfo {author} {\bibfnamefont {C.-S.}\ \bibnamefont {Tang}}, \
  and\ \bibinfo {author} {\bibfnamefont {Y.-C.}\ \bibnamefont {Chang}},\ }\href
  {\doibase 10.1103/PhysRevB.78.245312} {\bibfield  {journal} {\bibinfo
  {journal} {Phys. Rev. B}\ }\textbf {\bibinfo {volume} {78}},\ \bibinfo
  {pages} {245312} (\bibinfo {year} {2008})}\BibitemShut {NoStop}%
\bibitem [{\citenamefont {Smirnov}\ \emph {et~al.}(2008)\citenamefont
  {Smirnov}, \citenamefont {Bercioux}, \citenamefont {Grifoni},\ and\
  \citenamefont {Richter}}]{Smirnov2008a}%
  \BibitemOpen
  \bibfield  {author} {\bibinfo {author} {\bibfnamefont {S.}~\bibnamefont
  {Smirnov}}, \bibinfo {author} {\bibfnamefont {D.}~\bibnamefont {Bercioux}},
  \bibinfo {author} {\bibfnamefont {M.}~\bibnamefont {Grifoni}}, \ and\
  \bibinfo {author} {\bibfnamefont {K.}~\bibnamefont {Richter}},\ }\href
  {\doibase 10.1103/PhysRevLett.100.230601} {\bibfield  {journal} {\bibinfo
  {journal} {Phys. Rev. Lett.}\ }\textbf {\bibinfo {volume} {100}},\ \bibinfo
  {pages} {230601} (\bibinfo {year} {2008})}\BibitemShut {NoStop}%
\bibitem [{\citenamefont {Liang}\ \emph {et~al.}(2009)\citenamefont {Liang},
  \citenamefont {Yang},\ and\ \citenamefont {Wang}}]{Liang2009}%
  \BibitemOpen
  \bibfield  {author} {\bibinfo {author} {\bibfnamefont {F.}~\bibnamefont
  {Liang}}, \bibinfo {author} {\bibfnamefont {Y.~H.}\ \bibnamefont {Yang}}, \
  and\ \bibinfo {author} {\bibfnamefont {J.}~\bibnamefont {Wang}},\ }\href
  {\doibase 10.1140/epjb/e2009-00144-1} {\bibfield  {journal} {\bibinfo
  {journal} {The European Physical Journal B}\ }\textbf {\bibinfo {volume}
  {69}},\ \bibinfo {pages} {337} (\bibinfo {year} {2009})}\BibitemShut
  {NoStop}%
\bibitem [{\citenamefont {Smirnov}\ \emph {et~al.}(2009)\citenamefont
  {Smirnov}, \citenamefont {Bercioux}, \citenamefont {Grifoni},\ and\
  \citenamefont {Richter}}]{Smirnov2009}%
  \BibitemOpen
  \bibfield  {author} {\bibinfo {author} {\bibfnamefont {S.}~\bibnamefont
  {Smirnov}}, \bibinfo {author} {\bibfnamefont {D.}~\bibnamefont {Bercioux}},
  \bibinfo {author} {\bibfnamefont {M.}~\bibnamefont {Grifoni}}, \ and\
  \bibinfo {author} {\bibfnamefont {K.}~\bibnamefont {Richter}},\ }\href
  {\doibase 10.1103/PhysRevB.80.201310} {\bibfield  {journal} {\bibinfo
  {journal} {Phys. Rev. B}\ }\textbf {\bibinfo {volume} {80}},\ \bibinfo
  {pages} {201310} (\bibinfo {year} {2009})}\BibitemShut {NoStop}%
\bibitem [{\citenamefont {Scheid}\ \emph {et~al.}(2010)\citenamefont {Scheid},
  \citenamefont {Bercioux},\ and\ \citenamefont {Richter}}]{Scheid2010}%
  \BibitemOpen
  \bibfield  {author} {\bibinfo {author} {\bibfnamefont {M.}~\bibnamefont
  {Scheid}}, \bibinfo {author} {\bibfnamefont {D.}~\bibnamefont {Bercioux}}, \
  and\ \bibinfo {author} {\bibfnamefont {K.}~\bibnamefont {Richter}},\ }\href
  {\doibase https://doi.org/10.1016/j.chemphys.2010.05.001} {\bibfield
  {journal} {\bibinfo  {journal} {Chemical Physics}\ }\textbf {\bibinfo
  {volume} {375}},\ \bibinfo {pages} {276} (\bibinfo {year} {2010})},\ \bibinfo
  {note} {stochastic processes in Physics and Chemistry (in honor of Peter
  Hänggi)}\BibitemShut {NoStop}%
\bibitem [{\citenamefont {Lu}(2010)}]{Lu2010}%
  \BibitemOpen
  \bibfield  {author} {\bibinfo {author} {\bibfnamefont {J.-D.}\ \bibnamefont
  {Lu}},\ }\href {\doibase https://doi.org/10.1016/j.mee.2009.07.024}
  {\bibfield  {journal} {\bibinfo  {journal} {Microelectronic Engineering}\
  }\textbf {\bibinfo {volume} {87}},\ \bibinfo {pages} {216} (\bibinfo {year}
  {2010})}\BibitemShut {NoStop}%
\bibitem [{\citenamefont {Abdullah}\ \emph {et~al.}(2014)\citenamefont
  {Abdullah}, \citenamefont {Vick}, \citenamefont {Murphy},\ and\ \citenamefont
  {Hirohata}}]{Abdullah2014}%
  \BibitemOpen
  \bibfield  {author} {\bibinfo {author} {\bibfnamefont {R.~M.}\ \bibnamefont
  {Abdullah}}, \bibinfo {author} {\bibfnamefont {A.~J.}\ \bibnamefont {Vick}},
  \bibinfo {author} {\bibfnamefont {B.~A.}\ \bibnamefont {Murphy}}, \ and\
  \bibinfo {author} {\bibfnamefont {A.}~\bibnamefont {Hirohata}},\ }\href
  {\doibase 10.1088/0022-3727/47/48/482001} {\bibfield  {journal} {\bibinfo
  {journal} {Journal of Physics D: Applied Physics}\ }\textbf {\bibinfo
  {volume} {47}},\ \bibinfo {pages} {482001} (\bibinfo {year}
  {2014})}\BibitemShut {NoStop}%
\bibitem [{\citenamefont {Ang}\ \emph {et~al.}(2015)\citenamefont {Ang},
  \citenamefont {Ma},\ and\ \citenamefont {Zhang}}]{Ang2015}%
  \BibitemOpen
  \bibfield  {author} {\bibinfo {author} {\bibfnamefont {Y.~S.}\ \bibnamefont
  {Ang}}, \bibinfo {author} {\bibfnamefont {Z.}~\bibnamefont {Ma}}, \ and\
  \bibinfo {author} {\bibfnamefont {C.}~\bibnamefont {Zhang}},\ }\href
  {\doibase 10.1038/srep07872} {\bibfield  {journal} {\bibinfo  {journal}
  {Scientific Reports}\ }\textbf {\bibinfo {volume} {5}},\ \bibinfo {pages}
  {7872} (\bibinfo {year} {2015})}\BibitemShut {NoStop}%
\bibitem [{\citenamefont {Gomonay}\ \emph {et~al.}(2016)\citenamefont
  {Gomonay}, \citenamefont {Kläui},\ and\ \citenamefont
  {Sinova}}]{Gomonay2016}%
  \BibitemOpen
  \bibfield  {author} {\bibinfo {author} {\bibfnamefont {O.}~\bibnamefont
  {Gomonay}}, \bibinfo {author} {\bibfnamefont {M.}~\bibnamefont {Kläui}}, \
  and\ \bibinfo {author} {\bibfnamefont {J.}~\bibnamefont {Sinova}},\ }\href
  {\doibase 10.1063/1.4964272} {\bibfield  {journal} {\bibinfo  {journal}
  {Applied Physics Letters}\ }\textbf {\bibinfo {volume} {109}},\ \bibinfo
  {pages} {142404} (\bibinfo {year} {2016})},\ \Eprint
  {http://arxiv.org/abs/https://doi.org/10.1063/1.4964272}
  {https://doi.org/10.1063/1.4964272} \BibitemShut {NoStop}%
\bibitem [{\citenamefont {Chen}\ \emph {et~al.}(2019)\citenamefont {Chen},
  \citenamefont {Zhou}, \citenamefont {Cheng}, \citenamefont {Song},
  \citenamefont {Zhang}, \citenamefont {Wu}, \citenamefont {Ba}, \citenamefont
  {Li}, \citenamefont {Sun}, \citenamefont {You}, \citenamefont {Zhao},\ and\
  \citenamefont {Pan}}]{Chen2019}%
  \BibitemOpen
  \bibfield  {author} {\bibinfo {author} {\bibfnamefont {X.}~\bibnamefont
  {Chen}}, \bibinfo {author} {\bibfnamefont {X.}~\bibnamefont {Zhou}}, \bibinfo
  {author} {\bibfnamefont {R.}~\bibnamefont {Cheng}}, \bibinfo {author}
  {\bibfnamefont {C.}~\bibnamefont {Song}}, \bibinfo {author} {\bibfnamefont
  {J.}~\bibnamefont {Zhang}}, \bibinfo {author} {\bibfnamefont
  {Y.}~\bibnamefont {Wu}}, \bibinfo {author} {\bibfnamefont {Y.}~\bibnamefont
  {Ba}}, \bibinfo {author} {\bibfnamefont {H.}~\bibnamefont {Li}}, \bibinfo
  {author} {\bibfnamefont {Y.}~\bibnamefont {Sun}}, \bibinfo {author}
  {\bibfnamefont {Y.}~\bibnamefont {You}}, \bibinfo {author} {\bibfnamefont
  {Y.}~\bibnamefont {Zhao}}, \ and\ \bibinfo {author} {\bibfnamefont
  {F.}~\bibnamefont {Pan}},\ }\href {\doibase 10.1038/s41563-019-0424-2}
  {\bibfield  {journal} {\bibinfo  {journal} {Nature Materials}\ }\textbf
  {\bibinfo {volume} {18}},\ \bibinfo {pages} {931} (\bibinfo {year}
  {2019})}\BibitemShut {NoStop}%
\bibitem [{\citenamefont {Zangara}\ \emph {et~al.}(2019)\citenamefont
  {Zangara}, \citenamefont {Henshaw}, \citenamefont {Pagliero}, \citenamefont
  {Ajoy}, \citenamefont {Reimer}, \citenamefont {Pines},\ and\ \citenamefont
  {Meriles}}]{Zangara2019}%
  \BibitemOpen
  \bibfield  {author} {\bibinfo {author} {\bibfnamefont {P.~R.}\ \bibnamefont
  {Zangara}}, \bibinfo {author} {\bibfnamefont {J.}~\bibnamefont {Henshaw}},
  \bibinfo {author} {\bibfnamefont {D.}~\bibnamefont {Pagliero}}, \bibinfo
  {author} {\bibfnamefont {A.}~\bibnamefont {Ajoy}}, \bibinfo {author}
  {\bibfnamefont {J.~A.}\ \bibnamefont {Reimer}}, \bibinfo {author}
  {\bibfnamefont {A.}~\bibnamefont {Pines}}, \ and\ \bibinfo {author}
  {\bibfnamefont {C.~A.}\ \bibnamefont {Meriles}},\ }\href {\doibase
  10.1021/acs.nanolett.8b05114} {\bibfield  {journal} {\bibinfo  {journal}
  {Nano Letters}\ }\textbf {\bibinfo {volume} {19}},\ \bibinfo {pages} {2389}
  (\bibinfo {year} {2019})}\BibitemShut {NoStop}%
\bibitem [{\citenamefont {Huang}\ \emph {et~al.}(2022)\citenamefont {Huang},
  \citenamefont {Yang}, \citenamefont {Cheng}, \citenamefont {Lee},
  \citenamefont {Tseng}, \citenamefont {Wu}, \citenamefont {Pan}, \citenamefont
  {Che}, \citenamefont {Lai}, \citenamefont {Wang}, \citenamefont {Lin},\ and\
  \citenamefont {Tseng}}]{Huang2022}%
  \BibitemOpen
  \bibfield  {author} {\bibinfo {author} {\bibfnamefont {Y.-H.}\ \bibnamefont
  {Huang}}, \bibinfo {author} {\bibfnamefont {C.-Y.}\ \bibnamefont {Yang}},
  \bibinfo {author} {\bibfnamefont {C.-W.}\ \bibnamefont {Cheng}}, \bibinfo
  {author} {\bibfnamefont {A.}~\bibnamefont {Lee}}, \bibinfo {author}
  {\bibfnamefont {C.-H.}\ \bibnamefont {Tseng}}, \bibinfo {author}
  {\bibfnamefont {H.}~\bibnamefont {Wu}}, \bibinfo {author} {\bibfnamefont
  {Q.}~\bibnamefont {Pan}}, \bibinfo {author} {\bibfnamefont {X.}~\bibnamefont
  {Che}}, \bibinfo {author} {\bibfnamefont {C.-H.}\ \bibnamefont {Lai}},
  \bibinfo {author} {\bibfnamefont {K.-L.}\ \bibnamefont {Wang}}, \bibinfo
  {author} {\bibfnamefont {H.-J.}\ \bibnamefont {Lin}}, \ and\ \bibinfo
  {author} {\bibfnamefont {Y.-C.}\ \bibnamefont {Tseng}},\ }\href {\doibase
  https://doi.org/10.1002/adfm.202111653} {\bibfield  {journal} {\bibinfo
  {journal} {Advanced Functional Materials}\ }\textbf {\bibinfo {volume}
  {32}},\ \bibinfo {pages} {2111653} (\bibinfo {year} {2022})},\ \Eprint
  {http://arxiv.org/abs/https://onlinelibrary.wiley.com/doi/pdf/10.1002/adfm.202111653}
  {https://onlinelibrary.wiley.com/doi/pdf/10.1002/adfm.202111653} \BibitemShut
  {NoStop}%
\bibitem [{\citenamefont {V{\'e}lez}\ \emph {et~al.}(2022)\citenamefont
  {V{\'e}lez}, \citenamefont {Ruiz-G{\'o}mez}, \citenamefont {Schaab},
  \citenamefont {Gradauskaite}, \citenamefont {W{\"o}rnle}, \citenamefont
  {Welter}, \citenamefont {Jacot}, \citenamefont {Degen}, \citenamefont
  {Trassin}, \citenamefont {Fiebig},\ and\ \citenamefont
  {Gambardella}}]{Velez2022}%
  \BibitemOpen
  \bibfield  {author} {\bibinfo {author} {\bibfnamefont {S.}~\bibnamefont
  {V{\'e}lez}}, \bibinfo {author} {\bibfnamefont {S.}~\bibnamefont
  {Ruiz-G{\'o}mez}}, \bibinfo {author} {\bibfnamefont {J.}~\bibnamefont
  {Schaab}}, \bibinfo {author} {\bibfnamefont {E.}~\bibnamefont
  {Gradauskaite}}, \bibinfo {author} {\bibfnamefont {M.~S.}\ \bibnamefont
  {W{\"o}rnle}}, \bibinfo {author} {\bibfnamefont {P.}~\bibnamefont {Welter}},
  \bibinfo {author} {\bibfnamefont {B.~J.}\ \bibnamefont {Jacot}}, \bibinfo
  {author} {\bibfnamefont {C.~L.}\ \bibnamefont {Degen}}, \bibinfo {author}
  {\bibfnamefont {M.}~\bibnamefont {Trassin}}, \bibinfo {author} {\bibfnamefont
  {M.}~\bibnamefont {Fiebig}}, \ and\ \bibinfo {author} {\bibfnamefont
  {P.}~\bibnamefont {Gambardella}},\ }\href {\doibase
  10.1038/s41565-022-01144-x} {\bibfield  {journal} {\bibinfo  {journal}
  {Nature Nanotechnology}\ }\textbf {\bibinfo {volume} {17}},\ \bibinfo {pages}
  {834} (\bibinfo {year} {2022})}\BibitemShut {NoStop}%
\bibitem [{\citenamefont {Bercioux}\ and\ \citenamefont
  {Lucignano}(2015)}]{Bercioux2015}%
  \BibitemOpen
  \bibfield  {author} {\bibinfo {author} {\bibfnamefont {D.}~\bibnamefont
  {Bercioux}}\ and\ \bibinfo {author} {\bibfnamefont {P.}~\bibnamefont
  {Lucignano}},\ }\href {\doibase 10.1088/0034-4885/78/10/106001} {\bibfield
  {journal} {\bibinfo  {journal} {Reports on Progress in Physics}\ }\textbf
  {\bibinfo {volume} {78}},\ \bibinfo {pages} {106001} (\bibinfo {year}
  {2015})}\BibitemShut {NoStop}%
\bibitem [{\citenamefont {Faltermeier}\ \emph {et~al.}(2017)\citenamefont
  {Faltermeier}, \citenamefont {Budkin}, \citenamefont {Unverzagt},
  \citenamefont {Hubmann}, \citenamefont {Pfaller}, \citenamefont {Bel'kov},
  \citenamefont {Golub}, \citenamefont {Ivchenko}, \citenamefont {Adamus},
  \citenamefont {Karczewski}, \citenamefont {Wojtowicz}, \citenamefont {Popov},
  \citenamefont {Fateev}, \citenamefont {Kozlov}, \citenamefont {Weiss},\ and\
  \citenamefont {Ganichev}}]{Faltermeier2017}%
  \BibitemOpen
  \bibfield  {author} {\bibinfo {author} {\bibfnamefont {P.}~\bibnamefont
  {Faltermeier}}, \bibinfo {author} {\bibfnamefont {G.~V.}\ \bibnamefont
  {Budkin}}, \bibinfo {author} {\bibfnamefont {J.}~\bibnamefont {Unverzagt}},
  \bibinfo {author} {\bibfnamefont {S.}~\bibnamefont {Hubmann}}, \bibinfo
  {author} {\bibfnamefont {A.}~\bibnamefont {Pfaller}}, \bibinfo {author}
  {\bibfnamefont {V.~V.}\ \bibnamefont {Bel'kov}}, \bibinfo {author}
  {\bibfnamefont {L.~E.}\ \bibnamefont {Golub}}, \bibinfo {author}
  {\bibfnamefont {E.~L.}\ \bibnamefont {Ivchenko}}, \bibinfo {author}
  {\bibfnamefont {Z.}~\bibnamefont {Adamus}}, \bibinfo {author} {\bibfnamefont
  {G.}~\bibnamefont {Karczewski}}, \bibinfo {author} {\bibfnamefont
  {T.}~\bibnamefont {Wojtowicz}}, \bibinfo {author} {\bibfnamefont {V.~V.}\
  \bibnamefont {Popov}}, \bibinfo {author} {\bibfnamefont {D.~V.}\ \bibnamefont
  {Fateev}}, \bibinfo {author} {\bibfnamefont {D.~A.}\ \bibnamefont {Kozlov}},
  \bibinfo {author} {\bibfnamefont {D.}~\bibnamefont {Weiss}}, \ and\ \bibinfo
  {author} {\bibfnamefont {S.~D.}\ \bibnamefont {Ganichev}},\ }\href {\doibase
  10.1103/physrevb.95.155442} {\bibfield  {journal} {\bibinfo  {journal} {Phys.
  Rev. B}\ }\textbf {\bibinfo {volume} {95}},\ \bibinfo {pages} {155442}
  (\bibinfo {year} {2017})}\BibitemShut {NoStop}%
\bibitem [{\citenamefont {Faltermeier}\ \emph {et~al.}(2018)\citenamefont
  {Faltermeier}, \citenamefont {Budkin}, \citenamefont {Hubmann}, \citenamefont
  {Bel'kov}, \citenamefont {Golub}, \citenamefont {Ivchenko}, \citenamefont
  {Adamus}, \citenamefont {Karczewski}, \citenamefont {Wojtowicz},
  \citenamefont {Kozlov}, \citenamefont {Weiss},\ and\ \citenamefont
  {Ganichev}}]{Faltermeier2018}%
  \BibitemOpen
  \bibfield  {author} {\bibinfo {author} {\bibfnamefont {P.}~\bibnamefont
  {Faltermeier}}, \bibinfo {author} {\bibfnamefont {G.~V.}\ \bibnamefont
  {Budkin}}, \bibinfo {author} {\bibfnamefont {S.}~\bibnamefont {Hubmann}},
  \bibinfo {author} {\bibfnamefont {V.~V.}\ \bibnamefont {Bel'kov}}, \bibinfo
  {author} {\bibfnamefont {L.~E.}\ \bibnamefont {Golub}}, \bibinfo {author}
  {\bibfnamefont {E.~L.}\ \bibnamefont {Ivchenko}}, \bibinfo {author}
  {\bibfnamefont {Z.}~\bibnamefont {Adamus}}, \bibinfo {author} {\bibfnamefont
  {G.}~\bibnamefont {Karczewski}}, \bibinfo {author} {\bibfnamefont
  {T.}~\bibnamefont {Wojtowicz}}, \bibinfo {author} {\bibfnamefont {D.~A.}\
  \bibnamefont {Kozlov}}, \bibinfo {author} {\bibfnamefont {D.}~\bibnamefont
  {Weiss}}, \ and\ \bibinfo {author} {\bibfnamefont {S.~D.}\ \bibnamefont
  {Ganichev}},\ }\href {\doibase 10.1016/j.physe.2018.04.001} {\bibfield
  {journal} {\bibinfo  {journal} {Phys. E}\ }\textbf {\bibinfo {volume}
  {101}},\ \bibinfo {pages} {178} (\bibinfo {year} {2018})}\BibitemShut
  {NoStop}%
\bibitem [{\citenamefont {Sai}\ \emph {et~al.}(2021)\citenamefont {Sai},
  \citenamefont {Potashin}, \citenamefont {Szo{\l}a}, \citenamefont
  {Yavorskiy}, \citenamefont {Cywi{\'{n}}ski}, \citenamefont {Prystawko},
  \citenamefont {{\L}usakowski}, \citenamefont {Ganichev}, \citenamefont
  {Rumyantsev}, \citenamefont {Knap},\ and\ \citenamefont
  {Kachorovskii}}]{Sai2021}%
  \BibitemOpen
  \bibfield  {author} {\bibinfo {author} {\bibfnamefont {P.}~\bibnamefont
  {Sai}}, \bibinfo {author} {\bibfnamefont {S.~O.}\ \bibnamefont {Potashin}},
  \bibinfo {author} {\bibfnamefont {M.}~\bibnamefont {Szo{\l}a}}, \bibinfo
  {author} {\bibfnamefont {D.}~\bibnamefont {Yavorskiy}}, \bibinfo {author}
  {\bibfnamefont {G.}~\bibnamefont {Cywi{\'{n}}ski}}, \bibinfo {author}
  {\bibfnamefont {P.}~\bibnamefont {Prystawko}}, \bibinfo {author}
  {\bibfnamefont {J.}~\bibnamefont {{\L}usakowski}}, \bibinfo {author}
  {\bibfnamefont {S.~D.}\ \bibnamefont {Ganichev}}, \bibinfo {author}
  {\bibfnamefont {S.}~\bibnamefont {Rumyantsev}}, \bibinfo {author}
  {\bibfnamefont {W.}~\bibnamefont {Knap}}, \ and\ \bibinfo {author}
  {\bibfnamefont {V.~Y.}\ \bibnamefont {Kachorovskii}},\ }\href {\doibase
  10.1103/physrevb.104.045301} {\bibfield  {journal} {\bibinfo  {journal}
  {Phys. Rev. B}\ }\textbf {\bibinfo {volume} {104}},\ \bibinfo {pages}
  {045301} (\bibinfo {year} {2021})}\BibitemShut {NoStop}%
\bibitem [{\citenamefont {Betthausen}\ \emph {et~al.}(2012)\citenamefont
  {Betthausen}, \citenamefont {Dollinger}, \citenamefont {Saarikoski},
  \citenamefont {Kolkovsky}, \citenamefont {Karczewski}, \citenamefont
  {Wojtowicz}, \citenamefont {Richter},\ and\ \citenamefont
  {Weiss}}]{Betthausen2012}%
  \BibitemOpen
  \bibfield  {author} {\bibinfo {author} {\bibfnamefont {C.}~\bibnamefont
  {Betthausen}}, \bibinfo {author} {\bibfnamefont {T.}~\bibnamefont
  {Dollinger}}, \bibinfo {author} {\bibfnamefont {H.}~\bibnamefont
  {Saarikoski}}, \bibinfo {author} {\bibfnamefont {V.}~\bibnamefont
  {Kolkovsky}}, \bibinfo {author} {\bibfnamefont {G.}~\bibnamefont
  {Karczewski}}, \bibinfo {author} {\bibfnamefont {T.}~\bibnamefont
  {Wojtowicz}}, \bibinfo {author} {\bibfnamefont {K.}~\bibnamefont {Richter}},
  \ and\ \bibinfo {author} {\bibfnamefont {D.}~\bibnamefont {Weiss}},\ }\href
  {\doibase 10.1126/science.1221350} {\bibfield  {journal} {\bibinfo  {journal}
  {Science (New York, N.Y.)}\ }\textbf {\bibinfo {volume} {337}},\ \bibinfo
  {pages} {324} (\bibinfo {year} {2012})}\BibitemShut {NoStop}%
\bibitem [{\citenamefont {Matsubara}\ \emph {et~al.}(2022)\citenamefont
  {Matsubara}, \citenamefont {Kobayashi}, \citenamefont {Watanabe},
  \citenamefont {Yanase}, \citenamefont {Iwata},\ and\ \citenamefont
  {Kato}}]{Matsubara2022}%
  \BibitemOpen
  \bibfield  {author} {\bibinfo {author} {\bibfnamefont {M.}~\bibnamefont
  {Matsubara}}, \bibinfo {author} {\bibfnamefont {T.}~\bibnamefont
  {Kobayashi}}, \bibinfo {author} {\bibfnamefont {H.}~\bibnamefont {Watanabe}},
  \bibinfo {author} {\bibfnamefont {Y.}~\bibnamefont {Yanase}}, \bibinfo
  {author} {\bibfnamefont {S.}~\bibnamefont {Iwata}}, \ and\ \bibinfo {author}
  {\bibfnamefont {T.}~\bibnamefont {Kato}},\ }\href {\doibase
  10.1038/s41467-022-34374-7} {\bibfield  {journal} {\bibinfo  {journal}
  {Nature Communications}\ }\textbf {\bibinfo {volume} {13}},\ \bibinfo {pages}
  {6708} (\bibinfo {year} {2022})}\BibitemShut {NoStop}%
\bibitem [{\citenamefont {Bel'kov}\ and\ \citenamefont
  {Ganichev}(2008)}]{Belkov2008}%
  \BibitemOpen
  \bibfield  {author} {\bibinfo {author} {\bibfnamefont {V.~V.}\ \bibnamefont
  {Bel'kov}}\ and\ \bibinfo {author} {\bibfnamefont {S.~D.}\ \bibnamefont
  {Ganichev}},\ }\href {\doibase 10.1088/0268-1242/23/11/114003} {\bibfield
  {journal} {\bibinfo  {journal} {Semicond. Sci. Technol.}\ }\textbf {\bibinfo
  {volume} {23}},\ \bibinfo {pages} {114003} (\bibinfo {year}
  {2008})}\BibitemShut {NoStop}%
\bibitem [{\citenamefont {Lorke}\ \emph {et~al.}(1998)\citenamefont {Lorke},
  \citenamefont {Wimmer}, \citenamefont {Jager}, \citenamefont {Kotthaus},
  \citenamefont {Wegscheider},\ and\ \citenamefont {Bichler}}]{Lorke1998}%
  \BibitemOpen
  \bibfield  {author} {\bibinfo {author} {\bibfnamefont {A.}~\bibnamefont
  {Lorke}}, \bibinfo {author} {\bibfnamefont {S.}~\bibnamefont {Wimmer}},
  \bibinfo {author} {\bibfnamefont {B.}~\bibnamefont {Jager}}, \bibinfo
  {author} {\bibfnamefont {J.}~\bibnamefont {Kotthaus}}, \bibinfo {author}
  {\bibfnamefont {W.}~\bibnamefont {Wegscheider}}, \ and\ \bibinfo {author}
  {\bibfnamefont {M.}~\bibnamefont {Bichler}},\ }\href {\doibase
  https://doi.org/10.1016/S0921-4526(98)00121-5} {\bibfield  {journal}
  {\bibinfo  {journal} {Physica B: Condensed Matter}\ }\textbf {\bibinfo
  {volume} {249-251}},\ \bibinfo {pages} {312} (\bibinfo {year}
  {1998})}\BibitemShut {NoStop}%
\bibitem [{\citenamefont {Chepelianskii}\ \emph {et~al.}(2007)\citenamefont
  {Chepelianskii}, \citenamefont {Entin}, \citenamefont {Magarill},\ and\
  \citenamefont {Shepelyansky}}]{Chepelianskii2007}%
  \BibitemOpen
  \bibfield  {author} {\bibinfo {author} {\bibfnamefont {A.~D.}\ \bibnamefont
  {Chepelianskii}}, \bibinfo {author} {\bibfnamefont {M.~V.}\ \bibnamefont
  {Entin}}, \bibinfo {author} {\bibfnamefont {L.~I.}\ \bibnamefont {Magarill}},
  \ and\ \bibinfo {author} {\bibfnamefont {D.~L.}\ \bibnamefont
  {Shepelyansky}},\ }\href {\doibase 10.1140/epjb/e2007-00127-2} {\bibfield
  {journal} {\bibinfo  {journal} {The European Physical Journal B}\ }\textbf
  {\bibinfo {volume} {56}},\ \bibinfo {pages} {323} (\bibinfo {year}
  {2007})}\BibitemShut {NoStop}%
\bibitem [{\citenamefont {Sassine}\ \emph {et~al.}(2008)\citenamefont
  {Sassine}, \citenamefont {Krupko}, \citenamefont {Portal}, \citenamefont
  {Kvon}, \citenamefont {Murali}, \citenamefont {Martin}, \citenamefont
  {Hill},\ and\ \citenamefont {Wieck}}]{Sassine2008}%
  \BibitemOpen
  \bibfield  {author} {\bibinfo {author} {\bibfnamefont {S.}~\bibnamefont
  {Sassine}}, \bibinfo {author} {\bibfnamefont {Y.}~\bibnamefont {Krupko}},
  \bibinfo {author} {\bibfnamefont {J.-C.}\ \bibnamefont {Portal}}, \bibinfo
  {author} {\bibfnamefont {Z.~D.}\ \bibnamefont {Kvon}}, \bibinfo {author}
  {\bibfnamefont {R.}~\bibnamefont {Murali}}, \bibinfo {author} {\bibfnamefont
  {K.~P.}\ \bibnamefont {Martin}}, \bibinfo {author} {\bibfnamefont
  {G.}~\bibnamefont {Hill}}, \ and\ \bibinfo {author} {\bibfnamefont {A.~D.}\
  \bibnamefont {Wieck}},\ }\href {\doibase 10.1103/physrevb.78.045431}
  {\bibfield  {journal} {\bibinfo  {journal} {Phys. Rev. B}\ }\textbf {\bibinfo
  {volume} {78}},\ \bibinfo {pages} {045431} (\bibinfo {year}
  {2008})}\BibitemShut {NoStop}%
\bibitem [{\citenamefont {Kannan}\ \emph {et~al.}(2012)\citenamefont {Kannan},
  \citenamefont {Bisotto}, \citenamefont {Portal}, \citenamefont {Beck},\ and\
  \citenamefont {Jalabert}}]{Kannan2012}%
  \BibitemOpen
  \bibfield  {author} {\bibinfo {author} {\bibfnamefont {E.~S.}\ \bibnamefont
  {Kannan}}, \bibinfo {author} {\bibfnamefont {I.}~\bibnamefont {Bisotto}},
  \bibinfo {author} {\bibfnamefont {J.-C.}\ \bibnamefont {Portal}}, \bibinfo
  {author} {\bibfnamefont {T.~J.}\ \bibnamefont {Beck}}, \ and\ \bibinfo
  {author} {\bibfnamefont {L.}~\bibnamefont {Jalabert}},\ }\href {\doibase
  10.1063/1.4756786} {\bibfield  {journal} {\bibinfo  {journal} {Appl. Phys.
  Lett.}\ }\textbf {\bibinfo {volume} {101}},\ \bibinfo {pages} {143504}
  (\bibinfo {year} {2012})}\BibitemShut {NoStop}%
\bibitem [{\citenamefont {Motta}\ \emph {et~al.}(2021)\citenamefont {Motta},
  \citenamefont {Burger}, \citenamefont {Jiang}, \citenamefont
  {Gonz\'alez~Acosta}, \citenamefont {Jeli\ifmmode~\acute{c}\else \'{c}\fi{}},
  \citenamefont {Colauto}, \citenamefont {Ortiz}, \citenamefont {Johansen},
  \citenamefont {Milo\ifmmode \check{s}\else
  \v{s}\fi{}evi\ifmmode~\acute{c}\else \'{c}\fi{}}, \citenamefont {Cirillo},
  \citenamefont {Attanasio}, \citenamefont {Xue}, \citenamefont {Silhanek},\
  and\ \citenamefont {Vanderheyden}}]{Motta2021}%
  \BibitemOpen
  \bibfield  {author} {\bibinfo {author} {\bibfnamefont {M.}~\bibnamefont
  {Motta}}, \bibinfo {author} {\bibfnamefont {L.}~\bibnamefont {Burger}},
  \bibinfo {author} {\bibfnamefont {L.}~\bibnamefont {Jiang}}, \bibinfo
  {author} {\bibfnamefont {J.~D.}\ \bibnamefont {Gonz\'alez~Acosta}}, \bibinfo
  {author} {\bibfnamefont {i.~c. v.~L.}\ \bibnamefont
  {Jeli\ifmmode~\acute{c}\else \'{c}\fi{}}}, \bibinfo {author} {\bibfnamefont
  {F.}~\bibnamefont {Colauto}}, \bibinfo {author} {\bibfnamefont {W.~A.}\
  \bibnamefont {Ortiz}}, \bibinfo {author} {\bibfnamefont {T.~H.}\ \bibnamefont
  {Johansen}}, \bibinfo {author} {\bibfnamefont {M.~V.}\ \bibnamefont
  {Milo\ifmmode \check{s}\else \v{s}\fi{}evi\ifmmode~\acute{c}\else
  \'{c}\fi{}}}, \bibinfo {author} {\bibfnamefont {C.}~\bibnamefont {Cirillo}},
  \bibinfo {author} {\bibfnamefont {C.}~\bibnamefont {Attanasio}}, \bibinfo
  {author} {\bibfnamefont {C.}~\bibnamefont {Xue}}, \bibinfo {author}
  {\bibfnamefont {A.~V.}\ \bibnamefont {Silhanek}}, \ and\ \bibinfo {author}
  {\bibfnamefont {B.}~\bibnamefont {Vanderheyden}},\ }\href {\doibase
  10.1103/PhysRevB.103.224514} {\bibfield  {journal} {\bibinfo  {journal}
  {Phys. Rev. B}\ }\textbf {\bibinfo {volume} {103}},\ \bibinfo {pages}
  {224514} (\bibinfo {year} {2021})}\BibitemShut {NoStop}%
\bibitem [{Foo()}]{Footnote_Fig_1}%
  \BibitemOpen
  \href@noop {} {}\bibinfo {note} {Note that the measurements of the
  photosignal in $x$- and $y$-direction were performed on two samples with
  identical design and characteristic but different position of the contacts
  with respect to the triangles' orientation, see
  Fig.\,\ref{figA1}.}\BibitemShut {Stop}%
\bibitem [{\citenamefont {Dantscher}\ \emph {et~al.}(2017)\citenamefont
  {Dantscher}, \citenamefont {Kozlov}, \citenamefont {Scherr}, \citenamefont
  {Gebert}, \citenamefont {B\"arenf\"anger}, \citenamefont {Durnev},
  \citenamefont {Tarasenko}, \citenamefont {Bel'kov}, \citenamefont
  {Mikhailov}, \citenamefont {Dvoretsky}, \citenamefont {Kvon}, \citenamefont
  {Ziegler}, \citenamefont {Weiss},\ and\ \citenamefont
  {Ganichev}}]{Dantscher2017}%
  \BibitemOpen
  \bibfield  {author} {\bibinfo {author} {\bibfnamefont {K.-M.}\ \bibnamefont
  {Dantscher}}, \bibinfo {author} {\bibfnamefont {D.~A.}\ \bibnamefont
  {Kozlov}}, \bibinfo {author} {\bibfnamefont {M.~T.}\ \bibnamefont {Scherr}},
  \bibinfo {author} {\bibfnamefont {S.}~\bibnamefont {Gebert}}, \bibinfo
  {author} {\bibfnamefont {J.}~\bibnamefont {B\"arenf\"anger}}, \bibinfo
  {author} {\bibfnamefont {M.~V.}\ \bibnamefont {Durnev}}, \bibinfo {author}
  {\bibfnamefont {S.~A.}\ \bibnamefont {Tarasenko}}, \bibinfo {author}
  {\bibfnamefont {V.~V.}\ \bibnamefont {Bel'kov}}, \bibinfo {author}
  {\bibfnamefont {N.~N.}\ \bibnamefont {Mikhailov}}, \bibinfo {author}
  {\bibfnamefont {S.~A.}\ \bibnamefont {Dvoretsky}}, \bibinfo {author}
  {\bibfnamefont {Z.~D.}\ \bibnamefont {Kvon}}, \bibinfo {author}
  {\bibfnamefont {J.}~\bibnamefont {Ziegler}}, \bibinfo {author} {\bibfnamefont
  {D.}~\bibnamefont {Weiss}}, \ and\ \bibinfo {author} {\bibfnamefont {S.~D.}\
  \bibnamefont {Ganichev}},\ }\href {\doibase 10.1103/PhysRevB.95.201103}
  {\bibfield  {journal} {\bibinfo  {journal} {Phys. Rev. B}\ }\textbf {\bibinfo
  {volume} {95}},\ \bibinfo {pages} {201103} (\bibinfo {year}
  {2017})}\BibitemShut {NoStop}%
\bibitem [{\citenamefont {Ganichev}\ \emph {et~al.}(1993)\citenamefont
  {Ganichev}, \citenamefont {Prettl},\ and\ \citenamefont
  {Huggard}}]{Ganichev1993}%
  \BibitemOpen
  \bibfield  {author} {\bibinfo {author} {\bibfnamefont {S.~D.}\ \bibnamefont
  {Ganichev}}, \bibinfo {author} {\bibfnamefont {W.}~\bibnamefont {Prettl}}, \
  and\ \bibinfo {author} {\bibfnamefont {P.~G.}\ \bibnamefont {Huggard}},\
  }\href {\doibase 10.1103/physrevlett.71.3882} {\bibfield  {journal} {\bibinfo
   {journal} {Phys. Rev. Lett.}\ }\textbf {\bibinfo {volume} {71}},\ \bibinfo
  {pages} {3882} (\bibinfo {year} {1993})}\BibitemShut {NoStop}%
\bibitem [{\citenamefont {Ganichev}\ \emph {et~al.}(1995)\citenamefont
  {Ganichev}, \citenamefont {Yassievich}, \citenamefont {Prettl}, \citenamefont
  {Diener}, \citenamefont {Meyer},\ and\ \citenamefont {Benz}}]{Ganichev1995}%
  \BibitemOpen
  \bibfield  {author} {\bibinfo {author} {\bibfnamefont {S.~D.}\ \bibnamefont
  {Ganichev}}, \bibinfo {author} {\bibfnamefont {I.~N.}\ \bibnamefont
  {Yassievich}}, \bibinfo {author} {\bibfnamefont {W.}~\bibnamefont {Prettl}},
  \bibinfo {author} {\bibfnamefont {J.}~\bibnamefont {Diener}}, \bibinfo
  {author} {\bibfnamefont {B.~K.}\ \bibnamefont {Meyer}}, \ and\ \bibinfo
  {author} {\bibfnamefont {K.~W.}\ \bibnamefont {Benz}},\ }\href {\doibase
  10.1103/physrevlett.75.1590} {\bibfield  {journal} {\bibinfo  {journal}
  {Phys. Rev. Lett.}\ }\textbf {\bibinfo {volume} {75}},\ \bibinfo {pages}
  {1590} (\bibinfo {year} {1995})}\BibitemShut {NoStop}%
\bibitem [{\citenamefont {Ganichev}\ \emph {et~al.}(1985)\citenamefont
  {Ganichev}, \citenamefont {Terent'ev},\ and\ \citenamefont
  {Yaroshetskii}}]{Ganichev1985}%
  \BibitemOpen
  \bibfield  {author} {\bibinfo {author} {\bibfnamefont {S.~D.}\ \bibnamefont
  {Ganichev}}, \bibinfo {author} {\bibfnamefont {Y.~V.}\ \bibnamefont
  {Terent'ev}}, \ and\ \bibinfo {author} {\bibfnamefont {I.~D.}\ \bibnamefont
  {Yaroshetskii}},\ }\href@noop {} {\bibfield  {journal} {\bibinfo  {journal}
  {Pis'ma Zh. Tekh. Fiz.}\ }\textbf {\bibinfo {volume} {11}},\ \bibinfo {pages}
  {46} (\bibinfo {year} {1985})},\ \bibinfo {note} {[Sov. Tech. Phys. Lett.
  \textbf{11}, 20 (1989)]}\BibitemShut {NoStop}%
\bibitem [{\citenamefont {Danilov}\ \emph {et~al.}(2009)\citenamefont
  {Danilov}, \citenamefont {Wittmann}, \citenamefont {Olbrich}, \citenamefont
  {Eder}, \citenamefont {Prettl}, \citenamefont {Golub}, \citenamefont
  {Beregulin}, \citenamefont {Kvon}, \citenamefont {Mikhailov}, \citenamefont
  {Dvoretsky}, \citenamefont {Shalygin}, \citenamefont {Vinh}, \citenamefont
  {van~der Meer}, \citenamefont {Murdin},\ and\ \citenamefont
  {Ganichev}}]{Danilov2009}%
  \BibitemOpen
  \bibfield  {author} {\bibinfo {author} {\bibfnamefont {S.~N.}\ \bibnamefont
  {Danilov}}, \bibinfo {author} {\bibfnamefont {B.}~\bibnamefont {Wittmann}},
  \bibinfo {author} {\bibfnamefont {P.}~\bibnamefont {Olbrich}}, \bibinfo
  {author} {\bibfnamefont {W.}~\bibnamefont {Eder}}, \bibinfo {author}
  {\bibfnamefont {W.}~\bibnamefont {Prettl}}, \bibinfo {author} {\bibfnamefont
  {L.~E.}\ \bibnamefont {Golub}}, \bibinfo {author} {\bibfnamefont {E.~V.}\
  \bibnamefont {Beregulin}}, \bibinfo {author} {\bibfnamefont {Z.~D.}\
  \bibnamefont {Kvon}}, \bibinfo {author} {\bibfnamefont {N.~N.}\ \bibnamefont
  {Mikhailov}}, \bibinfo {author} {\bibfnamefont {S.~A.}\ \bibnamefont
  {Dvoretsky}}, \bibinfo {author} {\bibfnamefont {V.~A.}\ \bibnamefont
  {Shalygin}}, \bibinfo {author} {\bibfnamefont {N.~Q.}\ \bibnamefont {Vinh}},
  \bibinfo {author} {\bibfnamefont {A.~F.~G.}\ \bibnamefont {van~der Meer}},
  \bibinfo {author} {\bibfnamefont {B.}~\bibnamefont {Murdin}}, \ and\ \bibinfo
  {author} {\bibfnamefont {S.~D.}\ \bibnamefont {Ganichev}},\ }\href {\doibase
  10.1063/1.3056393} {\bibfield  {journal} {\bibinfo  {journal} {Journal of
  Applied Physics}\ }\textbf {\bibinfo {volume} {105}},\ \bibinfo {pages}
  {013106} (\bibinfo {year} {2009})}\BibitemShut {NoStop}%
\bibitem [{\citenamefont {Ganichev}(1999)}]{Ganichev1999}%
  \BibitemOpen
  \bibfield  {author} {\bibinfo {author} {\bibfnamefont {S.~D.}\ \bibnamefont
  {Ganichev}},\ }\href {\doibase 10.1016/s0921-4526(99)00637-7} {\bibfield
  {journal} {\bibinfo  {journal} {Phys. B}\ }\textbf {\bibinfo {volume}
  {273-274}},\ \bibinfo {pages} {737} (\bibinfo {year} {1999})}\BibitemShut
  {NoStop}%
\bibitem [{Note1()}]{Note1}%
  \BibitemOpen
  \bibinfo {note} {Note that application of an in-plane magnetic field does not
  affect the signals in the whole studied range $|\mu _0H|\leq \SI
  {2}{T}$.}\BibitemShut {Stop}%
\bibitem [{Note2()}]{Note2}%
  \BibitemOpen
  \bibinfo {note} {Note that the sample resistance does not depend on magnetic
  field, see Fig.\protect \tmspace +\thinmuskip {.1667em}\ref {figA2} in
  Appendix~\ref {appendixA}.}\BibitemShut {Stop}%
\bibitem [{Note3()}]{Note3}%
  \BibitemOpen
  \bibinfo {note} {We note that while in the theoretical consideration the
  current density $j$ is used, in the experiments, the electric current $J$ is
  measured, which is proportional to the current density $j$.}\BibitemShut
  {Stop}%
\bibitem [{\citenamefont {Olbrich}\ \emph {et~al.}(2014)\citenamefont
  {Olbrich}, \citenamefont {Golub}, \citenamefont {Herrmann}, \citenamefont
  {Danilov}, \citenamefont {Plank}, \citenamefont {Bel'kov}, \citenamefont
  {Mussler}, \citenamefont {Weyrich}, \citenamefont {Schneider}, \citenamefont
  {Kampmeier}, \citenamefont {Gr\"utzmacher}, \citenamefont {Plucinski},
  \citenamefont {Eschbach},\ and\ \citenamefont {Ganichev}}]{Olbrich2014}%
  \BibitemOpen
  \bibfield  {author} {\bibinfo {author} {\bibfnamefont {P.}~\bibnamefont
  {Olbrich}}, \bibinfo {author} {\bibfnamefont {L.~E.}\ \bibnamefont {Golub}},
  \bibinfo {author} {\bibfnamefont {T.}~\bibnamefont {Herrmann}}, \bibinfo
  {author} {\bibfnamefont {S.~N.}\ \bibnamefont {Danilov}}, \bibinfo {author}
  {\bibfnamefont {H.}~\bibnamefont {Plank}}, \bibinfo {author} {\bibfnamefont
  {V.~V.}\ \bibnamefont {Bel'kov}}, \bibinfo {author} {\bibfnamefont
  {G.}~\bibnamefont {Mussler}}, \bibinfo {author} {\bibfnamefont
  {C.}~\bibnamefont {Weyrich}}, \bibinfo {author} {\bibfnamefont {C.~M.}\
  \bibnamefont {Schneider}}, \bibinfo {author} {\bibfnamefont {J.}~\bibnamefont
  {Kampmeier}}, \bibinfo {author} {\bibfnamefont {D.}~\bibnamefont
  {Gr\"utzmacher}}, \bibinfo {author} {\bibfnamefont {L.}~\bibnamefont
  {Plucinski}}, \bibinfo {author} {\bibfnamefont {M.}~\bibnamefont {Eschbach}},
  \ and\ \bibinfo {author} {\bibfnamefont {S.~D.}\ \bibnamefont {Ganichev}},\
  }\href {\doibase 10.1103/PhysRevLett.113.096601} {\bibfield  {journal}
  {\bibinfo  {journal} {Phys. Rev. Lett.}\ }\textbf {\bibinfo {volume} {113}},\
  \bibinfo {pages} {096601} (\bibinfo {year} {2014})}\BibitemShut {NoStop}%
\bibitem [{\citenamefont {Danilov}\ \emph {et~al.}(2021)\citenamefont
  {Danilov}, \citenamefont {Golub}, \citenamefont {Mayer}, \citenamefont
  {Beer}, \citenamefont {Binder}, \citenamefont {Mönch}, \citenamefont
  {Min{\'{a}}r}, \citenamefont {Kronseder}, \citenamefont {Back}, \citenamefont
  {Bougeard},\ and\ \citenamefont {Ganichev}}]{Danilov2021}%
  \BibitemOpen
  \bibfield  {author} {\bibinfo {author} {\bibfnamefont {S.~N.}\ \bibnamefont
  {Danilov}}, \bibinfo {author} {\bibfnamefont {L.~E.}\ \bibnamefont {Golub}},
  \bibinfo {author} {\bibfnamefont {T.}~\bibnamefont {Mayer}}, \bibinfo
  {author} {\bibfnamefont {A.}~\bibnamefont {Beer}}, \bibinfo {author}
  {\bibfnamefont {S.}~\bibnamefont {Binder}}, \bibinfo {author} {\bibfnamefont
  {E.}~\bibnamefont {Mönch}}, \bibinfo {author} {\bibfnamefont
  {J.}~\bibnamefont {Min{\'{a}}r}}, \bibinfo {author} {\bibfnamefont
  {M.}~\bibnamefont {Kronseder}}, \bibinfo {author} {\bibfnamefont {C.~H.}\
  \bibnamefont {Back}}, \bibinfo {author} {\bibfnamefont {D.}~\bibnamefont
  {Bougeard}}, \ and\ \bibinfo {author} {\bibfnamefont {S.~D.}\ \bibnamefont
  {Ganichev}},\ }\href {\doibase 10.1103/physrevapplied.16.064030} {\bibfield
  {journal} {\bibinfo  {journal} {Phys. Rev. Applied}\ }\textbf {\bibinfo
  {volume} {16}},\ \bibinfo {pages} {064030} (\bibinfo {year}
  {2021})}\BibitemShut {NoStop}%
\bibitem [{\citenamefont {Belinicher}\ and\ \citenamefont
  {Sturman}(1980)}]{Belinicher1980}%
  \BibitemOpen
  \bibfield  {author} {\bibinfo {author} {\bibfnamefont {V.~I.}\ \bibnamefont
  {Belinicher}}\ and\ \bibinfo {author} {\bibfnamefont {B.~I.}\ \bibnamefont
  {Sturman}},\ }\href@noop {} {\bibfield  {journal} {\bibinfo  {journal} {Phys.
  Usp.}\ }\textbf {\bibinfo {volume} {23}},\ \bibinfo {pages} {199} (\bibinfo
  {year} {1980})},\ \bibinfo {note} {[\textit{Usp. Fiz. Nauk} \textbf{1980},
  \textit{130}, 415]}\BibitemShut {NoStop}%
\bibitem [{\citenamefont {Weber}\ \emph {et~al.}(2008)\citenamefont {Weber},
  \citenamefont {Golub}, \citenamefont {Danilov}, \citenamefont {Karch},
  \citenamefont {Reitmaier}, \citenamefont {Wittmann}, \citenamefont {Bel'kov},
  \citenamefont {Ivchenko}, \citenamefont {Kvon}, \citenamefont {Vinh},
  \citenamefont {van~der Meer}, \citenamefont {Murdin},\ and\ \citenamefont
  {Ganichev}}]{Weber2008}%
  \BibitemOpen
  \bibfield  {author} {\bibinfo {author} {\bibfnamefont {W.}~\bibnamefont
  {Weber}}, \bibinfo {author} {\bibfnamefont {L.~E.}\ \bibnamefont {Golub}},
  \bibinfo {author} {\bibfnamefont {S.~N.}\ \bibnamefont {Danilov}}, \bibinfo
  {author} {\bibfnamefont {J.}~\bibnamefont {Karch}}, \bibinfo {author}
  {\bibfnamefont {C.}~\bibnamefont {Reitmaier}}, \bibinfo {author}
  {\bibfnamefont {B.}~\bibnamefont {Wittmann}}, \bibinfo {author}
  {\bibfnamefont {V.~V.}\ \bibnamefont {Bel'kov}}, \bibinfo {author}
  {\bibfnamefont {E.~L.}\ \bibnamefont {Ivchenko}}, \bibinfo {author}
  {\bibfnamefont {Z.~D.}\ \bibnamefont {Kvon}}, \bibinfo {author}
  {\bibfnamefont {N.~Q.}\ \bibnamefont {Vinh}}, \bibinfo {author}
  {\bibfnamefont {A.~F.~G.}\ \bibnamefont {van~der Meer}}, \bibinfo {author}
  {\bibfnamefont {B.}~\bibnamefont {Murdin}}, \ and\ \bibinfo {author}
  {\bibfnamefont {S.~D.}\ \bibnamefont {Ganichev}},\ }\href {\doibase
  10.1103/PhysRevB.77.245304} {\bibfield  {journal} {\bibinfo  {journal} {Phys.
  Rev. B}\ }\textbf {\bibinfo {volume} {77}},\ \bibinfo {pages} {245304}
  (\bibinfo {year} {2008})}\BibitemShut {NoStop}%
\bibitem [{\citenamefont {Otteneder}\ \emph {et~al.}(2020)\citenamefont
  {Otteneder}, \citenamefont {Hubmann}, \citenamefont {Lu}, \citenamefont
  {Kozlov}, \citenamefont {Golub}, \citenamefont {Watanabe}, \citenamefont
  {Taniguchi}, \citenamefont {Efetov},\ and\ \citenamefont
  {Ganichev}}]{Otteneder2020}%
  \BibitemOpen
  \bibfield  {author} {\bibinfo {author} {\bibfnamefont {M.}~\bibnamefont
  {Otteneder}}, \bibinfo {author} {\bibfnamefont {S.}~\bibnamefont {Hubmann}},
  \bibinfo {author} {\bibfnamefont {X.}~\bibnamefont {Lu}}, \bibinfo {author}
  {\bibfnamefont {D.~A.}\ \bibnamefont {Kozlov}}, \bibinfo {author}
  {\bibfnamefont {L.~E.}\ \bibnamefont {Golub}}, \bibinfo {author}
  {\bibfnamefont {K.}~\bibnamefont {Watanabe}}, \bibinfo {author}
  {\bibfnamefont {T.}~\bibnamefont {Taniguchi}}, \bibinfo {author}
  {\bibfnamefont {D.~K.}\ \bibnamefont {Efetov}}, \ and\ \bibinfo {author}
  {\bibfnamefont {S.~D.}\ \bibnamefont {Ganichev}},\ }\href {\doibase
  10.1021/acs.nanolett.0c02474} {\bibfield  {journal} {\bibinfo  {journal}
  {Nano Lett.}\ }\textbf {\bibinfo {volume} {20}},\ \bibinfo {pages} {7152}
  (\bibinfo {year} {2020})}\BibitemShut {NoStop}%
\bibitem [{\citenamefont {Glazov}\ and\ \citenamefont
  {Golub}(2020)}]{Glazov2020}%
  \BibitemOpen
  \bibfield  {author} {\bibinfo {author} {\bibfnamefont {M.~M.}\ \bibnamefont
  {Glazov}}\ and\ \bibinfo {author} {\bibfnamefont {L.~E.}\ \bibnamefont
  {Golub}},\ }\href {\doibase 10.1103/PhysRevB.102.155302} {\bibfield
  {journal} {\bibinfo  {journal} {Phys. Rev. B}\ }\textbf {\bibinfo {volume}
  {102}},\ \bibinfo {pages} {155302} (\bibinfo {year} {2020})}\BibitemShut
  {NoStop}%
\bibitem [{Note4()}]{Note4}%
  \BibitemOpen
  \bibinfo {note} {Note that the traces were obtained from two separated arrays
  with identical parameters, which, however, may have slightly different
  characteristics, see Fig.\protect \tmspace +\thinmuskip {.1667em}\ref {figA1}
  in Appendix\protect \tmspace +\thinmuskip {.1667em}\ref
  {appendixA}.}\BibitemShut {Stop}%
\end{thebibliography}%
\end{document}